\documentclass{aa}  
\usepackage{rotating}
\usepackage{amssymb}
\usepackage{graphicx}
\usepackage{txfonts}
\usepackage{lscape}
\usepackage[bf,nooneline]{subfigure}
\bibpunct{(}{)}{;}{a}{}{,}

\def\lae{\mathrel{<\kern-1.0em\lower0.9ex\hbox{$\sim$}}}
\def\gae{\mathrel{>\kern-1.0em\lower0.9ex\hbox{$\sim$}}}

\begin{document} 

   \titlerunning {Field versus cluster ellipticals at $z\sim1.3$}
   \authorrunning {Saracco et al. }

\title{Cluster and field elliptical galaxies at $z\sim1.3$:}
\subtitle{The marginal role of the environment and the relevance of the 
 galaxy central regions}

   \author{P. Saracco\inst{1}\thanks{E-mail: paolo.saracco@brera.inaf.it.},  
A. Gargiulo$^{2}$, F. Ciocca$^{1,3}$, D. Marchesini$^{4}$
}

 \institute{INAF - Osservatorio Astronomico di Brera, Via Brera 28, 20121 Milano, Italy
\and INAF -  Istituto di Astrofisica Spaziale e Fisica Cosmica (IASF), 
Via E. Bassini 15, 20133 Milano, Italy 
\and Universit\'a degli Studi dell'Insubria, via Valleggio 11, 22100 Como,
Italy
\and Department of Physics and Astronomy, Tufts University, Medford, MA 02155, USA 
 }

   \date{Received 2016 May 9; accepted 2016 September 14}

% \abstract{}{}{}{}{} 
% 5 {} token are mandatory
 
  \abstract
  % context heading (optional)
  % {} leave it empty if necessary  
 %  {It is still matter of debate if the environment can affect the properties 
 %of galaxies at a given morphology.
%Recent simulations predict a clear dependence of the structure of 
%bulge-dominated galaxies with the environment where they belong. 
%In this paper we investigate in a coherent and homogeneous way the possible
%environmental effects on the population of elliptical galaxies at $z\sim1.3$
%and on their properties.
%} 
  % aims heading (mandatory)
  {}
   {The aim of this work is twofold: first, to assess whether the population of 
   elliptical galaxies in cluster at $z\sim1.3$ differs from the population in the field 
   and whether their intrinsic structure depends on the environment 
   where they belong; second, to constrain their properties 9 Gyr back 
   in time through the study of their scaling relations.}
    %methods heading (mandatory)
   {We compared a sample of 56 cluster elliptical galaxies selected 
from three clusters at $1.2<z<1.4$ with elliptical 
galaxies selected at comparable redshift in the GOODS-South field ($\sim30$), 
in the COSMOS area ($\sim180$), 
and in the CANDELS fields ($\sim220$). 
To single out the environmental effects, we selected cluster and field 
elliptical galaxies according to their morphology.
We compared physical and structural parameters of galaxies in the two 
environments and we derived the relationships between effective radius, surface
brightness, stellar mass, and stellar mass density $\Sigma_{R_e}$ 
within the effective radius and central mass density $\Sigma_{1kpc}$ , within 
1 kpc radius.}
  % results heading (mandatory)
   {We find that the structure and the properties of cluster elliptical galaxies 
   do not differ from those in the field: they are characterized by the same 
   structural parameters at fixed mass and they follow the same scaling relations.
   On the other hand, the population of field elliptical galaxies at $z\sim1.3$ 
   shows a significant lack of massive ($\mathcal{M}_*> 2 \times 10^{11}$ M$_\odot$)
    and large 
   ($R_e > 4-5$ kpc) elliptical galaxies with respect to the cluster. 
   Nonetheless, at $\mathcal{M}_*< 2 \times 10^{11}$ M$_\odot$, the two
   populations are similar. 
   The size-mass relation of cluster and field ellipticals at $z\sim1.3$ 
   clearly defines two different regimes, 
   above and below a transition mass $m_t\simeq2-3\times10^{10}$ M$_\odot$: 
   at lower masses the
   relation is nearly flat (R$_e\propto\mathcal{M}_*^{-0.1\pm0.2}$), the mean 
   radius is nearly constant at $\sim1$ kpc and, consequenly,
   $\Sigma_{R_e}\simeq\Sigma_{1kpc}$ while, at larger masses, the relation is 
   R$_e\propto\mathcal{M}_*^{0.64\pm0.09}$. 
   The transition mass marks the mass at which galaxies reach the maximum 
   stellar mass density. 
   Also the $\Sigma_{1kpc}$-mass relation follows two different regimes, above and below the 
   transition mass ($\Sigma_{1kpc}\propto{\mathcal{M}_*}^{0.64\ >m_t}_{1.07\ <m_t}$)
   defining a transition mass density $\Sigma_{1kpc}\simeq2-3\times10^3$ M$_\odot$ 
   pc$^{-2}$. 
   The effective stellar mass density $\Sigma_{R_e}$ does not correlate with mass; 
   dense/compact galaxies can be assembled over a wide mass regime, independently 
   of the environment. 
   %%%At small radii and/or for high stellar mass densities, elliptical galaxies at 
   %%%$z\sim1.3$ describe 
   %%%the same Zone of Exclusion described by the early-type galaxies in the local universe.
   The central stellar mass density, $\Sigma_{1kpc}$, 
   besides being correlated with the mass, is correlated to the 
   age of the stellar population: the higher the central stellar mass density, 
   the higher the mass, the older the age of the stellar population. 
   }
  % conclusions heading (optional), leave it empty if necessary 
   {While we found some evidence of environmental effects on the
elliptical galaxies as a population, we did not find differences between the 
intrinsic properties of  cluster and field elliptical galaxies at comparable 
redshift.
The structure and the shaping of elliptical galaxies
   at $z\sim1.3$ do not depend on the environment. However, a dense environment
   seems to be more efficient in assembling high-mass large ellipticals,
   much rarer in the field at this redshift. The correlation found between the
   central stellar mass density and the age of the galaxies beside the mass
   suggests a close connection of the central regions to the earliest 
   phases of formation.
%    the central regions retain memory of the initial
%   conditions, the outer regions of the subsequent story of the galaxy.
}
   \maketitle
%
%________________________________________________________________

\section{Introduction}
The existence of correlations among
some properties of the population of galaxies and the environment 
in which they reside is well established. 
%The environment in which a galaxy lives seems indeed to affect its 
%evolution and possibly its formation.
The composition of the population of galaxies, that is, its 
morphological mix, is different according to the environment where 
the population belongs. 
A clear example is the well-known morphology-density relationship, 
according to which early-type galaxies, originally classified as 
elliptical and lenticular galaxies, preferentially populate high-density 
environments and vice versa \citep{oemler74,dressler80,postman84}. 
This environmental effect has been confirmed both in the
local \citep[e.g.,][]{tran01, goto03,holden07,bamford09} 
and intermediate redshift Universe \citep[e.g.,][]{fasano00,treu03,smith05,
vanderwel07,pannella09,tasca09,tanaka12}.
In spite of the many observations supporting the above evidence,
the mechanisms responsible for this morphological segregation are still debated.

Galaxies in different environments can undergo different physical 
processes.
For instance, contrary to field galaxies, cluster galaxies are affected 
by the dense and hot intracluster medium. 
The ram pressure can overcome the gravitational forces 
keeping the gas anchored to the potential well, at least of the less massive 
galaxies, removing their gas and quenching their star formation.
Actually, the quenching efficiency seems to be higher in denser environments
\citep[e.g.,][]{haines13, vulcani15}.
This mechanism affects galaxies in a different way according to
their mass and shape
\citep[see e.g.,][for recent reviews]{boselli06, boselli14}.

%The environment seems also to affect the star formation timescale, 
%that appears shorter for galaxies in dense environments than in low-density 
%fields
%\citep[e.g.][]{thomas05,rettura10,tanaka13}.

Many observations suggest that the formation epoch of galaxies
depends mainly on their mass but, in the local universe, there is some
evidence that cluster early-type galaxies form earlier than field 
galaxies of the same mass 
\citep[e.g.,][]{kuntschner02,gebhardt03,thomas05}
and that the environment can play an important role in the late phases
of their evolution \citep{thomas10}.
On the other hand, some works seem to conclude that while galaxy mass 
regulates the timing of galaxy  formation, the environment regulates the 
timescale of their star formation histories \citep[e.g.,][]{tanaka10,rettura11}
and  controls the fraction of star-forming galaxies \citep[e.g.,][]{muzzin12}.

Actually, the star formation timescale appears shorter for galaxies in 
dense environments than for those in low-density fields
\citep[e.g.,][]{thomas05,rettura11,tanaka13}.
At intermediate redshift ($z\sim0.8$), however, the effect of the environment
becomes less evident and it seems to vanish, 
with cluster and field early-type galaxies characterized by similar 
stellar population properties \citep[e.g.,][]{lonoce14} following the same 
color evolution and scaling relations \citep[see][for a review]{renzini06}.  
Actually, 
it is not clear whether and when the environment affects the properties and 
hence the evolution of galaxies at a given morphology.

From the theoretical point of view, it is expected that field
and cluster bulge-dominated galaxies display different structures and
hence follow also different scaling relations.
Bulge-dominated galaxies should result from a sequence of major 
and minor mergers  through which most of their stellar mass is assembled
\citep[e.g.,][]{delucia06, khochfar11, shankar13}.
Minor mergers are considered very efficient in increasing the size 
of galaxies \citep[e.g.,][]{naab09,vandokkum10a}.
Since mergers are expected to be more frequent in denser environments,
they should produce larger galaxies than similarly massive
counterparts in the field \citep[e.g.,][]{shankar13}.  
Some recent simulations actually predict a clear environmental dependence
of the structure of bulge-dominated galaxies with their median size 
larger by a factor 1.5-3, moving from low to high-mass halos 
\citep{shankar14b}.

From an observational standpoint, many recent studies focused on the environmental 
dependence of the mass-size relation of early-type galaxies.
In the local universe, some works point toward the absence of an
environmental dependence for this relation 
\cite[e.g.,][]{guo09, weinmann09,huertas13b, shankar14b} while other studies 
suggest that cluster early-type galaxies are slightly smaller than their field
counterparts \citep[e.g.,][]{valentinuzzi10, poggianti13a}.  
Few studies at intermediate redshift point toward the
absence of environmental effect on the size distribution of early-type
galaxies, either morphologically- spectroscopically-, or color-selected
\citep[e.g.,][]{maltby10,rettura10,kelkar15}.
At higher redshift there are rather controversial results.
For instance, while \cite{raichoor12} find that morphologically selected
early-type galaxies in cluster at $z\sim1.2$ are more compact than in the 
field, the opposite is found by \cite{cooper12} at similar redshift 
and by \cite{papovich12} at slightly higher redshift, both works based
on different selection criteria and data quality.

Different selection criteria, different redshift ranges, different
quality of the data, and hence different accuracy in the derivation 
of the structural and physical parameters (size, stellar mass, age) 
of galaxies may be the reasons for, at least,  some of the above discrepancies.
A critical issue in this kind of analysis is, indeed, the morphological 
selection of galaxies instead of using selection criteria related to the
stellar population properties such as colors and/or star formation.
Since stellar population and structural evolution do not appear synchronous, 
criteria based on stellar population properties select galaxies with different 
morphological mixes at different redshift and in different environments.
A clear example is given by the significant different morphological mix 
observed in the red sequence population of cluster galaxies, 
largely populated by red disc-dominated (passive) galaxies at 
$z\sim1,$ and by ellipticals and lenticular in the local universe
\citep[e.g.,][]{depropris15, mei09,moran07}.
Analogously, the selection of passive galaxies based on color-color diagnostic
plots or on their low specific star formation rate (sSFR) produces
samples with a different mix of morphological types.
For instance, the fraction of elliptical galaxies in the passive galaxy
population is found to significantly change with mass at a given redshift
and in redshift at fixed mass \citep[e.g.,][]{moresco13, tamburri14, 
huertas13, huertas15} and consequently the mean properties of the 
sample vary \citep[e.g.,][]{bernardi10,mei12}.
The different composition of the samples thus selected 
%at different redshift, 
%in different environments and/or at different masses 
prevents the singling out of
possible environmental effects on the properties of a given morphological type.   

In this paper we aim to study in a coherent and homogeneous way the 
dependence of the population of elliptical galaxies, and of their properties, 
on the environment.
Here, we focus our attention on cluster and field elliptical galaxies at 
$z\sim1.3$, while we refer to a forthcoming paper for the environmental
effects on their  evolution.
We study a sample of 56 cluster elliptical galaxies selected 
in the three clusters: XMMJ2235-2557 at $z=1.39$ 
\citep{rosati09}, RDCS J0848+4453 at $z=1.27$ 
\citep{stanford97}, and XLSS-J0223-0436 
at $z=1.22$ \citep{andreon05,bremer06}. We compare their properties 
with those of a sample of 31 field elliptical galaxies 
selected in the GOODS-South field 
according to the same criteria. 
When possible, we make use also of a larger sample of
about 180 elliptical galaxies selected in the same way from the COSMOS catalog
at slightly lower redshift,  and of a sample of about 220 ellipticals
selected from CANDELS.
To single out the effect of the environment, we have tried to minimize all the
sources of uncertainty discussed above:
we selected galaxies in a narrow redshift range, $1.2<z<1.4$, to avoid
significant evolutionary effects;
we selected cluster and field elliptical galaxies on the basis 
of their morphology to compare samples with the same composition 
in the two environments;
we derived morphology and structural parameters from Hubble Space Telescope (HST) 
images at the same wavelength (with the exception of CANDELS).
Finally, the same wavelength coverage for all the galaxies allowed us to derive
their physical parameters (stellar mass, age) with the same
degree of uncertainty.

In Sec. 2 we describe the data and the samples.
In Sec. 3 we derive the structural (effective radius, surface brightness)
and the physical (stellar mass, absolute magnitude, and age)  parameters
for our galaxies.
In Sec. 4 we compare the population of cluster elliptical galaxies and
their properties with those in the field.
In Sec. 5 we derive the Kormendy relation of cluster and field ellipticals 
at $z\sim1.3$ while, in Sec. 6, we derive the size-mass relation.
Section 7 is focused on the stellar mass density of galaxies.
In Sec. 8, we summarize our results and present our conclusions.
Appendix \ref{ap:scaling} summarizes the best fitting relations reported
in the text obtained with the least squares method and reports also
those obtained using the orthogonal regression. 

Throughout this paper we use a standard cosmology with
$H_0=70$ Km s$^{-1}$ Mpc$^{-1}$, $\Omega_m=0.3$, and $\Omega_\Lambda=0.7$.
All the magnitudes are in the Vega system, unless otherwise specified.

%__________________________________________________________________

\section{Data description and samples' definition}
The samples of galaxies used are covered by multiwavelength data in the 
range $0.38-8.0$ $\mu$m obtained from the Hubble Space Telescope (HST) 
and Spitzer archival data, ground-based ESO-VLT archival data
and with observations obtained at the Large Binocular Telescope (LBT).

%The samples are described in detail in 
%Ciocca et al. (2016, in prep.), \cite{saracco14} and 
%Saracco et al. (2016, in prep) for what concern the samples
%of elliptical galaxies in the clusters XMMJ2235-2557, RDCSJ0848+4453 and 
%XLSSJ0223-0436 respectively while in \cite{tamburri14} for what concern
%the sample of field ellipticals selected in the GOODS-South region. 

%Below, we summarize the data used in this analysis.

\subsection{Cluster sample} 
Cluster elliptical galaxies have been selected in the three clusters 
XMMJ2235-2557 at $z=1.39$ \citep{rosati09}, RDCS J0848+4453 at $z=1.27$ 
\citep{stanford97}, and XLSS-J0223-0436 at $z=1.22$ \citep{andreon05,bremer06}
according to the criteria described in \cite{saracco14}.
Briefly, using Sextractor \citep{bertin96}, we first detected on the ACS-F850LP 
image all the sources in the region surrounding each cluster. 
Then, we selected all the galaxies brighter than $z_{850}<24$ 
within a projected radius D$\le1$ Mpc from the cluster center.
The completeness at this magnitude limit is 100\% in all the three
cluster fields.
According to the $i_{775}-z_{850}$ color that traces well
the redshift of galaxies, we selected all the galaxies within $\pm0.2$ from 
the peak in the distribution centered at the mean color 
(typically $\langle i_{775}-z_{850}\rangle\sim 1.1$ at $z\sim1.3$)
of the cluster members spectroscopically confirmed 
\citep[see e.g., Fig. 1 in][for cluster RDCS J0848+4453 as example]{saracco14}.
%Then, according to the $i_{775}-z_{850}$ color distribution,  
%we selected those galaxies having a color within $\sim\pm0.2$ mag
% from the color of the cluster members spectroscopically
%confirmed, tipically $\langle i_{775}-z_{850}\rangle\sim 1.1$ at $z\sim1.3$,
%and defining the narow peak centered at this color.
Of these galaxies, we selected only those we classified as elliptical:
17 ellipticals in the cluster XMMJ2235-2557, 
16 ellipticals in the cluster RDCSJ0848+4453,  and 23 ellipticals
in the cluster XLSSJ0223-0436, summing up to 56 
cluster ellipticals  in the redshift range $1.2<z<1.4$.

Spectroscopic redshift are available for $\sim$60\% of the selected cluster
elliptical galaxies.
For the remaining 40\%, the photometric redshift we estimated is consistent
with the redshift of the cluster. 
{  The photometric redshift accuracy is $\sigma_{\Delta z/(1+z_s)}=0.04$
(0.02 using the normalized median absolute deviation).
The comparison between photometric and spectroscopic redshift
for the galaxies'
spectroscopically confirmed cluster members is shown in Appendix \ref{ap:cluster}.}
Using the sample of field elliptical galaxies
in the GOODS-S region as control sample, we expect about six galaxies 
out of the 56 to be non cluster members. 
Hence, the analysis presented in this work is not affected by
 uncertainties related to cluster membership.
The data available and used for each cluster are summarized in \ref{ap:cluster}.
\subsection{Field sample} 
{GOODS-South:}  Our main sample of field ellipticals is composed of 31 
ellipticals at $1.2<z<1.45$ (27 at $1.2<z<1.4$)  in the GOODS-South field.
They have been extracted from the sample of \cite{tamburri14} who
morphologically classified the 1302 galaxies
brighter than K(AB)=22 in that field, identifying 247 ellipticals in the 
redshift range 0.1-2.5.
The completeness at this K-band magnitude limit is 100\% and assures the
same completeness to a magnitude  F850LP fainter than 24. 
There are 31 elliptical galaxies with F850LP$\le24$ in the redshift range 
$1.2<z<1.45$. 
%The completeness at this K-band magnitude limit is 100\% and assures the
%same completeness to a magnitude  F850LP fainter than 24. 
Spectroscopic redshifts are available for 22 ($\sim70\%$) of them while
the remaining nine galaxies have photometric redshift as derived by 
\cite{santini09} and independently checked by \cite{tamburri14}.

For this sample, the multiwavelength data used is 
composed of the deep optical HST-Advanced Camera for Surveys (ACS) observations 
in the filters F435W, F606W, F775W, and F850LP ($>$40000 s) described in 
\cite{giavalisco04}, VLT U, J, H, and Ks bands and Spitzer Infrared Array Camera 
(IRAC) observations described in \cite{grazian06} and in \cite{santini09}.
A detailed description of this data set and of the sample of field
ellipticals is provided by \cite{tamburri14}.

{COSMOS:}  The sample of field ellipticals we selected in the COSMOS field 
is in the redshift range $1.0<z<1.2$, slightly lower than the redshift of
cluster and field samples for the reasons below.
The shallower HST-ACS observations \citep[$\sim$ 2000 s,][]{koekemoer07}
of this field and the incompleteness affecting the catalogs from which we extracted 
the sample, prevented us from using it in all the comparisons.
We discuss the different cuts applied to this sample comparison by comparison. 

To construct this sample, we crossmatched the catalog of \cite{davies15} 
including all the available spectroscopic redshifts in the COSMOS area, with
the catalog of structural parameters based on the HST-ACS images in the F814W 
filter of \cite{scarlata07}.
The crossmatch produced a sample of $\sim$110100 galaxies in the magnitude 
range $19<F814W_{AB}<24.8$ over the 1.6 deg$^2$ of the HST-ACS COSMOS field.
We then selected all the galaxies (3425) brighter than F814W$_{AB}<23.5$ mag, 
%for which structural parameters and morphological classification are still reliable 
%\citep{sargent07}, and 
with elliptical morphological classification 
\citep[parameter Type=1 in the][catalog]{scarlata07}.
However, of these 3425 elliptical galaxies, $\sim1000$ do not have a reliable
morphological classification and effective radius measurement 
(parameter R\_GIM2D$<0$).  
Actually, in this catalog, effective radius and morphological classification 
are given and considered reliable for values of R$_e$ larger than 
$\sim$0.17 arcsec. 
In the redshift range $1.0<z<1.2$ , there are 178 elliptical galaxies with
reliable effective radius and morphological classification, 
20\% of which have spectroscopic redshift.
In the redshift range $1.2<z<1.45$,
there are 60 elliptical galaxies, all of them without spectroscopic redshift and 
reliable measurements.
For these reasons, for our purposes, we have considered  
the sample of ellipticals at $1.0<z<1.2$.
We discuss the various limits and biases of this sample when we use 
it in the comparisons if necessary.    

{  {CANDELS:}  This sample is composed of 224 elliptical galaxies in the redshift
range $1.2<z<1.4$.
Even if the morphology and structural parameters of these galaxies 
are derived in the F160W band instead of the F850LP band, we considered 
also this large data set since it can improve the statistic and reduce 
the effect of the cosmic variance.
To homogenize the CANDELS data to our data, we compared the structural 
and the physical parameters making use of the elliptical
galaxies in common selected in the GOODS-South field.
The comparison is shown in Appendix \ref{ap:candels}.

To construct this sample we started from the master catalog available at the 
Rainbow\footnote{https://rainbowx.fis.ucm.es/Rainbow\_navigator\_public/.}
database that includes the multiwavelength data of the 5 CANDELS and 3D-HST 
fields (GOODS-S, GOODS-N, COSMOS, UDS, and EGS) from 
\cite{koekemoer11, grogin11, guo13, skelton14},
and \cite{brammer12} summing up  about 207000 sources.
Photometric redshifts and stellar parameters are those presented in \cite{dahlen13} and
\cite{santini15}.
The structural parameters are those presented in \cite{vanderwel12}, while
the morphological classification is described in \cite{huertas15}.

We first selected in each CANDELS field all the galaxies in the  redshift range $1.2<z<1.4$.
We considered the photometric redshift when the spectroscopic one was not available. 
We then selected those galaxies for which GALFIT provided a good fit to their 
surface brightness profile in the F160W band \citep[parameter gfit\_f\_h=0;][]{vanderwel12}
and for which the morphological classification pointed to a spheroid, according to the 
dominant class parameter \citep[visualHCPG\_dom\_class=0;][]{huertas15}.
Finally,  according to the F160W-band magnitude distribution of our control sample of 
galaxies in the GOODS-S field, we selected those galaxies brighter than $F160W_{AB}<22.5$. 
The resulting sample is composed of 224 elliptical galaxies, a negligible fraction ($<5\%$) 
of which with spectroscopic redshift.
}

\section{Morphology, surface brightness profile, and Spectral Energy Distribution fitting}
Morphology and structural parameters have been derived homogeneously on 
the ACS-F850LP image both for field and cluster galaxies, with the exception
of the COSMOS and CANDELS samples.
The morphological classification is independent of the \S'ersic index and
it is solely based on the visual inspection of the images of the galaxies
and on the inspection of the residuals of the fitting to their luminosity profile.
A galaxy is classified as elliptical/spheroidal if it has a regular shape 
with no hint of disc on the F850LP image and no irregular or structured 
residuals resulting from the profile fitting with a single \S'ersic component.
{  In Appendix \ref{ap:samples} (Fig. \ref{fig:thumb}) we show the F850LP images 
for a representative sample of
elliptical galaxies in the field and cluster samples.}

The effective radius R$_e$ [kpc] ($r_e$ [arcsec]) has been derived by 
fitting a S\'ersic profile, 
\begin{equation}
I(R)=I_e exp\left[-b_n\left[\left({R\over R_e}\right)^{1/n}-1\right]\right]
\label{sersic}
\end{equation}
to the observed light profile in the ACS-F850LP image 
($\lambda_{rest}\simeq4000$ \AA\ at $z\sim1.3$), both for field
and for cluster ellipticals.
The two-dimensional fitting was performed using \texttt{Galfit} 
software \citep[v. 3.0.4,][]{peng02}.
In the following analysis, we always refer to the quantities resulting from 
the \S'ersic profile fitting (Eq. \ref{sersic}) unless explicitly
stated otherwise, since the use of a de Vaucouleurs profile ($n=4$) 
introduces a significant bias on the structural parameters that 
depends on the intrinsic \S'ersic index of the galaxy 
\citep[e.g.,][]{donofrio08, taylor10, raichoor12}.
The effective radius has been derived as $r_e=a_e\sqrt{b/a}$, 
where $a_e$ is the semi-major axis of the projected elliptical 
isophote  containing half of the total light provided by \texttt{Galfit,} 
and $b/a$ is the axial ratio.
For all the galaxies, the fit to the surface brightness profile extends
over more than five magnitudes and, apart from the largest cluster 
galaxies, up to $>3$R$_e$.

The morphological classification of the COSMOS sample is based on a principal 
component analysis of the F814W images optimized through a visual inspection of 
all the galaxies analyzed \citep{scarlata07}.
The structural parameters of the COSMOS sample are based on
a \S'ersic profile fitting as in our analysis even if performed
with \texttt{GIM2D} instead of \texttt{Galfit}, as described in \cite{sargent07}.
\cite{damjanov15} find a good agreement between the effective radius
provided by the two methods, \texttt{GIM2D} and \texttt{Galfit}, when based 
on a single \S'ersic best fitting component, as in our case.

{  The morphological classification of the CANDELS sample is a visual-like 
classification performed on the WFC3-F160W images for sources brighter than 
F160W$_{AB}<24.5$ mag. 
It is based on the deep learning method  described in \cite{huertas15} 
calibrated on the pure visual morphological classification performed on F160W-band 
images for a smaller sample by \cite{kartaltepe15}.
The structural parameters are based on a \S'ersic profile fitting performed with 
\texttt{Galfit} on the WFC3-F160W images as described in \cite{vanderwel12}.
According to the comparison between the effective radii in the F850LP band 
and those in the F160W band for the 31 galaxies in common in the GOODS-S field 
(see appendix \ref{ap:candels}), we did not apply any scaling to the 
effective radii of the CANDELS sample.
}

Stellar mass $\mathcal{M}_*$, B-band absolute magnitude $M_B$ , and
mean age of the stellar populations were derived 
by fitting with the software \texttt{hyperz} \citep{bolzonella00}, 
the observed spectral energy distribution (SED) of each 
galaxy at its redshift with Bruzual and Charlot models 
\citep[][BC03]{bruzual03}.
In the fitting, we adopted the Chabrier \citep{chabrier03} initial mass function 
(IMF), four exponentially declining star formation histories (SFHs) $e^{-t/\tau}$ 
with e-folding time $\tau$= [0.1, 0.3, 0.4, 0.6] Gyr and solar metallicity Z$_\odot$. 
Extinction A$_V$ has been considered and treated as a free parameter
in the fitting, allowing it to vary in the range $0<A_V<0.6$ mag. 
We adopted the extinction curve of \cite{calzetti00}.
Dealing with properties related to the light profile of galaxies,
%among the different stellar masses provided by the BC03 models,
we considered the mass $\mathcal{M}_*$ of the BC03 models,
the net mass in stars at the age of their observation after having 
returned the gas to the interstellar medium (ISM). 

{  Since both the field and cluster samples are magnitude-limited (F850$_{lim}<24$ mag)
and selected in the same narrow redshift range,
the minimum stellar mass at which the sample is complete depends on the M/L ratio.
According to the method used by \cite{pozzetti10}, we estimated for each galaxy
the limiting mass log($\mathcal{M}_{lim}$)=log($\mathcal{M}_*$)+0.4(F850-F850$_{lim}$)
that a galaxy would have if its F850LP magnitude was equal to the limiting magnitude.
Then, considering the distribution of the values of $\mathcal{M}_{lim}$ , we
considered as minimum mass $\mathcal{M}_{min}\simeq8\times10^9$ M$_\odot$,
the mass above which 95\% of them lie \citep[see also][]{saracco14}.
The comparisons made in the following sections take into account this mass limit
even if the number of galaxies at lower masses is negligible.}

The $M_B$  absolute magnitudes have been derived using
the observed apparent magnitude in the filter F850LP (F814W for COSMOS)
sampling $\lambda_{rest}\sim4000$ \AA\ at the redshift of the galaxies.
The color k-correction term that takes the different
filters response (F850LP vs B) into account  
was derived  from the best-fitting template.

For homogeneity with our cluster and field galaxy samples, and to account 
for the new spectroscopic redshifts in the catalog of \cite{davies15}, 
we derived the stellar masses and the other physical parameters of the 178
elliptical galaxies in the COSMOS sample by applying the same
fitting procedure described above using the multiwavelength UVISTA photometry 
of \cite{muzzin13}.

  For the CANDELS sample, we used the stellar masses provided in the 
master catalog (derived with the Chabrier IMF) and scaled by a factor of 1.12, 
as resulting from the comparison with those we derived for the 31
galaxies in common in the GOODS-S sample
(see appendix \ref{ap:candels} for the comparison).

In Tables \ref{clsample} and \ref{fisample} we list the basic properties 
of elliptical galaxies in cluster and in the GOODS-South field respectively.
For each galaxy we report right ascension and declination, 
%spectroscopic redshift,
best-fitting apparent magnitude $F850_{850}^{fit}$ and structural parameters 
$b/a$, $n$, R$_e$ [kpc], B-band surface brightness (see below), luminosity
evolution correction to $z=0$, age of the best fitting template, 
stellar mass, stellar mass within 1 kpc radius (see \S\ 7), and B-band 
absolute magnitude.

\section{Cluster and field ellipticals at $z\sim1.3$: environmental effects}
Cluster and field elliptical galaxies could be physically similar, have similar 
structures, (e.g., have the same radius at fixed mass) but follow different 
mass, age, and effective radius distributions simply because galaxies with a given 
property could be more frequent in one environment with respect to the other.
On the contrary, cluster and field ellipticals could share the same
distributions but be physically different at fixed mass, for example smaller or 
larger, denser or less dens, older or younger if belonging to one environment 
instead of to the other.
%Galaxies with a given property could be more frequent in an environment 
%with respect to the other.
%Analogously, cluster and field elliptical galaxies could share the same
%distributions but be physically different, e.g. smaller/larger (denser/less 
%dens) or older/younger at fixed mass if belonging to an environment instead
%of to the other.
In this section we address these two different kinds of environmental 
effects, trying to assess whether a population differs from the one in the 
other environment and/or whether the intrinsic properties of elliptical galaxies 
at $z\sim1.3$ depend on the environment where they belong.

\begin{figure}
\centering
%\vskip -8truecm
\includegraphics[width=9.truecm]{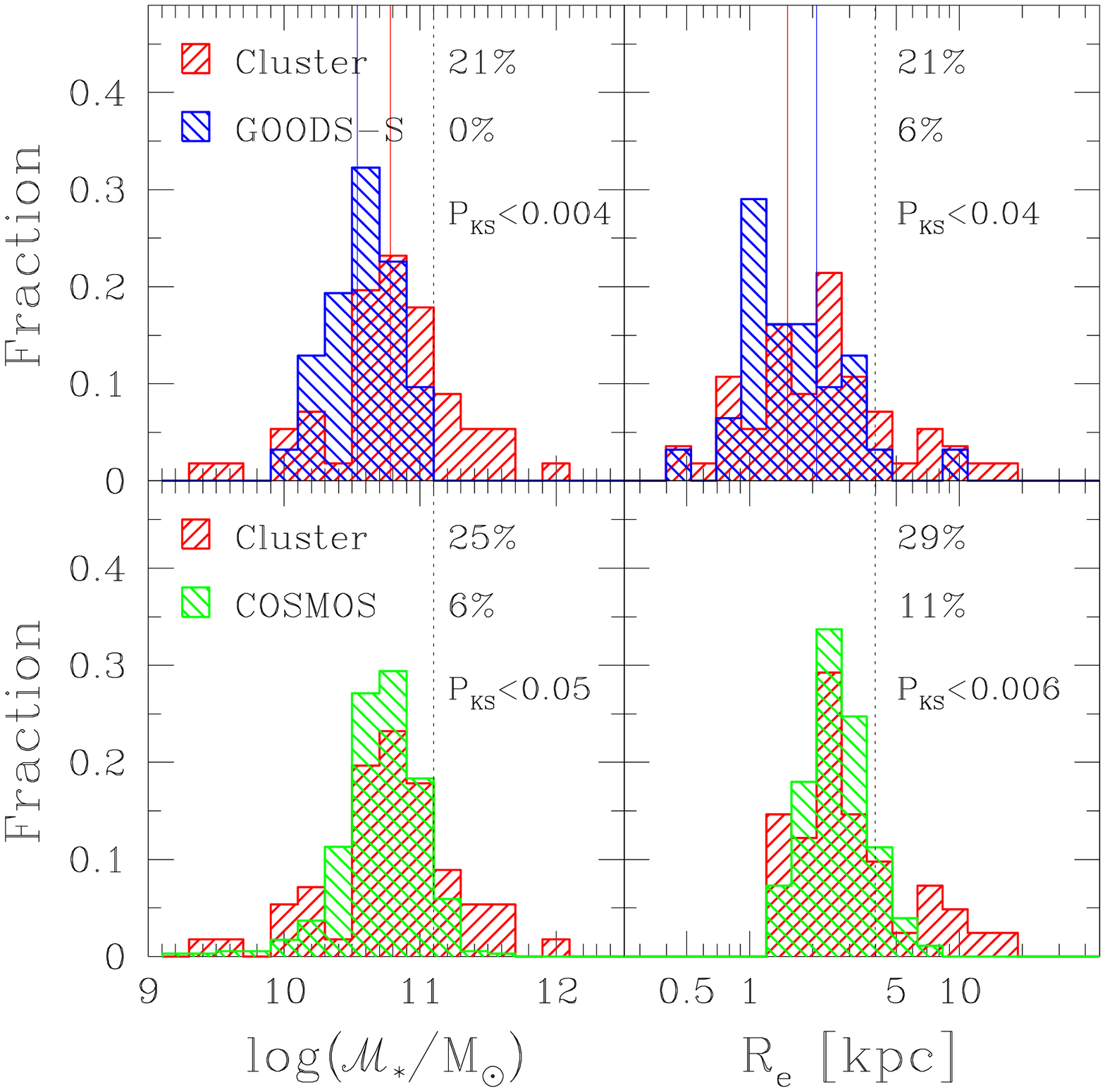}
\includegraphics[width=9.truecm]{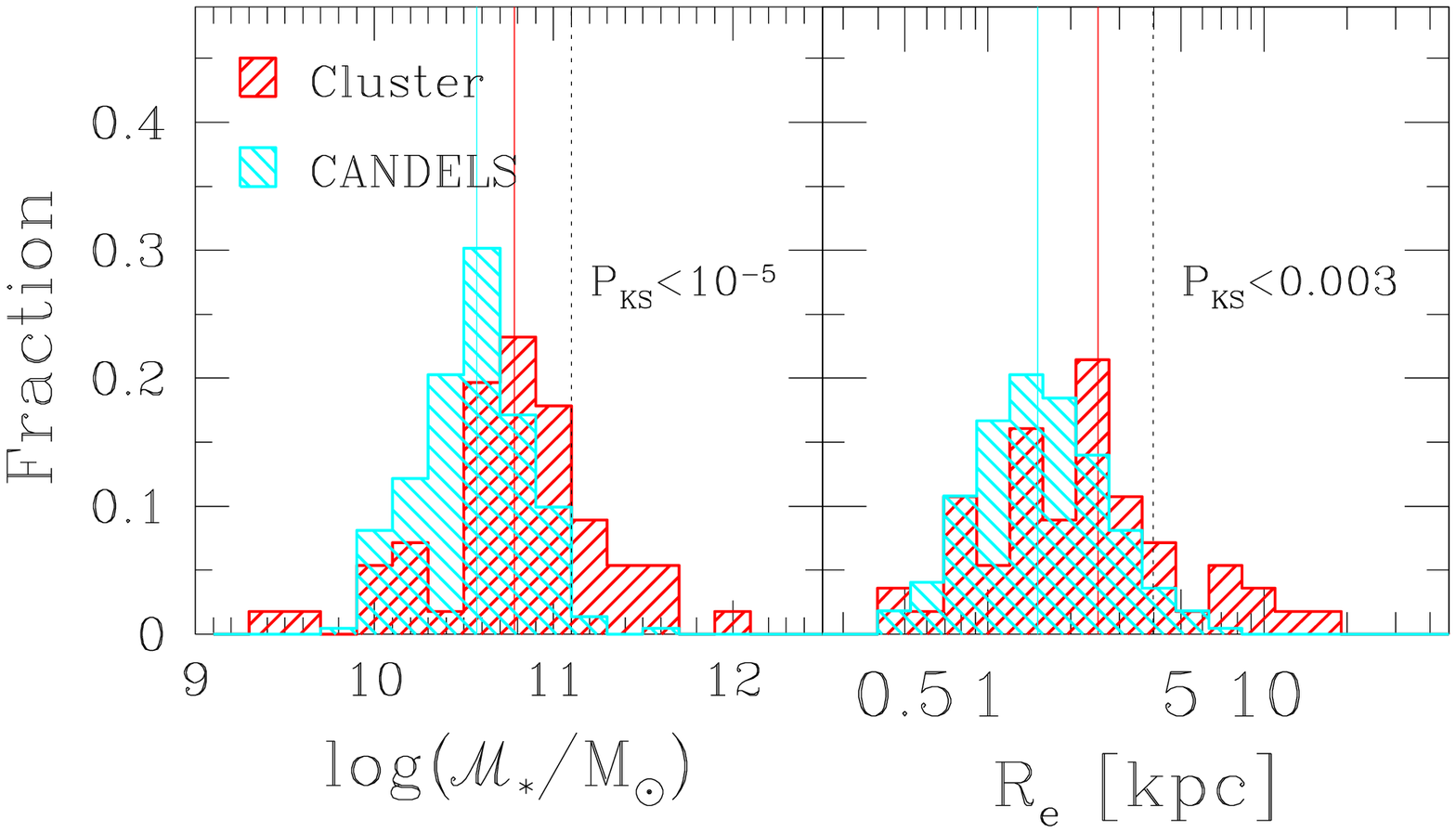}
\caption{Distribution of stellar mass and effective radius for cluster 
and field elliptical galaxies. 
Upper panel - Distribution of the stellar mass $\mathcal{M}_*$ (left) and of
the effective radius R$_e$ (right) for the 56 cluster elliptical galaxies  
(red shaded histograms) and the 31 elliptical galaxies in the GOODS-South 
field (blue shaded histograms).  
The lower panels show the comparison with the 
field elliptical galaxies selected in the COSMOS area (green histograms).  
{  The colored solid line marks the median values of the distributions
(see text).}
The dotted line marks the reference values log$\mathcal{M}_*=11.1$ $M_\odot$ 
and R$_e$=4 kpc above which the distributions are significantly different.
   Lower panel - Same as in the upper panel but in this case the distributions
of the cluster ellipticals are compared with those of the
elliptical galaxies selected from CANDELS (cyan histograms; see \S\ 2).
 }
\label{histos1}
\end{figure}

\begin{figure}
\includegraphics[width=9.truecm]{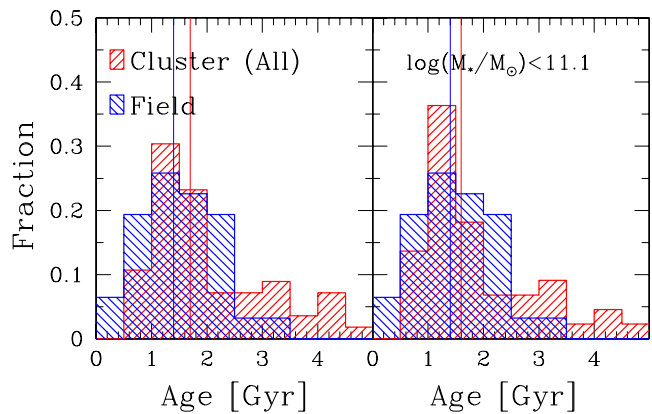}
 \caption{Age distribution of cluster and field ellipticals.
Left - The distribution of the age of the stellar population
of field elliptical galaxies (blue shaded histogram) is compared
with the age distribution of all the cluster ellipticals 
(red shaded histogram).
{  The colored solid lines mark the median values of the distributions
(see text).}
Right - Same as left panel but in this case the comparison is made
by considering field and cluster ellipticals in the same mass range, that is, by excluding cluster galaxies with masses larger than 
log$(\mathcal{M}_*)=11.1$ M$_\odot$. 
}
\label{histo_age}
\end{figure}

\begin{figure}
\includegraphics[width=9.truecm]{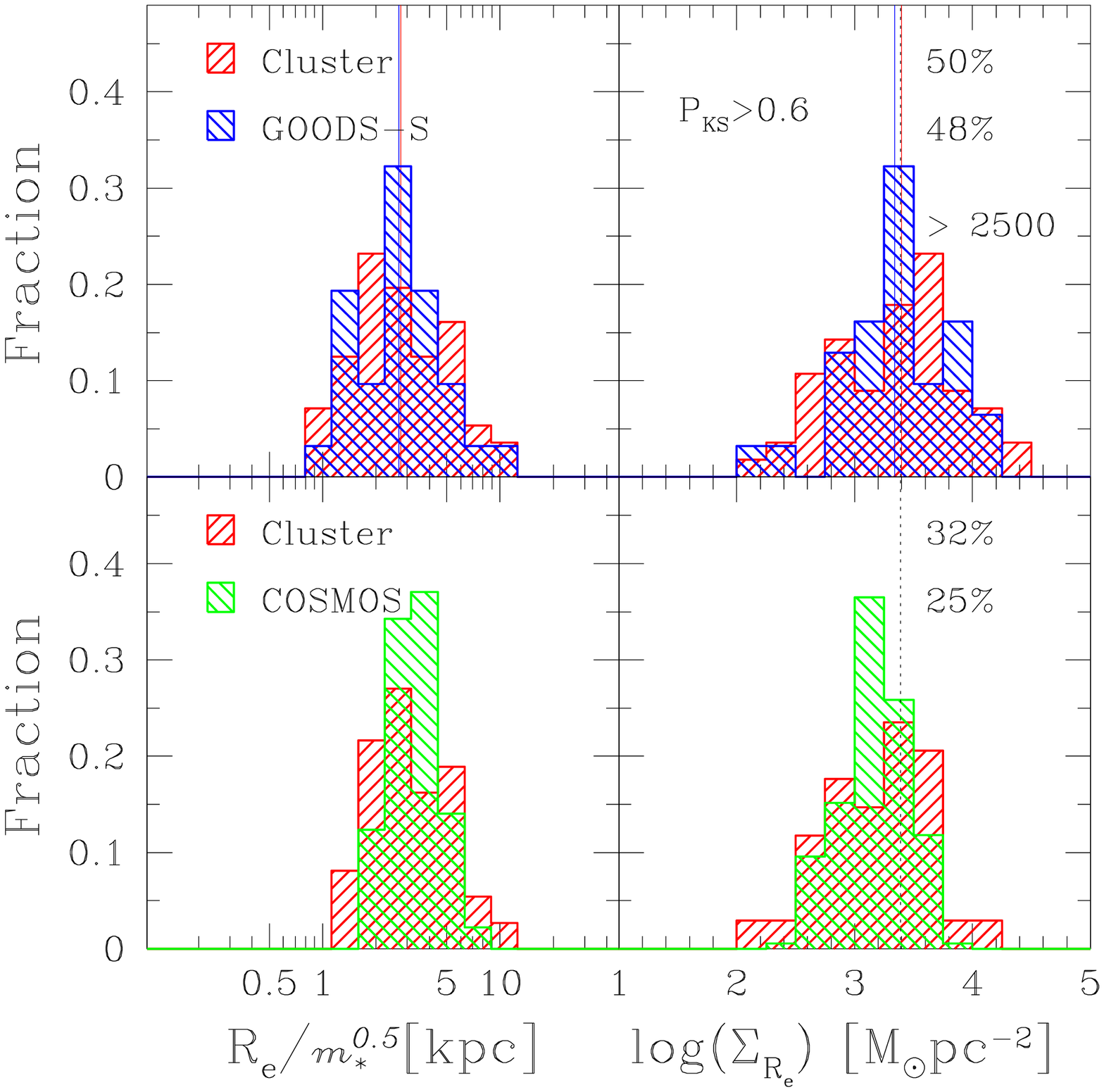}
\includegraphics[width=9.truecm]{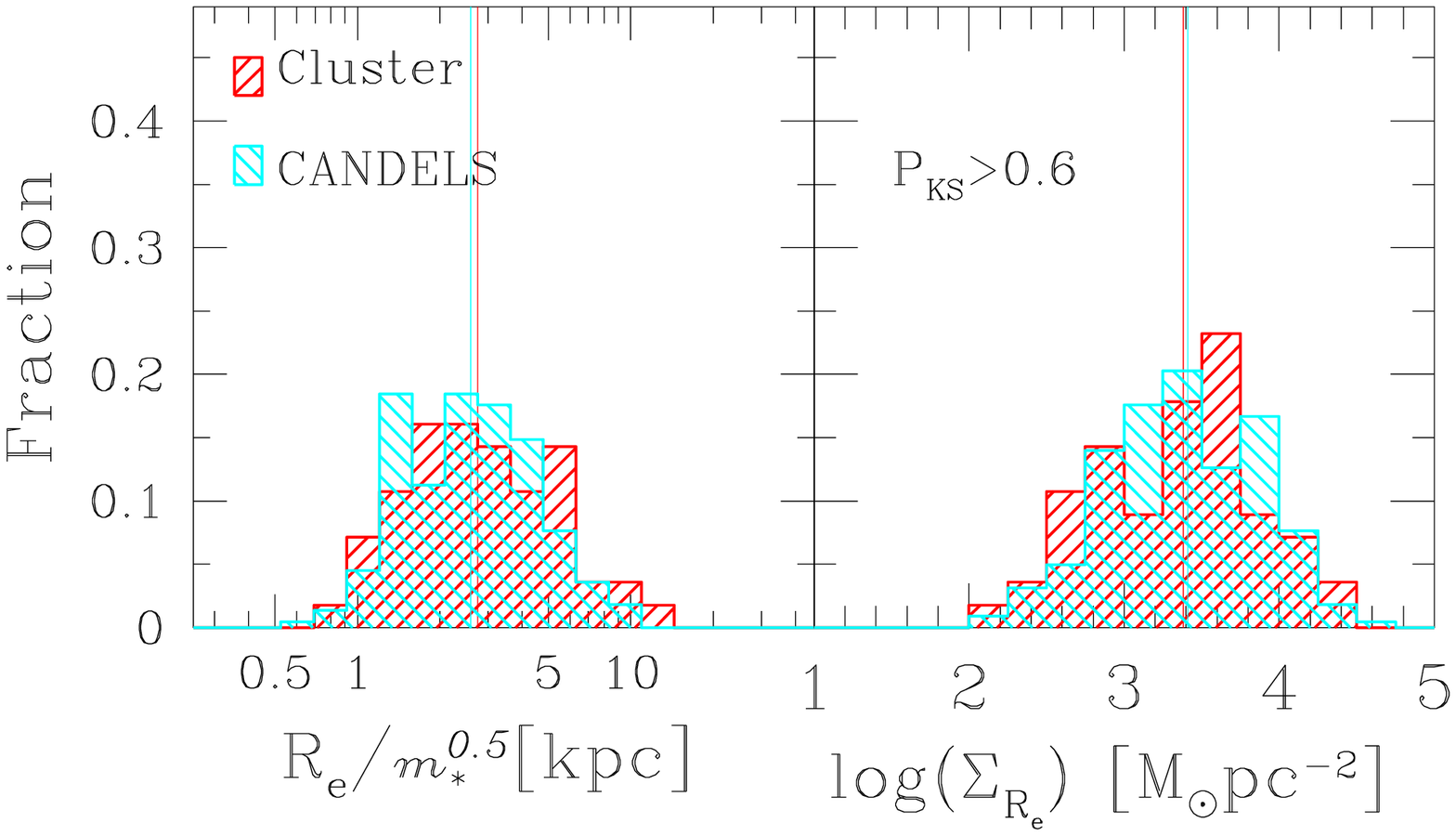}
\caption{Distribution of mass-normalized radius and stellar mass 
density for cluster and field elliptical galaxies. 
Upper panel - Distributions of the mass normalized radius  (left) and
of the effective stellar mass density $\Sigma_{R_e}$ (right) for cluster (red
histograms) and field elliptical galaxies in the GOODS-South field 
(blue histograms) and in the COSMOS area  (green histograms). 
The effective radius R$_e$ is normalized to the stellar mass  $m_*^{0.5}$  
where $m_*=\mathcal{M}_*/10^{11}$ M$_\odot$.
{  Colored solid lines mark the median values of the distributions.}
 The black dotted line marks the median value $\Sigma_{R_e}=2500$ 
 M$_\odot$ pc$^{-2}$ of the elliptical galaxies in cluster and in the GOODS-S
 field.
{  Lower panel - Same as in the upper panel but the comparison is made with the 
field  elliptical galaxies selected in the CANDELS area (cyan histograms).} 
 }
\label{histos2}
\end{figure}

\subsection{The population of elliptical galaxies and their dependence 
on the environment}
{  Before comparing field and cluster samples, we verified that no 
significant cluster-to-cluster variation in the distributions of ages, 
effective radii, S\'ersic index, and stellar masses were present. 
However, we note that, as  can be seen from Table \ref{clsample},  the
fraction of galaxies with masses log(M$_*$/M$_\odot$)>11  in cluster XMMJ2235,
the one at highest redshit, is higher than in the other two clusters, 
even if this difference is not statistically significant.
The absence of significant cluster-to-cluster variations is reflected
in the agreement among the Kormendy relations (see \S\ 5).}

In Fig. \ref{histos1} the distributions of the stellar mass and of the 
effective radius of cluster galaxies are compared with those of 
field galaxies.
The distributions 
%of the stellar mass and of the effective 
%radius of galaxies shown in Fig. \ref{histos1} 
show that elliptical galaxies 
in cluster reach higher stellar masses and larger radii than field ellipticals. 
%in the GOODS-South region. 
Indeed, there is a lack of massive ($\mathcal{M}_*>2\times10^{11}$ M$_\odot$ and 
large (R$_e>4-5$ kpc) galaxies in the GOODS-S field at $z\sim1.3$ with  respect to 
the cluster environment at the same redshift (upper panel of Fig. \ref{histos1}),
in agreement with what was found by  \cite{raichoor11} at this redshift.
This lack is statistically significant at more than 3$\sigma$ as shown
by the Kolmogorov-Smirnov (KS) test that provides a probability 
P$_{KS}\simeq0.004$ that the observed distributions of stellar mass 
are extracted from the same parent population and a probability  
P$_{KS}\simeq0.04$ ($\sim2\sigma$) for the distributions of the effective radius.
Only 6\% of the elliptical galaxies in the field are larger than 
R$_e>4$ kpc while they are 21\% in cluster.
A consequence of this is that the median effective radius R$_e$ is 
slightly larger in cluster (2.1 kpc) than in the field (1.5 kpc).
The smaller number of large galaxies ($>4-5$ kpc) in the field is due 
to the lack of galaxies with mass higher than $\sim2\times10^{11}$ M$_\odot$.
The median stellar mass is larger in the cluster sample 
($6\times 10^{10}$ M$_\odot$) than in the field 
($3.5\times 10^{10}$ M$_\odot$) and while 21\% of elliptical galaxies 
in cluster have masses $\mathcal{M}_*>1.2\times10^{11}$ M$_\odot$, 
there are no galaxies exceeding this mass in the field.
%A  Kolmogorov-Smirnov (KS) test provides a significance at more than
%3$\sigma$  of these differences.

These differences can be better assessed by comparing the two-dimensional
distributions [R$_e,\mathcal{M}_*$] of galaxies using the generalized KS test. 
Once compared in the mass range 
$\mathcal{M}_*<2\times10^{11}$ M$_\odot$, cluster and field
galaxies follow the same [R$_e,\mathcal{M}_*$] distribution (P$_{2dKS}=0.16$) with
the same median effective radius (1.7 kpc), while considering
the whole mass range they differ at $\sim3\sigma$ (P$_{2dKS}=0.01$).
This confirms that the two distributions differ because of the lack 
(excess) of massive and large galaxies in the field (in cluster).
Contrary to that found by \cite{papovich12} at $z\sim1.6$, 
we do not find a lack of small galaxies (R$_e<1$ kpc) in cluster with
respect to the field. 

{  The lack of large and massive galaxies in the field at $z\sim1.3$
is not a peculiarity of the GOODS-South region but is an actual 
property of the field environment, as confirmed by the comparison  
with the COSMOS  and the CANDELS samples shown in the middle and in the lowest 
panels of Fig. \ref{histos1} respectively.} 
%In the middle panels of Fig. \ref{histos1}  the distributions of 
%cluster elliptical galaxies are compared with those  
%in the COSMOS area at $1.0<z<1.2$, while in the lower panels
%with those in the CANDELS fields.
%It is worth to remind that the effective radius for galaxies in the 
%COSMOS sample are given for 178 galaxies for values
%larger than $\sim0.17$ arcsec \citep[see][]{scarlata07} that, at $z\simeq1.2$ 
%corresponds to $\sim1.3$ kpc.
We note that the effective radii for galaxies in the COSMOS sample are given for values
larger than $\sim0.17$ arcsec \citep[see][]{scarlata07} that, at $z\simeq1.2,$ 
corresponds to $\sim1.3$ kpc.
Hence, to compare the distributions of R$_e$ and of the
quantities involving R$_e$ (normalized effective radius and mass density) 
with the COSMOS sample, in the cluster sample we selected only galaxies with 
R$_e>1.3$ kpc. 
%{The effective radii of CANDELS galaxies are in the F160W band (see appendix
%\ref{ap:candels}).

  The lack of galaxies with mass higher than $2-3\times10^{11}$ M$_\odot$ 
is significant both in the COSMOS area ($\sim2\sigma$ significance, 
P$_{KS}\simeq0.05$) and in the CANDELS fields ($>5\sigma$, P$_{KS}<10^{-5}$).
Analogously, elliptical galaxies with effective radius larger than $4-5$ kpc are 
extremely rare at high significance level both in the COSMOS area 
($\sim3\sigma$, P$_{KS}\simeq0.006$) and in the CANDELS fields ($>3\sigma$, 
P$_{KS}\simeq0.003$) with respect to the cluster environment.

The comparison of the two-dimensional distributions [R$_e,\mathcal{M}_*$]
confirms that the COSMOS and CANDELS samples differ (P$_{2dKS}=0.02$ and
P$_{2dKS}=2\times10^{-4}$ respectively) from the cluster sample for 
the high-mass galaxies, 
since they do not differ (P$_{2dKS}=0.12$ and P$_{2dKS}=0.04$) when only 
galaxies with $\mathcal{M}_*<2\times10^{11}$ M$_\odot$ are considered.
 A lack of massive galaxies in the field with respect to the cluster 
environment is not surprising given the known correlation between halo mass
and  stellar mass of the galaxies populating it 
\citep[e.g.,][]{lin04,beutler13,shankar14a}.
For instance, \cite{huertas13b} show that galaxies more massive than
$10^{11}$ M$_\odot$ are two to three times more frequent in halos with masses of
10$^{14}$ M$_\odot$ than in halos one order of magnitude less massive.
This may be due to the higher frequency of merging expected in  higher
density regions.
Hence, it is expected that the field environment is less rich in high-mass 
and, consequently, large galaxies than the cluster environment, 
as seen also in the local universe.
In practice, some kind of environmental dependence in the
population of (elliptical) galaxies in the two environments is expected.

In Fig. \ref{histo_age} the distributions of the mean age of the stellar
population derived from the SED fitting are shown for cluster and field
(GOODS-S) elliptical galaxies. 
It is well known that this age does not represent the age of the bulk of 
the stars, that is the formation redshift $z_{form}$, but rather the age of the 
stars producing the dominant light \citep[e.g.,][]{greggio11}.  
Hence, only old ages, compared to the age of the universe at the 
redshift of the galaxy, are indicative of or coincident with the formation 
redshift $z_{form}$ while, in the other cases, they may be indicative of 
the last burst of star formation.
Our goal is not to determine the age of the bulk of the stars in 
elliptical galaxies at $z\sim1.3$ but, rather, to assess whether
the stellar populations show differences correlated with the environment
where the galaxies reside. 
Hence, this kind of luminosity-weighted age is suited to this purpose.
 
In the left panel of  Fig. \ref{histo_age}, the distributions of the two 
entire samples are compared.
It is well known that a correlation between age and stellar mass exists:
the larger the mass, the older the age. 
Hence, in the right panel, to delete the effect of the tail
at high masses of cluster galaxies, we compare the distributions of 
galaxies selected in the same stellar mass range 
log$\mathcal{M}_*<11.1$ M$_\odot$.
The median age of galaxies in the field is med(Age)$_f\simeq1.4$ Gyr, 
slightly lower than in cluster (med(Age)$_c\simeq1.7$ Gyr).
Figure \ref{histo_age} shows that the oldest elliptical galaxies at $z\sim1.3$
are in cluster and that there is a fraction of galaxies with old ages, older than 
$\sim3.5-4$ Gyr ($z_{form}>5$), not present in the field.
As shown in the right panel of Fig. \ref{histo_age}, this excess of old galaxies in cluster
remains, even when the two samples are compared in the same mass range.
Thus, it seems that there is a lack of elliptical galaxies
in the field which are as old as the oldest elliptical galaxies in cluster.
However, this lack is not statistically significant (P$_{KS}\simeq0.09$).
 
%This could mean that or the oldest cluster elliptical galaxies  
%have actually formed their stars before than the oldest ellipticals
%in the field, 
%or that these latter have had a longer star formation or experienced 
%subsequent episodes of star formation at later times.
%However, it is worth to note that the KS test provides a probability 
%P$_{KS}\simeq0.09$, hence this difference is only marginally significant.
 
The above comparisons show that the population of elliptical galaxies in 
cluster at $z\sim1.3$ differs from the one in the field.
There is a significant lack of massive  and consequently 
large elliptical galaxies in the field with respect to the
cluster environment, suggesting that the assembly of massive 
($\mathcal{M}_*>2-3\times10^{11}$ M$_\odot$) and large 
(R$_e>4-5$ kpc) elliptical galaxies at $z\sim1.3$ 
is a prerogative of the cluster environment.
Moreover, we find hints of the fact that the oldest elliptical galaxies
at this redshift are in cluster, in agreement with what has been derived
from the study of the elliptical galaxies in the local universe 
\citep[e.g.,][]{thomas05,thomas10}. 
%, even if cluster 
%and field elliptical galaxies do not differ in their structure.
%We will show this last point more quantitatively later in \SS\ 4.2.
%Moreover, the above comparisons suggest that the assembly of ultramassive 
%($\mathcal{M}_*>2-3\times10^{11}$ M$_\odot$) and consequently large 
%(R$_e>4-5$ kpc) elliptical galaxies at $z\sim1.3$ 
%is a prerogative of the cluster environment.
These results suggest that either the assembly of massive galaxies takes place 
earlier or it is more efficient and faster in cluster than in the field
\citep[see also][]{menci08,rettura10,rettura11}.

\subsection{Structural properties of elliptical galaxies in cluster and in the field} 
%It is important to asses whether the intrinsic properties of elliptical
%galaxies in cluster differ from those in the field , due to different 
%structural properties of the galaxies.
%In particular, there are no evidence of denser/more compact galaxies in 
%an environment with  respect to the other.
%This is clearly shown 
In Fig. \ref{histos2} the distributions
of the mass-normalized radius R$_e/m_*^a$ 
and of the mass surface density $\Sigma_{R_e}$ of cluster galaxies are 
compared with those of field galaxies.
Following \cite{newman12} and \cite{cimatti12} 
(see also \cite{huertas13}; \cite{delaye14}; \cite{allen15}), 
we defined the mass-normalized radius as R$_e/m_*^a$ 
where $m_*$ is the stellar mass of the galaxy in units of 
$10^{11}$ M$_\odot$, that is, 
$m_*=\mathcal{M}_*/10^{11}$ [M$_\odot$], and $a=0.5$ 
is the best fitting slope of the size-mass relation of our samples of 
elliptical galaxies (see \S\ 6).
The mass-normalized radius thus defined removes the dependence of the
radius from the mass allowing the comparison of the size distributions
of galaxies also in the case of different mass distributions.
This allows us to determine whether cluster galaxies tend to be
smaller or larger, denser or less dense than field galaxies independently
of their mass.   
It is worth noticing that this is equivalent to comparing the 
effective stellar mass surface density of galaxies, perhaps a  more
meaningful quantity, defined as
$\Sigma_{R_e}=\mathcal{M}_*/2\pi R_e^2$, being the mass normalized radius
defined above the inverse of the square-root of $\Sigma_{R_e}$.

{  The distributions of the normalized effective radius of the cluster 
and field elliptical galaxies shown in Fig. \ref{histos2},
belong to the same parent population (P$_{KS}\geq 0.6$).
The median value is med(R$_e/m_*^{0.5}$)=2.7$\pm0.2$ kpc both in cluster 
and in the field (GOODS-S, COSMOS and CANDELS).}
The same result, that there is no difference in the compactness (in
the size at fixed stellar mass) of galaxies
in cluster and in the field, is provided by comparing the mass 
surface density distributions (right panels). 
We note that the median value of the stellar mass surface density is
$\Sigma_{R_e}\sim2500$ M$_\odot$ pc$^{-2}$ both for cluster and
for field galaxies; that is, $\sim$50\% of galaxies are denser than 
this value both in cluster and in the field.
The distributions and the above results remain unchanged both considering the 
samples in the same mass range (log$\mathcal{M}_*<11.1$ M$_\odot$) or
in the whole mass range. 
Hence,  cluster and field ellipticals do not show different structural properties.

%\paragraph{Comparison with the literature}
At redshift similar to our study,  
 \cite{delaye14} find that at $1.1<z<1.6$
field galaxies in the mass range $3\times10^{10}-10^{11}$
M$_\odot$ are 10\%-30\% smaller than cluster galaxies.
However, the authors themselves notice that the apparent greater compactness of field 
galaxies may be due to the different wavelengths at which their effective
radii were estimated.  
In particular, they used near-IR (optical rest-frame) size from CANDELS 
for field galaxies while using optical (U/B rest-frame, as in our study) size 
from ACS observations for cluster galaxies.
The apparent size of galaxies can be
different at different wavelengths (see, e.g., \cite{labarbera10}) 
and, at $z>1$, galaxies seen in the F160W filter appear 10\%-20\% smaller 
than in the F850LP filter 
\citep[e.g.,][; see also appendix \ref{ap:candels}]{cassata11,gargiulo12}.
Therefore, the apparent smaller size of field galaxies they found could be
due to this effect.
Supporting this hypothesis, we note that the median value of the 
mass normalized radius  we measured in the ACS-F850LP filter both for cluster 
and field ellipticals, perfectly agrees with the one they derive for cluster 
galaxies ($\sim2.8$ kpc for $a=0.57$) on the ACS images.

\cite{raichoor12} find that the population of smaller size elliptical galaxies 
at $z\sim1.3$ is more numerous in denser environments (groups and clusters) 
than in the field environment, even if galaxies do not show different 
structures  at fixed mass.
In contrast,  \cite{papovich12} find that, at fixed stellar mass, field 
galaxies at $z\sim1.6$ are smaller, hence denser or more compact,
than cluster galaxies.
In line with our results, \cite{newman14} \citep[see also][]{rettura10} find no
significant difference between the size of field
and cluster galaxies at $z\sim1.8$ in the same mass range.
Analogously, \cite{allen15} find no 
significant differences between the mass normalized radius of field and 
cluster quiescent galaxies at $z\sim2$, for which they estimate a mean mass 
normalized semi-major axis $a_{1/2,maj}\sim2.0$ kpc in the F160W filter.
\cite{allen15} normalized the mass using $5\times10^{10}$ M$_\odot$ and the radius 
using $a=0.76$.
By scaling our effective radius according to these values, we obtain  
med(R$_e/m_*^{0.76}$)$\simeq1.9$ kpc.
Assuming a mean axis ratio $b/a=0.7$ (see Tables \ref{clsample} and
\ref{fisample}), that is a factor $(b/a)^{-1/2}\simeq1.2$ and
a correction of 10\% for the longer wavelength,
our median radius translates into a median semi-major axis of $\sim2.1$ kpc
in the F160W filter, in agreement with the semi-major axis
estimated by \cite{allen15} at $z\sim2$.

At lower redshift than our study,
\cite{huertas13} find no significant differences
in the structural properties of field and cluster galaxies up
to $z\sim1$.
 \cite{kelkar15} find no significant difference 
in the size distribution of cluster and field galaxies at $0.4<z<0.8$ both for a given 
morphology and for a given B-V color, ruling out average size differences 
larger than 10\%-20\%.
\cite{shankar14b} 
studied the dependence of the median sizes of central galaxies on host 
halo mass using a sample of galaxies at $z<0.3$ extracted from the Sloan
Digital Sky Survey (SDSS). 
They found no difference between the structural properties
of early-type galaxies in different environments, at fixed stellar mass.

\begin{figure}
\includegraphics[width=9.0truecm]{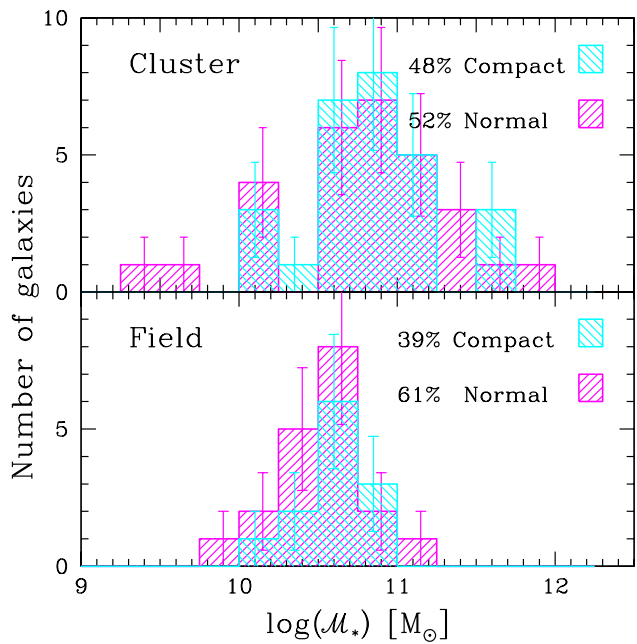}
\caption{Stellar mass distribution of compact and normal galaxies
in field and cluster environment. Upper panel - The distribution of the stellar
mass of normal elliptical galaxies (magenta histogram) in cluster is compared
to the one of compact galaxies (cyan histogram) defined according to the criterion
log$(\mathcal{M}_*/R_e^{1.5})>10.3$.  
Lower panel - Same as for the upper panel but for field elliptical galaxies.
}
\label{histo_comp}
\end{figure}

We also considered the definition of a compact/dense galaxy based on 
the criterion log$\mathcal{M}_*/R_e^{1.5}>10.3$ M$_\odot$ kpc$^{-1.5}$
(e.g., \cite{barro13}; \cite{poggianti13}; \cite{damjanov15}).   
Figure \ref{histo_comp} shows the stellar mass distributions of cluster and
field elliptical galaxies classified as compact/dense and normal 
according to this criterion.
The fraction of compact galaxies in the field  (39($\pm11$)\%) is slightly 
lower than in cluster (48($\pm9$)\%),  even if this difference is not 
statistically significant.
Hence we find that also the fraction of compact galaxies thus defined is similar in 
the two different environments, confirming the previous results.
 
We have shown that at fixed stellar mass, elliptical galaxies 
in cluster and in the field at $1.2<z<1.4$ are characterized by the same 
structural parameters.
Elliptical galaxies in the field have a similar density to those
in cluster; their stellar mass density and their structural
parameters are not dependent on the environment where they reside.  
What we see is that ultramassive, ($\mathcal{M}_*>2-3\times10^{11}$ M$_\odot$) large 
(R$_e>4-5$ kpc) early-type galaxies at $z\sim1.3$ are more abundant in the 
cluster environment than in the field environment.
We could suppose that in the field, the assembly of such massive early-type galaxies
takes place over a longer time or that it starts later than in cluster, and that
the lack of ultramassive elliptical galaxies in the field is filled up at lower redshift.
However, this is not the case.
Indeed, the number density of ultramassive dense elliptical galaxies in the field 
seems to be constant over the redshift range $0.2<z<1.5$ \citep{gargiulo16}.
This suggests that both the populations of high-mass elliptical galaxies, 
in cluster and in the field, are mostly formed at that redshift.
This could be explained if the assembly of massive galaxies is more efficient in 
cluster than in the field.
Clear evidences in favor of this has still not emerged from observations, however, 
some analyses point toward a faster star formation
in cluster than in the field \citep[e.g.,][]{rettura11,guglielmo15} and others show that
the stellar mass distribution in the central parts of the mass distribution in clusters are already in place at 
$z\sim1$ (\citealt{vanderburg15}, see also \citealt{vulcani16}).
Irrespective of this, the fact that cluster and field elliptical galaxies are 
characterized by the same structural parameters suggests that the formation processes 
acting in the two environments are the same.

\section{The Kormendy relation of cluster and field elliptical galaxies at $z=1.3$}
\begin{figure}
\includegraphics[width=9truecm]{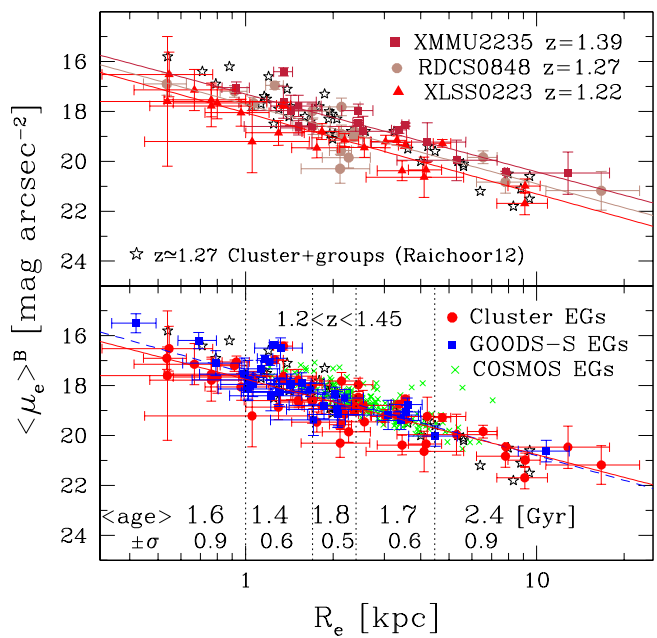}
\caption{Kormendy relation in the rest-frame B-band for cluster and field elliptical 
galaxies at $z\simeq1.3$. 
{Upper panel -} 
 The B-band surface brightness of the 17 galaxies in the cluster 
XMMU2235 at $z\sim1.39$ (dark red filled squares), of the 16 galaxies in the 
RDCS0848 at $z\sim1.27$ (brick red filled circles), and of the 23 galaxies in the 
cluster XLSS0223 at $z\sim1.22$ (red filled triangles) are plotted as a function 
of their effective radius R$_e$ at the redshift of the clusters.
The three colored lines are the best fitting Kormendy relations to the three
clusters reported in Table \ref{fitkr}.
The open symbols are the sample of early-type galaxies in the Linx clusters
and groups studied by \cite{raichoor12}. 
 %The black line represent the Kormendy relation at $z=0$ (eq. \ref{krz0}).
{\em Lower panel:} The surface brightness of the 56 cluster galaxies (red
filled circles), of the 31 field galaxies at $1.2<z<1.45$ in the GOODS-South 
region (blue filled squares), and of the 178 galaxies at $1.0<z<1.2$ in the 
COSMOS area (green filled triangles) are plotted as a function of their 
effective radius. 
The red solid line and the blue dashed line are the best fitting relations obtained
for cluster and GOODS-S field galaxies respectively, and are reported in Table
\ref{fitkr}.
At the bottom of the figure, the mean age $<age>$ [Gyr] of galaxies 
in the different intervals of effective radius together with its dispersion ($\sigma$) is reported.
}
\label{kr_cl_field2}
\end{figure}

Elliptical galaxies both in field and in clusters follow the scaling relation  
\begin{equation}
\langle\mu\rangle_e = \alpha + \beta \log(R_{e}),
\label{kr}
\end{equation}
(Kormendy 1977; KR hereafter),
a relation between the logarithm of the effective radius R$_e$ [kpc] 
and the mean surface brightness $\langle\mu\rangle_e$  within R$_e$.
The slope parameter $\beta\sim 3$ in the B-band is found not to vary 
out to $z\sim1$ \citep[e.g.,][]{hamabe87, schade96, ziegler99, labarbera03, 
reda04, diserego05, saracco09, saracco14}
as also shown from the studies of the fundamental plane of galaxies
\citep[e.g.,][]{graham06a, donofrio08, gargiulo09, saglia10, jorgensen14, zahid15}. 
Possibly, a slightly steeper slope is found at higher stellar masses  
($\mathcal{M}>3-4\times10^{11}$ M$_\odot$) typically for brightest cluster galaxies 
\citep[BCGs; e.g.,][]{bai14, bildfell08, vdlinden07}.  
The different slope could suggest a disparity in the 
M/L or a break in the homology at these masses 
\citep[e.g.,][]{vdlinden07}.
Actually, at these large stellar masses, galaxies seem to deviate from 
the scaling relations defined by galaxies with lower masses, suggesting that the 
mass accretion at such large masses may take place according to different 
mechanisms \citep[e.g.,][]{bernardi11,bernardi14}.
Contrary to the slope, the zero point $\alpha$ of the KR is found to vary 
with the redshift  of the galaxies according to the expected luminosity 
evolution.

In this section, we derive and compare the Kormendy relation of cluster and 
field elliptical galaxies at $z\sim1.3$ in order to asses whether the
environment plays a role in shaping this relation. 
We consider here the B-band rest-frame since morphological 
parameters have been derived in the F850LP filter sampling 
$\lambda_{rest}\simeq 4000$ \AA\ at $z\sim1.3$. 
For each galaxy we derived the mean effective surface brightness 
\begin{equation}
\langle\mu\rangle_e^B =M_B(z)+2.5log(R_e^2)+38.57,
\label{murest}
\end{equation}
where $M_B(z)$ is the absolute magnitude of the galaxy
in the rest-frame B-band at redshift $z$, and
$R_e$ is in [kpc], after correcting for the cosmological dimming
term 10log(1+z).
The surface brightnesses thus obtained are reported in Table 1.

In the upper panels of Fig. \ref{kr_cl_field2}, the rest-frame B-band surface 
brightness of the elliptical galaxies in the cluster XMMJ2235 at $z\simeq1.39$,
RDCS0848 at $z\simeq1.27,$ and in the cluster XLSS0223 at $z\simeq1.22$ 
is plotted as a function of their effective radius R$_e$. 
The solid colored lines are the best fitting Kormendy relations we obtained
using the least square fit for the three clusters and are summarized in Table 
\ref{fitkr}.

\begin{table}
\caption{Kormendy relation of cluster and field ellipticals at $z\sim1.3$}
{\small
\centerline{
\begin{tabular}{ccccc}
\hline
\hline
  ID & $z$& Ngal & $\alpha$ & $\beta$  \\ 
     &      &          &                       \\
\hline
XMMJ2235& 1.39&17 & $17.3\pm0.2$ & $3.1\pm0.4$  \\
RDCS0848& 1.27&16 & $17.7\pm0.2$ & $3.2\pm0.5$  \\
XLSS0223& 1.22&23 & $17.9\pm0.1$ & $3.1\pm0.3$  \\
\hline
%<Cluster>&             1.30&56 & $17.8\pm0.1$ & $3.0\pm0.2$  \\
<Cluster>&[1.2-1.4]&56 & $17.7\pm0.1$ & $3.0\pm0.2$   \\
<Field>  &[1.2-1.4]&31 & $17.5\pm0.2$ & $3.3\pm0.3$   \\
<All>    &[1.2-1.4]&87 & $17.6\pm0.1$ & $3.2\pm0.2$   \\
\hline
\end{tabular}
}
}
\label{fitkr}
\end{table}

The relations found are in agreement between them
and they are described by the same slope $\beta\simeq3.1$ (see Table \ref{fitkr}).
The brightening by about 0.4 mag of the zeropoint $\alpha$
of the KR of the XMMU2235 cluster, reflects
the significantly higher redshift, in terms of luminosity evolution, of this 
cluster with respect to the other two.
Indeed, by computing for each galaxy of the XMMU2235 cluster the 
luminosity evolution since $z=1.39,$ we found that, on average, galaxies would 
get fainter by $0.35(\pm0.1)$ mag at $z=1.25$ ($\Delta t\sim$0.4 Gyr), in 
agreement with the different zeropoint.
The distribution of the galaxies in the [$\mu_e$, R$_e$] plane and the
relevant best fitting relations do not show significant differences
among the populations of elliptical galaxies (EGs)  in these three clusters.
So, we have defined the KR for cluster ellipticals at $z\sim1.3$ considering the 
whole sample of 56 cluster galaxies.
The resulting best fitting KR for the whole sample is 
\begin{equation}
\langle\mu\rangle_e^B=17.7(\pm0.1)+3.0(\pm0.2)log(R_e)\hskip 0.3truecm Cluster\\\
z\in[1.2-1.4], 
\label{krmean}
\end{equation}
in good agreement with that found by \cite{jorgensen14} and
\cite{raichoor12} at the same redshift (see appendix A for the orthogonal fit).
The slope of the KR we find at $z\sim1.3$ agrees also with the slope
found at intermediate redshift \citep{saglia10, jorgensen13, zahid15} and in the 
local universe \citep[e.g.,][]{jorgensen95a, donofrio08, gargiulo09, labarbera10}.

In the lower panel of Fig. \ref{kr_cl_field2}, the surface brightness
of the 31 field elliptical galaxies in the redshift range $1.2<z<1.45$ 
in the GOODS-S region is plotted as a function of their effective radius 
and superimposed onto the sample of cluster galaxies at 
the same redshift. 
The resulting best fitting KR to field elliptical galaxies is
\begin{equation}
\langle\mu\rangle_e^B=17.5(\pm0.2)+3.3(\pm0.3)log(R_e)\hskip 0.3truecm Field\\\
z\in[1.2-1.4]. 
\label{krfield}
\end{equation}
The slope is slightly steeper than the one for cluster galaxies even if the 
difference is not statistically significant.
We verified that this small difference is due to the massive and
large (R$_e>4-5$ kpc) cluster galaxies not present in the field
(see previous section), which tend to slightly flatten the relation.
For comparison, we also plotted the $178$ EGs selected in the COSMOS area
at $1.0<z<1.2$ with measured effective radius.
This sample is highly incomplete for radii $<1.3$ kpc 
since measurements of radii are available in the catalog for values
larger than $\sim0.17$ arcsec (see \S\ 2.2). 

The distribution of the galaxies in the [$\mu_e$,R$_e$] plane and the
best fitting relations found show that cluster and 
field elliptical galaxies at $\sim1.3$ follow the same size-surface 
brightness relation.
We find no evidence statistically significant of a dependence of the KR on the 
environment  at $z\sim1.3$, in agreement with what has been previously found by other 
authors \citep{rettura10, raichoor12} at similar redshift.
Analogously, we find that the zeropoint of the KR of early-type galaxies at 
this redshift is about two magnitudes brighter than at $z=0,$ as also  
found by \cite{holden05}, \cite{rettura10},
\cite{raichoor12}, and \citet{saracco14} on samples of cluster and 
group galaxies at similar redshift.

At the bottom of Fig. \ref{kr_cl_field2}, we report the mean age of galaxies
in different intervals of effective radius.
As can be seen, there is a variation of the age of galaxies along the Kormendy relation,
as previously noticed by \cite{raichoor12}: the age tends to increase going toward 
larger and lower surface brightness galaxies.
This small age gradient is related to the age-mass relation
in which higher mass galaxies (larger and usually with lower surface brightness)
are older than lower mass ones.
%Given the existence of a small age gradient, it follows that a variation of 
%the M/L along the relation will take place as well and that it will evolve
%differentially woth time.

\section{The size-mass relation}

\begin{figure}
%\hskip -0.2truecm
\includegraphics[width=9.truecm]{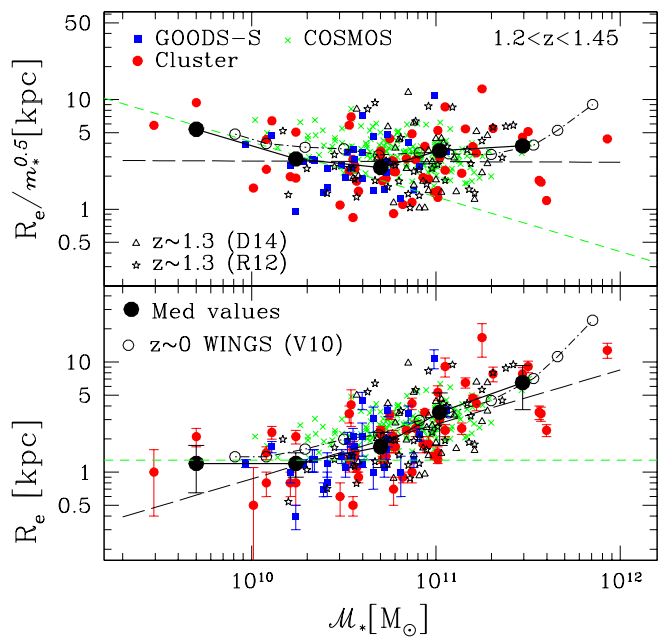}
 \caption{Effective radius-stellar mass relation for elliptical galaxies.
Upper panel - The mass normalized radius R$_e/m_*^{0.5}$ is plotted 
as a function of the stellar mass $\mathcal{M}_*$. 
The filled symbols are the cluster ellipticals of our sample (red circles),
the field ellipticals we selected at $1.2<z<1.45$
in the GOODS-S field (blue squares), and the ellipticals at $1.0<z<1.2$ in 
the COSMOS field (green crosses). 
Open symbols are data taken from \cite{raichoor12} (R12, stars), 
from \cite{delaye14} (D14, triangles), and from 
\cite{valentinuzzi10} (V10, circles; see also \cite{poggianti13a}). 
The green short-dashed line is $y=1.3[kpc]/m_*^{0.5}+c$, the line below
which there are no measured effective radii in the COSMOS
catalog \citep{sargent07}. We report errorbars only in the lower panel
just for clarity in the plot.
Lower panel - The effective radius R$_e$ is plotted versus the stellar
mass. Symbols are as in the upper panel.
The big black filled triangles are the median values of our sample of cluster
and field ellipticals at $z\sim1.3$. 
The long-dashed line is the best fitting relation
R$_e\propto\mathcal{M}_*^{0.5}$ over the whole mass range.
}
\label{size_mass}
\end{figure}

Figure \ref{size_mass} shows the size-mass relation for cluster
and field elliptical galaxies at $z\sim1.3  $ . 
In the upper panel, we show the mass normalized effective radius 
R$_e/m_*^{0.5}$ (see \S\ 4) as a function of stellar mass, while in the lower
panel we show the effective radius versus the mass.
Superimposed onto our data, we plot for comparison the early-type galaxies, in
groups and in cluster at $z\sim1.27$, of the sample of \cite{raichoor11,
raichoor12} (R12).
They derived stellar masses adopting the Salpeter IMF.
Thus, we scaled their masses by a factor 1.7 to match them to the 
Chabrier IMF, according to the recipe of \cite{longhetti09} and to that
found for elliptical galaxies at this redshift \citep{saracco14, tamburri14}.
Besides this sample, we also plotted the cluster early-type galaxies belonging to the 
sample of \cite{delaye14} (D14) in the redshift range $1.22<z<1.4$ that have 
stellar masses based on Chabrier IMF, as for our data. 
These two samples cover almost the same mass range as that covered by our data,
extending down to $2\times10^{10}$ M$_\odot$ at low masses and up to 
$2\times10^{11}$ M$_\odot$ at high masses.
The effective radii for both the samples have been derived on the ACS-F850LP 
images, as for our data.
As can be seen, the agreement among the different samples is very good.

In the mass range $\sim10^{10}-10^{11}$ M$_\odot$, well covered by field
and cluster ellipticals,  
galaxies in the two environments do not show differences in
the size-mass plane.
%In this mass range, we do not find significant difference in 
%the median size-mass relation of galaxies in the two environments.
The median effective radius of cluster ellipticals is $1.4\pm0.7$ kpc 
for  $\mathcal{M}_*\in[10^{10}-5\cdot 10^{10}]$ M$_\odot$ and $2.1\pm0.6$ kpc 
for $\mathcal{M}_*\in[5\cdot 10^{10}-10^{11}]$ M$_\odot$
 , to be compared with $1.3\pm0.4$ kpc 
and $2.0\pm0.8$ kpc for field ellipticals in the same mass ranges
(the quoted errors are the median absolute deviation). 
Similar results, that there is no evidence of a dependence of the size-mass relation 
on the environment, are found by \cite{rettura10}, \cite{raichoor12}, \cite{newman14}, and 
\cite{allen15} at redshift comparable or higher than ours, and by 
\cite{kelkar15} and \cite{huertas13} at intermediate redshift ($0.4<z<1$).
To our knowledge, at intermediate and high redshift, \cite{cooper12}, 
\cite{lani13}, and \cite{strazzullo13} find a correlation between the size 
of massive quiescent galaxies and the environment where they reside:
galaxies in denser environments are much larger (less dense) then galaxies 
in low density environments.
However, in these latter works, the selection of galaxies is based on the
Sersic index and/or on UVJ colors instead of on morphology, 
making a reliable comparison difficult.
Selections based on colors or Sersic index provide samples
with a different mix of morphological types \citep[e.g.,][]{mei12,tamburri14}.
It has been clearly shown that the size of galaxies at fixed mass is a 
strong function of morphology \citep[e.g.,][]{mei12,bernardi14}.
The different fractions of morphological types in the 
different environments that these selection criteria provide, 
could give rise to different results \citep[see also the discussion in][]{newman14}.    

In the local universe, where the statistic is higher than at higher redshift,
the results seem to confirm the absence of strong dependence on the environment.
\cite{poggianti13a}, using the cluster galaxies in the Wide-field 
Nearby Galaxy-clusters Survey (WINGS)
\citep{fasano06, valentinuzzi10} and the field galaxies of the Padova-Millennium 
Galaxy and Group Catalog \citep[PM2GC][]{calvi11}, find that the mass-size relation
of cluster galaxies lies slightly ($\sim1\sigma$) below the relation for field
galaxies. 
They note that, for a given mass, the fraction of smaller (i.e., denser) 
galaxies in cluster is slightly higher than the fraction in the field. 
Perhaps, this reflects the larger fraction of lenticular galaxies present
in the cluster sample than in the field sample, since lenticular galaxies
tend to be smaller than ellipticals at fixed stellar mass 
\citep[e.g.,][]{maltby10,bernardi14}.
\cite{huertas13b} using the Sloan Digital Sky Survey data
\citep[SDSS,DR7][]{abazajian09} find no differences between the size-mass
of early-type galaxies in the two environments, a result emerging also
from the analysis of \cite{shankar14b} and, previously, from 
the analysis of \cite{maltby10} and \cite{rettura10}.
Thus, it seems that if differences in the size-mass relation of elliptical
galaxies depending on the environment are present, 
they have to be small and due to differences in the 
population of galaxies rather than in the structural 
properties of galaxies at fixed type and mass.

\begin{figure*}
\begin{center}
%\vskip -2.5truecm
\includegraphics[width=12.truecm]{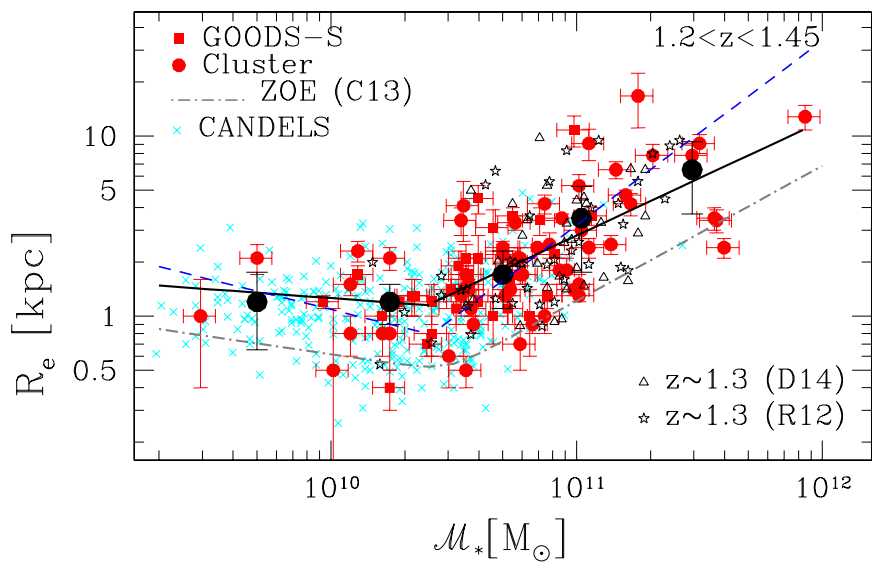}
 \caption{Effective radius-stellar mass relation for elliptical galaxies.
Filled symbols are field (squares) and cluster (circles) ellipticals of our
sample. 
Empty symbols are as Fig. \ref{size_mass}.  
The thick solid lines are the best fitting relations obtained with the
least square fit for
values lower and higher than $\mathcal{M}_*=2.5\times10^{10}$ M$_\odot$,
R$_e=26\times\mathcal{M}_*^{-0.13\pm0.2}$ , and
R$_e=2.77\times10^{-7}\mathcal{M}_*^{0.64\pm0.09}$ respectively.
The blue dashed line is the best fitting relations obtained in the
two mass ranges using the orthogonal fit (see appendix A), R$_e\propto\mathcal{M}_*^{-0.3\pm0.2}$ 
and   R$_e\propto\mathcal{M}_*^{1.0\pm0.1}$ respectively.
The gray dot-dashed line is the relation 
$R_e=0.53(\mathcal{M}_*/3\times10^{10})^{-0.2}(0.5+0.5(\mathcal{M}_*/3\times10^{10})^{0.8})^{0.119}$
defining the Zone of Exclusion (ZOE) as obtained by \cite{cappellari13}
and scaled by a factor $<R_e/R_e^{maj}>=0.76$ (see text).
Open symbols represent the samples of \cite{raichoor12} (stars, R12)
and of \cite{delaye14} (triangles, D14). 
{  Cyan crosses are elliptical galaxies selected from CANDELS}.
}
\end{center}
\label{zoe}
\end{figure*}
\begin{figure*}
\centering
\includegraphics[width=13truecm]{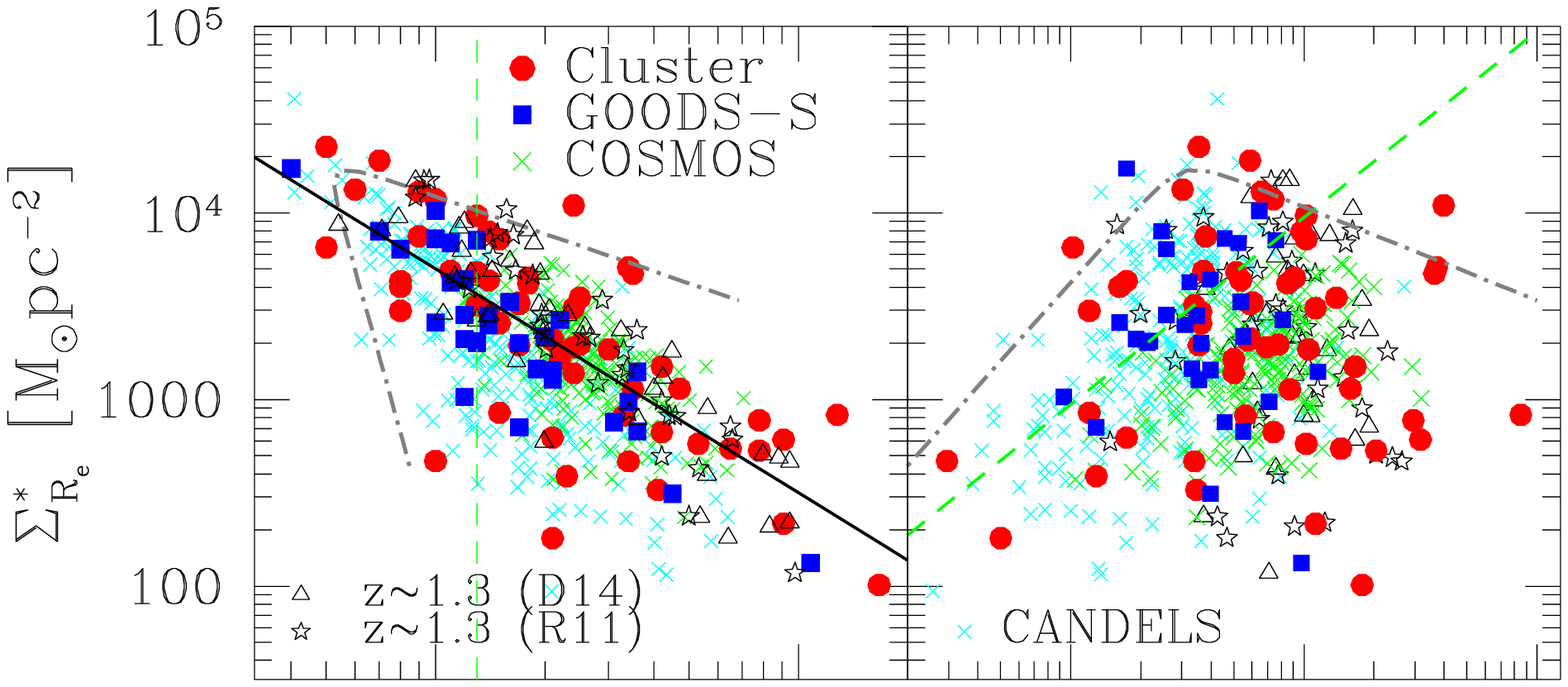}
\includegraphics[width=13truecm]{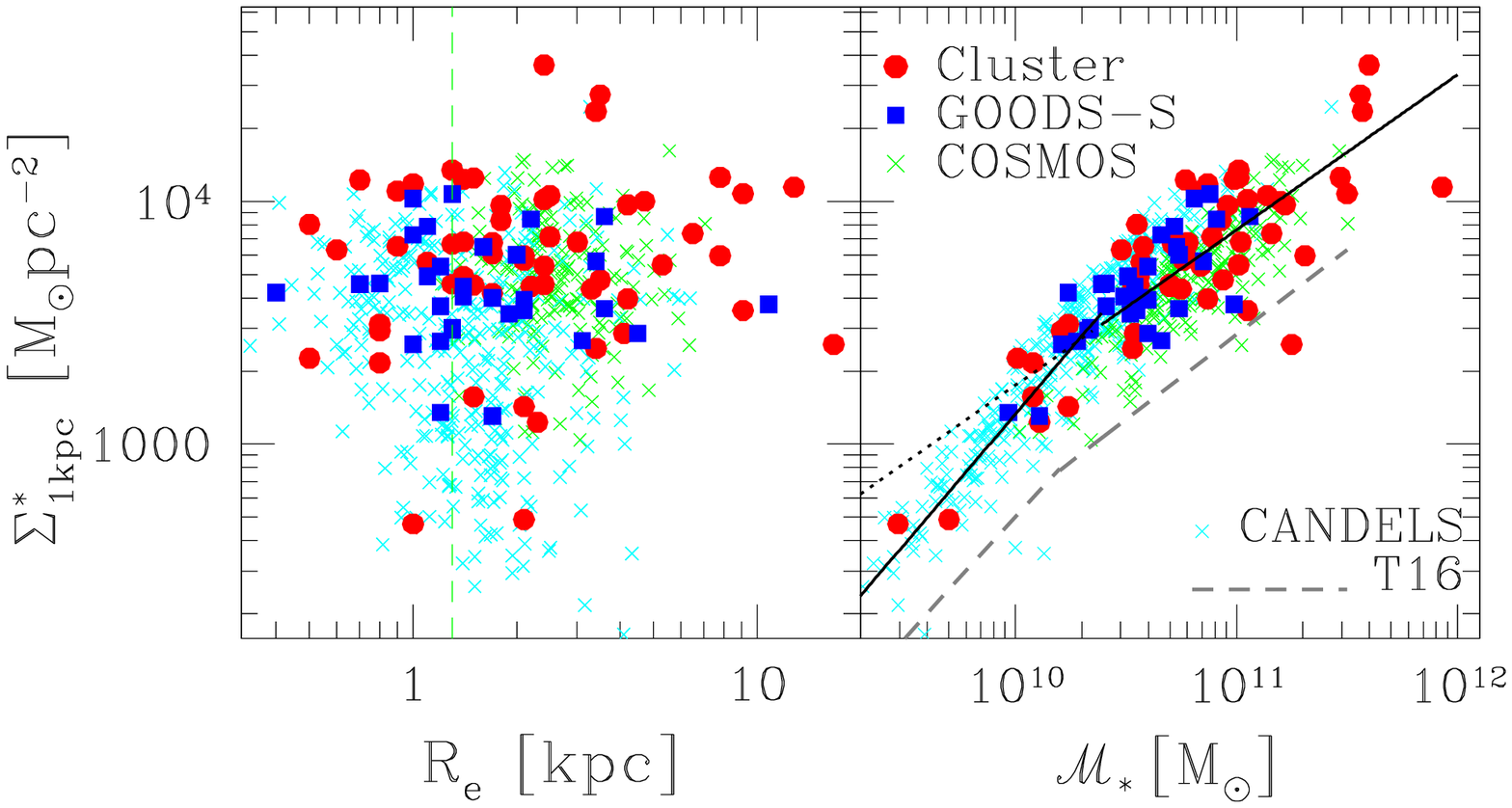}
\caption{Stellar mass surface density versus 
effective radius and stellar mass. 
Upper panel -
The effective stellar mass density
 within the effective radius $\Sigma_{R_e}=\mathcal{M}_*/2\pi R_e^2$,
  is plotted as a function of 
the effective radius R$_e$ (left) and of the stellar mass $\mathcal{M}_*$ 
(right). Blue filled squares are field ellipticals in the GOODS-S regions,
green crosses are those in the COSMOS area, and red filled circles 
are cluster ellipticals. 
The black solid line is the best-fitting relation 
$\Sigma_{Re}\propto R_e^{-1.2\pm0.1}$ (see appendix A for the orthogonal fit).
The green dashed line in the right panel shows the stellar mass density
as a function of mass for a constant value of R$_e=1.3$ kpc (marked on the
left panel),
corresponding to the limiting value $\sim0.17$ arcsec adopted for
the effective radius in the COSMOS catalog of \cite{scarlata07}. 
The gray dot-dashed line marks the Zone of Exclusion (ZOE) as defined
in \S\ 6.1.
The open stars are the early-type galaxies selected by 
\cite{raichoor11} (R11) in groups and clusters at $z\sim1.27$. 
The open triangles are the cluster elliptical galaxies 
at $1.22<z<1.4$ studied by \cite{delaye14} (D14).
{  The cyan crosses are the elliptical galaxies from CANDELS}.
Lower panel -
The central stellar mass  density calculated within 1 kpc radius 
($\Sigma_{1kpc}$) is plotted as a function of the effective radius (left) 
and of the stellar mass (right).
Symbols are as in the upper panel.
%Blue filled squares are field ellipticals in the GOODS-S regions,
%green crosses are ellipticals in the COSMOS field and red filled circles 
%are cluster ellipticals. 
%The green dashed line marks the value R$_e=1.3$ kpc,
%corresponding to the limiting value $\sim0.17$ arcsec adopted for
%the effective radius in the COSMOS catalog of \cite{scarlata07}.
The solid black line is the best-fitting broken power law  
$\Sigma_{1kpc}\propto\mathcal{M}_*^{(0.64\pm0.06)}$ for 
$\mathcal{M}_*>m_t$ and $\Sigma_{1kpc}\propto\mathcal{M}_*^{(1.07\pm0.1)}$
for $\mathcal{M}_*<m_t$, with $m_t=2.5\times10^{10}$ M$_\odot$.
The gray dashed line is the relation found by \cite{tacchella16} with their
model of mass growth and quenching, arbitrarily normalized. 
{  Errorbars are not shown to observe the agreement among the different samples
and the relations defined by the data.} 
}
\label{rhoe_mass_re}
\end{figure*}

Even if the following issue is not the focus of the present paper,
we believe that it deserves to be at least mentioned.
By fitting the size-mass relation  to the whole sample of 
cluster and field elliptical galaxies over the
whole mass range with a single power law of the form 
R$_e=b\mathcal{M}_*^a$
, we find
\begin{equation}
log(R_e)=0.5(\pm0.06)log\mathcal{M}_*-5.0(\pm0.7),
\label{eq_size}
\end{equation}
using the least square method (see appendix A for the orthogonal fit).
Figure \ref{size_mass} shows clearly that a single power law 
does not provide a good fit to the data over the whole mass range considered
and that the relation tends to be curved consistently with other relations
\citep[see, e.g.,][for a review]{graham13}.
In particular, the observed median size relation deviates significantly from 
a single power law and changes the slope at a transition mass  
$m_t\simeq2-3\times10^{10}$ M$_\odot$: at lower masses, the relation tends to 
flatten, while at masses larger than this the relation tends to steepen.
The changing of some relationships  in correspondence with this
value of mass scale has been previously noted in the local universe 
\citep[e.g.,][]{kauffmann03,shankar06}.  
The changing slope of the size-mass relation and the flattening at low masses 
in our data at $z\sim1.3$ is better visible in Fig. 7.
This feature is visible in all the samples that extend down to at least 
$10^{10}$ M$_\odot$ 
\citep[e.g.,][]{valentinuzzi10, maltby10, cappellari13, lani13, kelkar15}
and it has been previously noticed in the local samples of early
type galaxies and discussed by 
\cite{bernardi11,bernardi14}. 
Considering a broken power law and fixing the transition mass at 
$m_t\sim2.5\times10^{10}$ M$_\odot$ , we find that the effective radius of 
elliptical galaxies scales with stellar mass to be
\begin{equation}
R_e=26\times\mathcal{M}_*^{-0.13\pm0.2}\\\\\\\\\\\\ for \\\ \mathcal{M}_*<m_t
\end{equation}
and 
\begin{equation}
R_e=2.77\times10^{-7}\mathcal{M}_*^{0.64\pm0.09} \\\\\\\ for
\\\ \mathcal{M}_*>m_t.
\end{equation}
in the case of the least square fit (see appendix A for orthogonal fit).
Hence, the effective radius remains nearly constant
at $\sim1$ kpc and  the relation is nearly flat for stellar masses 
lower than $\sim3\times10^{10}$ M$_\odot$, while it systematically increases
at larger masses.
It follows that, at masses $<3\times10^{10}$ M$_\odot$,  the stellar mass 
density $\Sigma_{R_e}$ of galaxies increases rapidly with mass up to the transition 
mass  where it reaches the maximum. 
Indeed, as can be seen from Fig. \ref{rhoe_mass_re}, the highest values 
of $\Sigma_{R_e}$ are for galaxies with masses close to $3\times10^{10}$ M$_\odot$.
At higher masses, the radius increases as R$_e\propto\mathcal{M}_*^{0.64}$
and the stellar mass density decreases.
Given the similarity, this behavior has to be related to the zone of 
exclusion (ZOE) empirically defined by \cite{bender92} and by \cite{burstein97} for 
galaxies at small sizes and at large densities.
This is shown in Fig. 7 where we also reproduce
the best fitting ZOE relation obtained by \cite{cappellari13} in the
local universe from the 
ATLAS$^{3D}$ sample.
They define the relation in their eq. (4) by using the major axis 
R$_e^{maj}$ instead of the effective radius.
We derive R$_e$, by re-scaling the major axis R$_e^{maj}$ of eq. (4)
of a factor $<R_e/R_e^{maj}>=0.76$, the mean value derived from 
the ATLAS$^{3D}$ sample \citep{cappellari13a}.
We do not further investigate here the meaning of the transition mass. 
%and we refer to a furthcoming paper for a larger discussion. 

\section{On the effective and the central stellar mass densities of elliptical galaxies}

\subsection{The effective stellar mass density}
In the upper panels of Fig. \ref{rhoe_mass_re}, the effective stellar mass density 
$\Sigma_{R_e}=\mathcal{M}_*/2\pi R_e^2$ of elliptical galaxies at 
$z\sim1.3$ is shown as a function of their effective radius (left) and 
of their stellar mass  (right).
Superimposed onto our data, the samples of \cite{raichoor11,raichoor12} 
(R11), of \cite{delaye14} and {  the one selected from CANDELS} are also shown.

The correlation between the effective mass density and the effective
radius is a different way to represent the Kormendy relation
being $\Sigma_{R_e}$ related to $<\mu_e>=-2.5log<I_e>$ through the mass to 
light ratio of the galaxy.
Actually, the definition of $\Sigma_{R_e}$ here and commonly adopted does
not take into account the presence of color gradients 
observed in most of the early-type galaxies at high redshift
\citep[e.g.,][Ciocca et al. 2016, in prep.]{gargiulo11, gargiulo12, guo11, chan16}.
Color gradients imply that the M/L is not radially constant and that, 
consequently, the stellar mass profile does not follow exactly the surface 
brightness profile.
Hence, 50\% of the stellar mass will not be contained within the half-light 
radius R$_e$, and the fraction of the stellar mass within R$_e$ will be slightly 
but systematically different than 0.5.
Indeed, the observed color gradients of elliptical galaxies at these redshift
are systematically in one direction or null, both for field ellipticals
\citep{gargiulo11,gargiulo12,guo11} and for cluster ellipticals (Ciocca et al
2016).
%Since the color gradients observed in elliptical galaxies at these redshift
%are sistematically negative or null, 
Hence, this would reflect in a small offset of the relation along
$\Sigma_{R_e}$ and, since the variation can slightly differ from galaxy to 
galaxy, this would contribute to the observed scatter of the relation.

The correlation shown in Fig. \ref{rhoe_mass_re} is best fitted by the relation 
\begin{equation}
log(\Sigma_{Re})=-1.2(\pm0.1)\times log(R_e)+3.67(\pm0.05),
\end{equation} 
where $\Sigma_{Re}$ [M$_\odot$ pc$^{-2}$] and R$_e$ [kpc].
 It is worth noting that the best fitting Kormendy relation obtained for the
 whole sample  (see Table \ref{fitkr}, $<$All$>$)
 being $-2.5log<I_e>\propto3.2 R_e$ , would imply a slope -1.28 of the 
 $\Sigma_{R_e}-R_e$ relation if the M/L was constant along the Kormendy relation.
However, given the small age gradient (see \S\ 6), it follows that a variation of 
the M/L along the relation takes place, generating
the small difference in the slope.

No significant difference is found by fitting field and cluster galaxies 
separately, as in the case of the Kormendy relation.
Field and cluster galaxies follow the same $\Sigma_{Re}$-R$_e$
relation and occupy the same region of the plane, apart from the lack of massive and 
large galaxies in the field with respect to the cluster environment.
The same slope is obtained by fitting the sample of \cite{raichoor11} 
($b\simeq-1.3$), while a slightly steeper slope ($b\simeq-1.5$) is found 
for the sample of \cite{delaye14}.
We note that the slope of this relation ($\sim-1.2$) seems to
be constant over a wide redshift range:
studying the mass fundamental plane 
of early type galaxies at $z\sim0$, \cite{hyde09} find a slope
$b\simeq-1.19$; \cite{zahid15} find this slope 
 for quiescent galaxies at $z\sim0.7$ \citep[see also][]{bezanson15}, 
 while \cite{bezanson13} find that this slope well reproduces compact massive 
 galaxies out to $z\sim2$.
%We study and discuss the evolution of the $\Sigma_{Re}\propto R_e^{b}$
 %relation and its implications in a forthcoming paper.

It is important to note that Fig. \ref{rhoe_mass_re}, besides showing that
cluster and field elliptical galaxies occupy the same locus in the 
$\Sigma_{Re}-\mathcal{M}_*$  plane (upper right-hand panel), 
shows that $\Sigma_{Re}$ is not correlated to the stellar mass of the galaxy.
Galaxies with high or low effective stellar mass densities can be realized 
independently of their stellar mass, that is, dense/compact galaxies can be
assembled in any mass regime independently of
the environment in which they reside, cluster or field.
Actually, the data show that not all the values of $\Sigma_{Re}$ can be realized 
for a given mass, according to the ZOE relation (see Fig. \ref{zoe}) that defines 
the locus that galaxies occupy in the [$\Sigma_{R_e}$-R$_e$,$\mathcal{M}^*$] planes.
As noted in the previous section, the highest values of effective stellar mass
density are reached by galaxies with stellar mass close to the transition mass 
$\sim3\times10^{10}$ M$_\odot$ that, in our sample is 
$\Sigma_{R_e}\sim22600$ M$_\odot$pc$^{-2}$ for a galaxy of mass
$\sim3.5\times10^{10}$ M$_\odot$.

\subsection{The central stellar mass density}
In Fig. \ref{rhoe_mass_re} (lower panels) the central stellar mass density
$\Sigma_{1kpc}$ defined as the mass density within 1 kpc radius, is shown 
as a function of the effective radius (left) and of the stellar mass (right).
Following \cite{saracco12}, we derived the stellar 
mass interior to 1 kpc multiplying the luminosity within this fixed
radius by the mass-to-light ratio of the galaxy 
\begin{equation}
\mathcal{M}_{1kpc}={L_{1kpc}}\times\left ({\mathcal{M}_*\over L_{tot}}\right )_{gal}
={\gamma(2n,x)\over \Gamma(2n)}\times\mathcal{M}_*,
\end{equation}
where 
\begin{equation}
L_{1kpc}=2\pi I_eR_e^2  n {e^{b_n}\over{b_n^{2n}}}\gamma(2n,x)
\label{l1}
\end{equation}
is the luminosity within the central region of 1 kpc, 
L$_{tot}$ is the total luminosity obtained by replacing in Eq. \ref{l1} 
$\gamma(2n,x)$ with the complete gamma function $\Gamma(2n)$
\citep{ciotti91}, and $\mathcal{M}_*$ is the stellar mass of the 
galaxy obtained from the SED fitting.
In Eq. \ref{l1}, $n$ is the S\'ersic's index, $x=b_n(R_1/R_e)^{1/n}$ , and 
$\gamma(2n,x)$ is the incomplete gamma function.
We assumed the analytic expression $b_n=1.9992n-0.3271$ 
\citep{capaccioli89} to approximate the value  of $b_n$. 

Figure \ref{rhoe_mass_re} shows that the central stellar mass density 
$\Sigma_{1kpc}$ is
tightly correlated with the stellar mass of galaxies, as already shown by
\cite{saracco12}, while it is independent of the effective radius,
at the opposite of the effective mass density.
The correlation is best fitted by the relation
\begin{equation} 
log(\Sigma_{1kpc})=0.64(\pm0.06)\times log\mathcal{M}_*-3.2(\pm0.6),  
\end{equation}
in good agreement with what was previously found for early-type 
galaxies both at $1<z<2$ and in the local universe \citep{saracco12}, 
for massive elliptical galaxies in cluster at $z\sim0.5$ \citep{bai14},
but also for star forming galaxies at $z\sim2.2$ \citep{tacchella15} and 
in the local universe \citep{fang13}.
Also in this case, cluster and field ellipticals follow the same
relation.

However, it can be seen that also the $\Sigma_{1kpc}$-mass relation follows
two different regimes, above and below 
the transition mass:
at masses $m_t<2.5\times10^{10}$ M$_\odot$ the relation tends to 
steepen with $\Sigma_{1kpc}\propto\mathcal{M}_*^{1.07\pm0.1}$.
The transition mass defines a transition central mass density  
$\Sigma_{1kpc}\simeq2-3\times10^3$ M$_\odot$ pc$^{-2}$.
It is interesting to note that the presence of two different regimes and the 
slopes that we find agree with 
that found by \cite{tacchella16} in their simulations.
They find a flattening of the $\Sigma_{1kpc}$-mass relation when the galaxies 
reach  masses $\mathcal{M}_*>10^{10.2}$ M$_\odot$  (see their fig. 8).
The two different regimes they found should represent 
two different phases in the stellar mass growth of the galaxies.   

\begin{figure}
\includegraphics[width=9truecm]{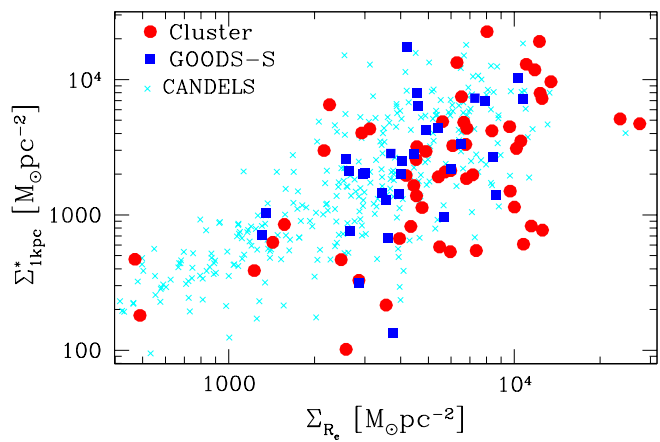}
\caption{Central versus 
effective stellar mass density. 
The central stellar mass  density calculated within 1 kpc radius 
($\Sigma_{1kpc}$) is plotted as a function of the effective stellar
mass density ($\Sigma_{R_e}$). 
Blue filled squares are field ellipticals in the GOODS-S regions,
red filled circles are cluster ellipticals. 
Cyan crosses are elliptical galaxies from CANDELS.  
{  Errorbars are not shown to observe the agreement among the different samples
and the relationship between the quantities.} 
}
\label{rho1_rhoe}
\end{figure}

We note that high-mass galaxies are centrally denser than
lower mass ones but this is not true when the effective radius 
is considered.
Indeed, as shown in Fig. \ref{rho1_rhoe} \citep[see also][]{saracco12}, 
the correlation between the central and 
the effective stellar mass density is characterized by a very large scatter and,
as shown by Fig. \ref{rhoe_mass_re}, the same value of effective mass
density can be realized over a wide range of stellar mass, especially low
density values. 
Hence, it is the galaxy central region which is strictly connected to the total 
stellar mass of the galaxy and, consequently,  to its main mass growth.
The tight linear relation between the central mass density and the stellar mass 
suggests a direct connection between the central regions and the earliest
phases of formation during which most of the stellar mass is formed and assembled.

Contrary to the central density, the effective mass density and hence the 
regions at larger radii, are very weakly dependent on 
the mass suggesting that they are not good tracers of the growth of the bulk
of the mass of the galaxy.   
The effective radius and hence the effective density, 
may have been strongly affected by episodes of relaxation, 
of inside-out quenching \citep{tacchella16}, and of minor mass accretion 
(minor mergers) that may have taken place after the growth of the bulk of
the mass and before $z\sim1.3$.
This could explain the large scatter in the $\Sigma_{1kpc}-\Sigma_{R_e}$
relation.  
The scaling of the central mass density with mass suggests
that the process of assembly and of shaping of the elliptical
is invariant with mass, at least in the mass range considered.  

\begin{figure*}
%\hskip -0.5truecm
%\includegraphics[width=12truecm]{mean_age_med.ps}
\includegraphics[width=9.8truecm]{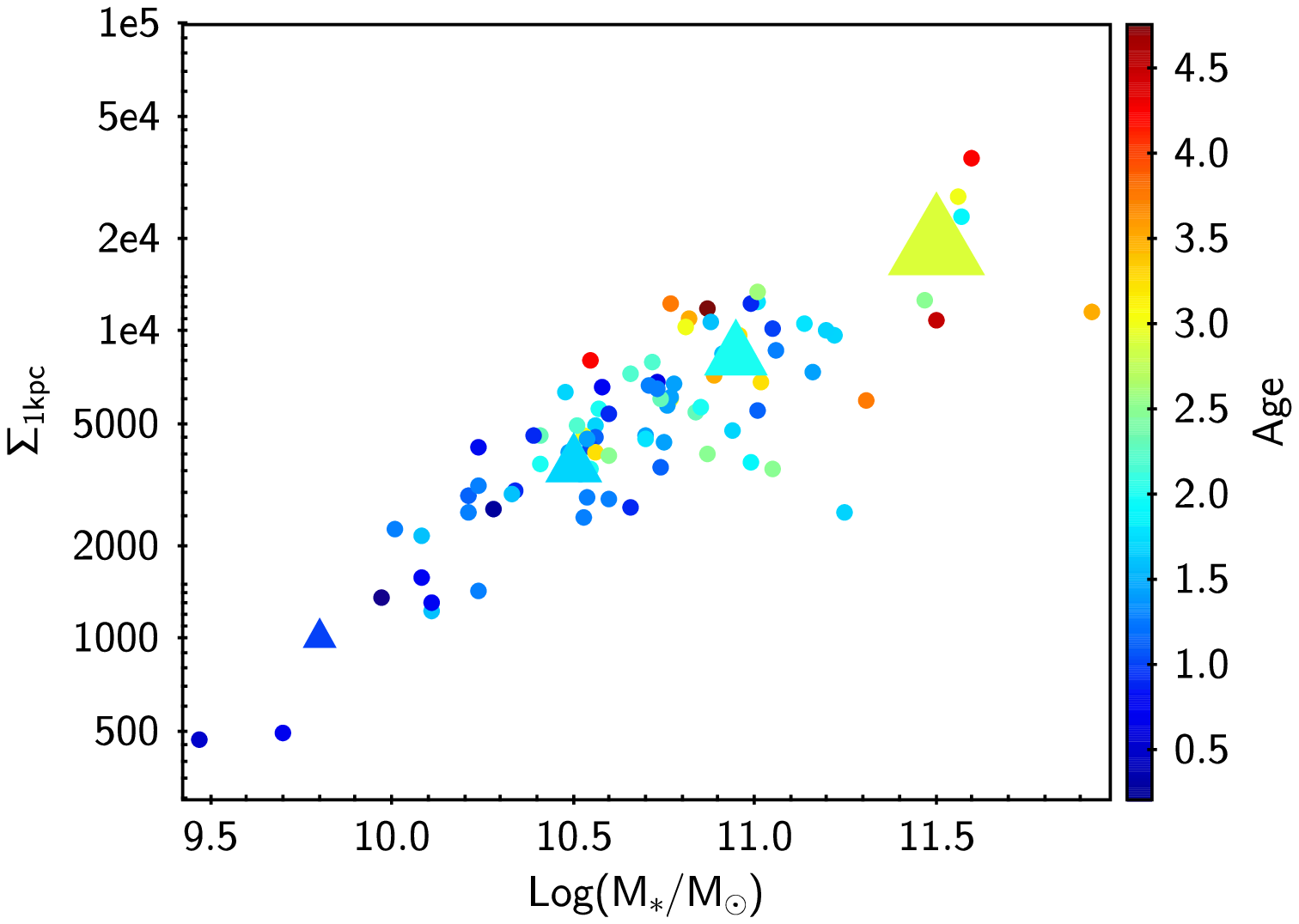}
%\hskip -0.6truecm
\includegraphics[width=9.8truecm]{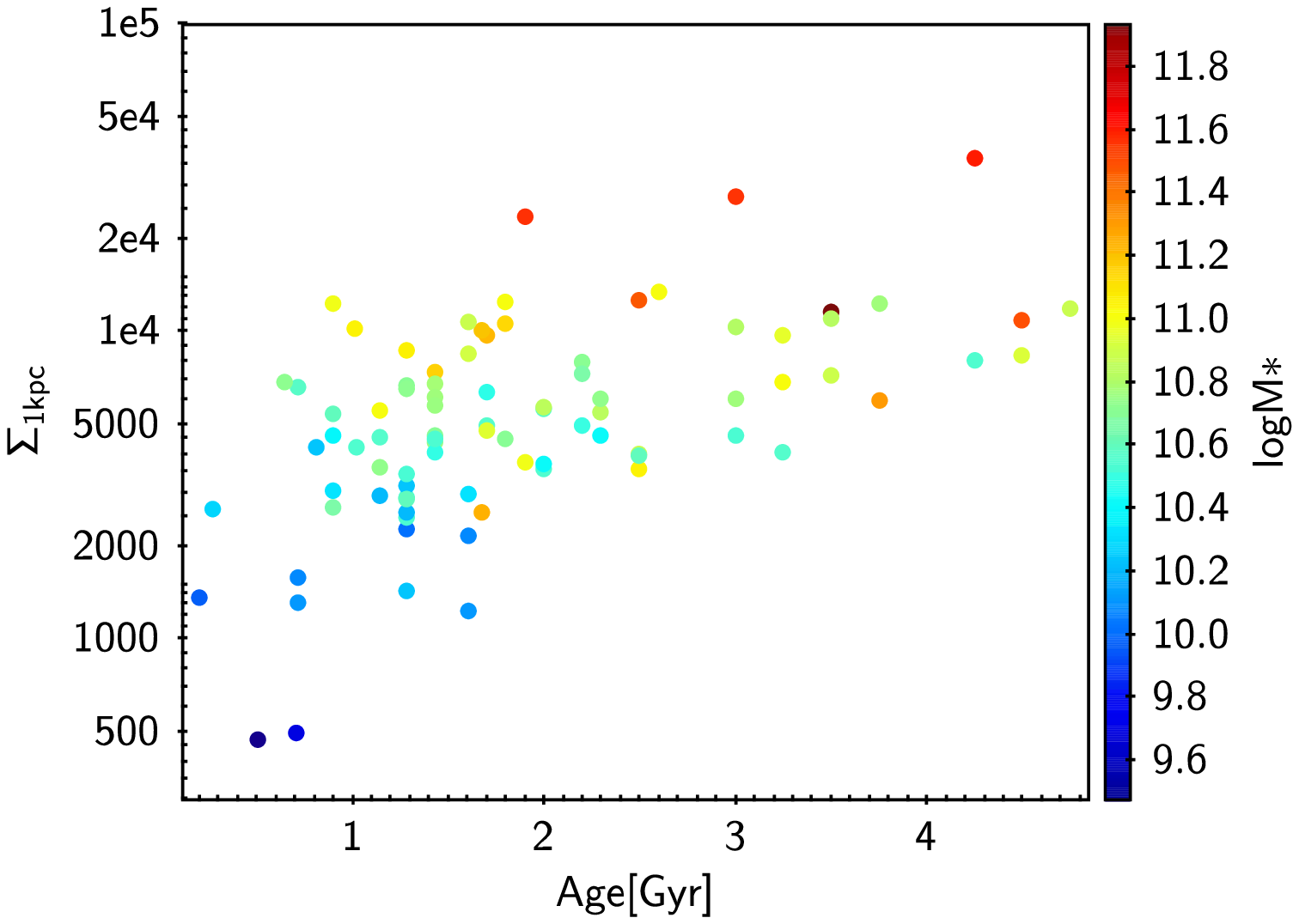}
 \caption{Central stellar mass density $\Sigma_{1kpc}$ versus 
 stellar mass (left) and versus age (right).
Left - The surface stellar mass density $\Sigma_{1kpc}$ within 1 kpc 
radius is plotted as a function of the stellar mass of the galaxies.
The different colors mark the different mean age of the stellar population 
of the galaxies resulting from the best fitting to their SED. 
The large colored triangles represent the median age in four bins
of stellar mass.
Right - The central stellar mass density of cluster and field elliptical
galaxies at $z\sim1.3$ is plotted as a function of the age of their
stellar population. 
Symbols are colored as a function of stellar mass.  
{  Errorbars are not shown to highlight the relationship between the quantities.} 
}
\label{c1_age}
\end{figure*}

Figure \ref{c1_age} shows the central stellar mass density as a function
of the stellar mass for field and cluster elliptical galaxies. 
Symbols have different colors according to the age resulting from the
best fitting to their SED.
The large symbols represent the median age of galaxies in four different 
mass bins.
The figure shows that the higher the central mass density, the higher the mass, 
 the older the stellar population.
 The correlation between central density and age is better shown in the left
 panel of the figure.
The higher central mass density of high-mass galaxies reflects the
higher normalization of their mass profile with respect to lower mass ones 
and, again, the close connection with the main episode of star formation.
The older age may suggest that higher mass galaxies underwent less subsequent
episodes of star formation at later times or that their massive burst 
takes place earlier than lower mass ones.

\section{Summary and conclusions} 
In this paper we focused our investigation on the dependence of the 
population of elliptical galaxies at $z\sim1.3$ and of their properties on 
the environment.
We constructed two main samples of elliptical galaxies at the same redshift,
the first one  composed of 56 galaxies selected in three clusters at 
$1.2<z<1.4$, the second one  composed of 31 galaxies selected in the
GOODS-South field at the same redshift.
A third and larger ($\sim180$ galaxies) sample of field galaxies has been 
extracted from the COSMOS catalogs at slightly lower redshift ($1.0<z<1.2$)
and used when it has been possible in some of the comparisons.

The selection of galaxies has been made on the basis of a pure
morphological criterion based on the visual inspection of their
luminosity profile in the ACS-F850LP image and of the residuals 
resulting from the profile fitting with a regular Sersic profile.
The narrow redshift range adopted minimized the evolutionary effects,
while the morphological selection of galaxies produced samples with the 
same composition, allowing us to single out the effect of 
the environment at a given morphology. 

We compared the physical and structural properties of the population
of elliptical galaxies in the two environments and we derived and compared
the relationships among effective radius, surface brightness, stellar
mass, and stellar mass density.
Our main results can be summarized as follows:
\begin{itemize}
\item 
The structure and the properties of cluster elliptical galaxies do not 
differ from those in the field. 
Cluster and field elliptical galaxies have the same median effective
radius, the same mass normalized radius, and the same stellar mass density at 
fixed mass. 
\item 
Cluster and field elliptical galaxies at $z\sim1.3$ follow the same 
Kormendy (size-surface brightness) relation with a slope $\beta\simeq3.0$.
They also follow the same size-mass relation and the same size-mass density 
relations.
The comparison of our data with those from the literature at the same redshift
and with comparable selection criteria shows excellent agreement.
\item  
The population of cluster elliptical galaxies differs from the one in the field
for high-mass and large elliptical galaxies.
Indeed,  there is a significant lack of massive 
($\mathcal{M}_*>2\times10^{11}$) M$_\odot$ and large (R$_e>4-5$ kpc) elliptical 
galaxies in the field with respect to the cluster.
   Nonetheless, at $\mathcal{M}_*< 2 \times 10^{11}$ M$_\odot$, the two
   populations are similar. 
There seems to be also a lack of elliptical galaxies in the field which are as old as
the oldest ellipticals in cluster.
\end{itemize} 
Hence, we do not find a dependence of the structure of elliptical galaxies
on the environment where they belong, contrary to what has been predicted by some
recent simulations.
The above results show that the structure and the shaping of elliptical 
galaxies at $z\sim1.3$ do not depend on the environment where they belong.
However, they suggest that a dense environment is more efficient in assembling
high-mass elliptical galaxies.

From the study of the scaling relations, we obtained the following
results:
\begin{itemize}
\item
The size-mass relation is only coarsely best fitted by a 
single power law of the form R$_e\propto\mathcal{M}_*^{0.5}$.
We find a transition mass $m_t\sim2-3\times 10^{10}$ M$_\odot$,
in agreement with local studies, that defines
two different regimes of the size-mass relation.
Above this mass, the relation is steeper 
(R$_e\propto\mathcal{M}_*^{0.64}$), while below this mass the relation gets flat 
(R$_e\propto\mathcal{M}_*^{-0.13}$), and the effective radius is
nearly constant at $\sim$1 kpc.
The transition mass marks the mass at which a galaxy can reach the maximum
stellar mass density within the effective radius.
\item
The stellar mass density within the effective radius is tightly
correlated with the effective radius.
This size-effective mass density relation is best fitted by 
$\Sigma_{R_e}\propto R_e^{-1.2}$ in agreement with the Kormendy relation.
Galaxies with high or low effective stellar mass densities can be realized 
independently of the environment in which they reside
and almost independently of their mass.
The data show that a galaxy cannot occupy any locus in the 
[$\Sigma_{Re},\mathcal{M}_*$] plane but that a Zone of Exclusion, 
as the one defined by the early-type galaxies in the local universe, 
exists at $z\sim1.3$.
\item
The central stellar mass density within 1 kpc radius is tightly correlated
with mass and well fitted by $\Sigma_{1kpc}\propto\mathcal{M}_*^{0.64}$,
as previously found, at masses larger than $m_t$.
At lower masses, the relation steepens with $\Sigma_{1kpc}\propto\mathcal{M}_*^{1.07}$. 
We find that the central mass density is also correlated with the 
age of the stellar population such that  the higher
the central stellar mass density, the older the age, the higher the mass.
\end{itemize}
High mass galaxies are characterized by correspondingly high central stellar 
mass densities and old stellar population. 
The scaling of the central mass density with mass suggests
that the process of assembly and of shaping of elliptical galaxies 
does not depend on the mass, at least in the mass range considered.  
These correlations taken all together, are indicative of the close
connection of the central regions of the galaxies to the earliest phases  
of formation. 
The central regions of elliptical galaxies most probably store the information 
on their assembly, retain memory of the initial conditions, and are strictly 
connected to the bulk of their stellar mass growth.
The outer regions (effective stellar mass density and effective radius) instead, 
store the information on the (subsequent) events that the galaxy may have 
experienced and that have affected the outer structure, but not its mass.

\section*{Acknowledgments}
We are grateful to the anonymous referee for the constructive comments and
suggestions.
PS wishes to thank Stefano Andreon for the usefull comments.
This work is based on data collected at the European Southern 
Observatory (ESO) telescopes (Prog. ID 084.A-0214(A), 60.A-9284(H))
and with the NASA/ESA 
Hubble Space Telescope, obtained from the 
data archive at ESO and at the Space Telescope Science Institute which is 
operated by the Association of Universities for Research in Astronomy. 

This work is also based on observations carried out at the Large Binocular 
Telescope (LBT; Prog. ID 19\_2014). 
The LBT is an international collaboration among institutions in the 
United States, Italy, and Germany. 
We acknowledge the support from the LBT-Italian Coordination Facility for the
execution of observations, data distribution, and reduction.

This work made use of the Rainbow Cosmological Surveys Database, which is 
operated by the Universidad Complutense de Madrid (UCM), partnered with the 
University of California Observatories at Santa Cruz (UCO/Lick,UCSC)

\nocite{}
\bibliographystyle{aa}
\bibliography{paper_cl_field_r1}

\begin{appendix}

\section{Data description}
\subsection{Cluster data}
\label{ap:cluster}
We describe here the data available and used for the three clusters.

{XMMJ2235-2557} - The HST data are composed of optical ACS 
observations in the filters F775W (5060 s) and F850LP (6240 s), and 
of near-IR WFC3 observations in the filter F160W (1200 s) 
described in \cite{rosati09} and in \cite{strazzullo10}.  
The ACS and the WFC3 images used here have a pixel scale of 0.05 
arcsec/pixel and a resolution of FWHM$_{F850}\simeq0.11$ arcsec and of 
FWHM$_{F160}\simeq0.2$ arcsec.
VLT data include VIMOS U-band observations 
\citep[21000 s, FWHM$\simeq0.8$ arcsec,][]{nonino09}, 
and HAWKI J- and K-band observations 
\citep[$\sim10500$ s,][]{lidman08, lidman13} characterized
by a FWHM$_J\simeq0.5$ arcsec and FWHM$_K\simeq0.35$ arcsec.
Spitzer data are composed of fully co-added IRAC mosaics (0.6 arcsec/pixel) 
in the four bandpasses 3.6$\mu$m, 4.5$\mu$m, 5.8$\mu$m ($\simeq2100$ s), 
and 8.0 $\mu$m ($\simeq1900$ s). 
Spectroscopic redshifts for five galaxies have been derived from \cite{rosati09}.
A detailed description of this data set and of this sample is provided 
by Ciocca et al. (2016, in prep).
 
{RDCSJ0848+4453} - The HST data includes ACS observations (0.05 arcsec/pixel) 
in the filters  F775W (7300 s) and F850LP (12200 s, FWHM$\simeq0.12$ arcsec) 
described in \cite{postman05}, and of near-IR NICMOS-NIC3 observations (0.2
arcsec/pixel) in the filter F160W (11200 s; FWHM$\simeq0.22$ arcsec)
described in \cite{vandokkum01}. 
LBT data are composed of LBC observations in the  U, B, V, and R filters
\citep[$\sim$14000 s each; see][]{saracco14} and are characterized by a 
FWHM$\sim1.0$ arcsec.
Spitzer fully co-added mosaics (0.6 arcsec/pixel) in the four IRAC 
bandapasses 3.6$\mu$m and 5.8$\mu$m ($\sim$2200 s), 4.5$\mu$m, and
8.0 $\mu$m ($\simeq1900$ s) complete this data set. 
Spectroscopic redshifts are available for 12 galaxies from \cite{jorgensen14}
and \cite{vandokkum01}. 
A complete and detailed description of this data set is provided by 
\cite{saracco14}.

{XLSSJ0223-0436} - The data set used for this cluster is composed 
of HST-ACS images  in the F775W (2000 s) and F850LP (6000 s) filters, 
proprietary LBT-LBC observations (Prog. ID 19\_2014) 
in the V and I filters ($\sim 6300$ s and 
$\sim4400$ s respectively, FWHM$\simeq0.7$ arcsec), VLT-HAWKI observations 
 in the J and K filters ($\sim 3000$ s each, FWHM$\simeq 0.35$ arcsec), and
Spitzer fully co-added mosaics (0.6 arcsec/pixel) in the four IRAC 
bandapasses 3.6$\mu$m, 4.5$\mu$m, 5.8$\mu$m, and 8.0 $\mu$m.
Spectroscopic redshift are available for about 14 galaxies from 
\cite{bremer06} and Saracco et al. (2016, in prep.), where  a detailed 
description of this data set is also given.

\begin{figure}
\label{fig:z_zspec}
\begin{center}
\includegraphics[width=7.truecm]{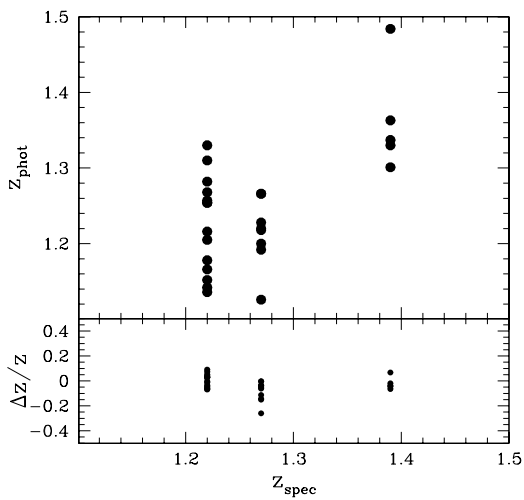}
 \caption{Comparison between photometric and spectroscopic redshifts
 for the elliptical cluster members.
}
\end{center}
\end{figure}

{  In Fig. \ref{fig:z_zspec} we show the comparison between the photometric
redshift and the spectrosopic redshift for the elliptical galaxies
spectroscopically confirmed cluster members.
The photometric redshift accuracy is $\sigma_{\Delta z/(1+z_s)}=0.04$
(0.02 using the normalized median absolute deviation).}
 
\subsection{CANDELS data sample}
\label{ap:candels}
In order to homogenize as much as possible the parameters of CANDELS
galaxies with those of our samples, we considered the 31 elliptical galaxies
selected in the GOODS-S field in common with the CANDELS catalog.
We first checked the agreement between the morphological classification and
between the main parameters used in this analysis: 
the stellar mass and the effective radius.

As far as the morphological classification is concerned, all the 31 galaxies we classified 
as ellipticals in the F850LP images have dominant class=0, that is, they are 
classified spheroids also in the CANDELS catalog.
Four of them have the flag parameter value on the goodness
of the Galfit fit parameter\_gfit\_f\_h=1  suggesting that
the fitting to their profile in the F160W band is not very good.

\begin{figure}
\includegraphics[width=4.4truecm]{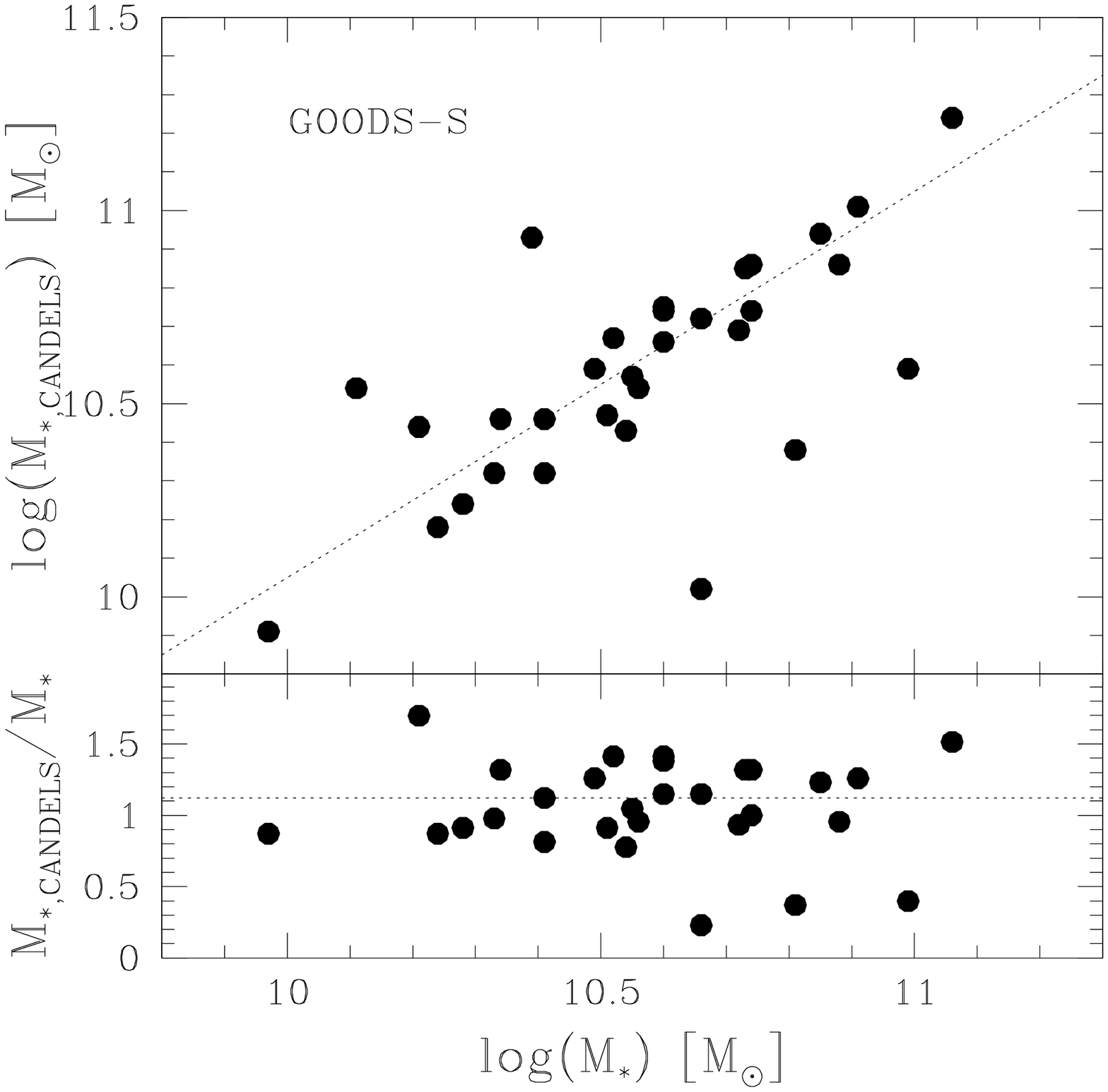}
\includegraphics[width=4.4truecm]{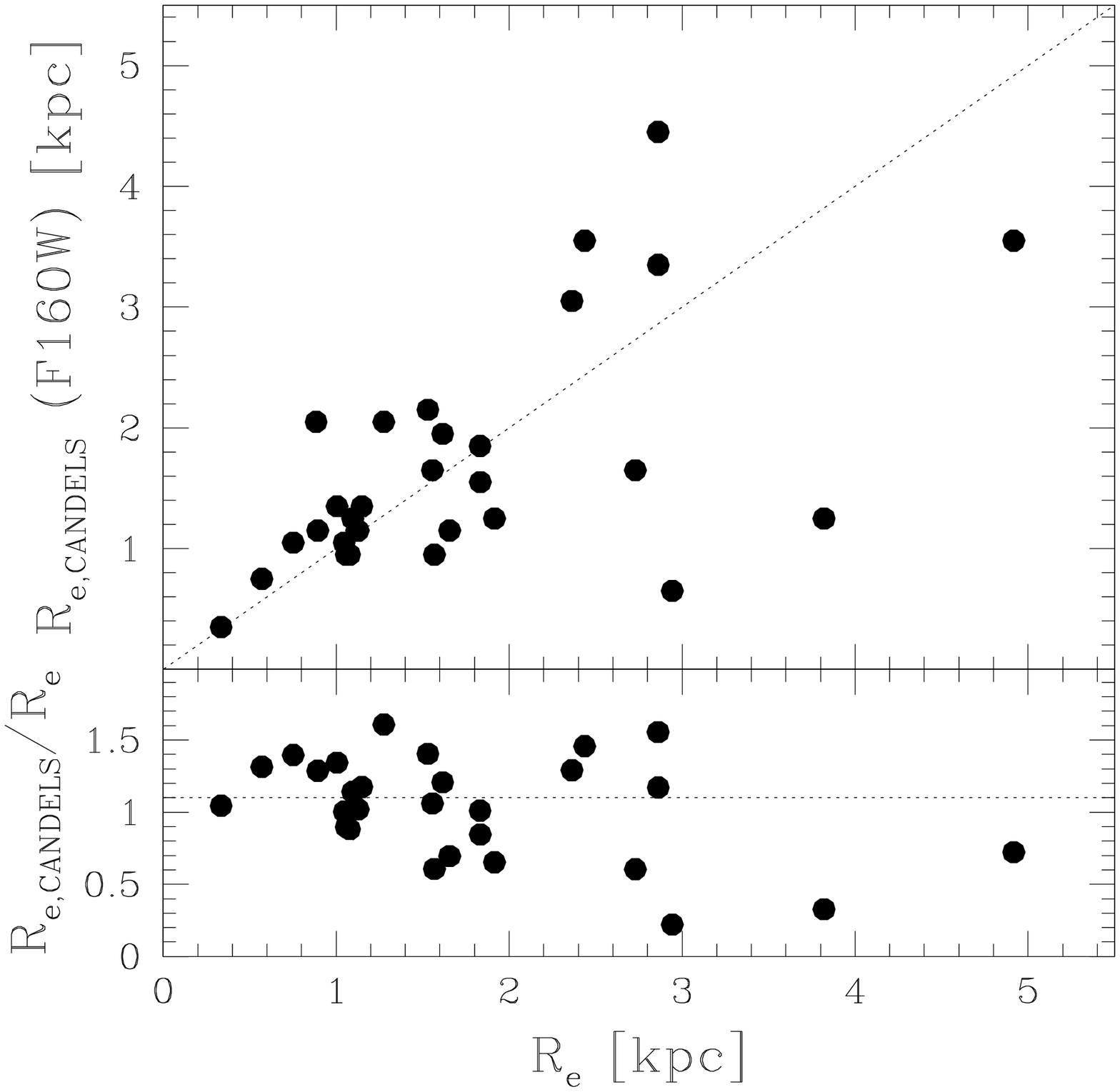}
 \caption{ Left - The stellar mass from CANDELS catalog is compared with the
 stellar mass we estimated for the 31 galaxies of the GOODS-S field in common.
 Right - The effective radius in the F160W band from CANDELS catalog is plotted
 as a function of the effective radius we estimated in the F850LP band for
 the 31 galaxies in the GOODS-S field. 
}
\label{fig:cand_comp}
\end{figure}

In Fig. \ref{fig:cand_comp} (left panel) we compare the stellar masses
of CANDELS catalog with our estimate for the 31 galaxies in common.
The agreement is rather good; we estimate a small offset of about 
a factor 0.05dex in the sense that the stellar masses in the CANDELS catalog
seems to be slightly larger than our estimates.
We applied this small scaling to the CANDELS data.
 
In the right panel of Fig. \ref{fig:cand_comp} we compare  
the effective radii in the F160W band from CANDELS catalog with those we 
estimated in the F850LP band.
The agreement is rather good at small radii while the scatter is large
at larger radii.
We do not clearly detect the 10\%-20\% offset between the effective radii estimated 
in the two bands found in other works 
\citep[e.g.,][]{gargiulo12,cassata11}.
Possibly, a trend with the radius (the larger the F850LP band radius, the
lower the ratio R$_e$(F160)/R$_e$(850)) is present, however, the small
statistic and the large scatter do not allow us to confirm this.
For these reasons, we did not apply any scaling to the effective radii 
of CANDELS sample. 
%In any case, even a scaling of a factor 1.2 would not affect the results.
 
\section{Cluster and field data samples}
\label{ap:samples}

\begin{figure*}
\label{fig:thumb}
\centering
%\vskip -8truecm
\includegraphics[width=2.truecm]{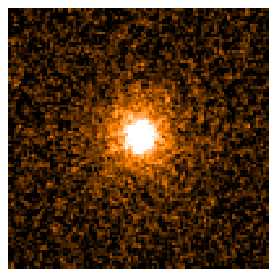}
\includegraphics[width=2.truecm]{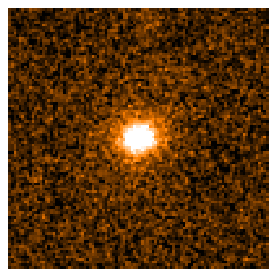}
\includegraphics[width=2.truecm]{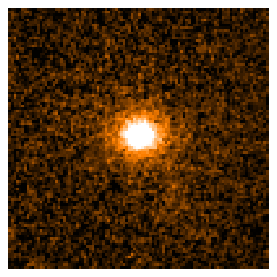}
\includegraphics[width=2.truecm]{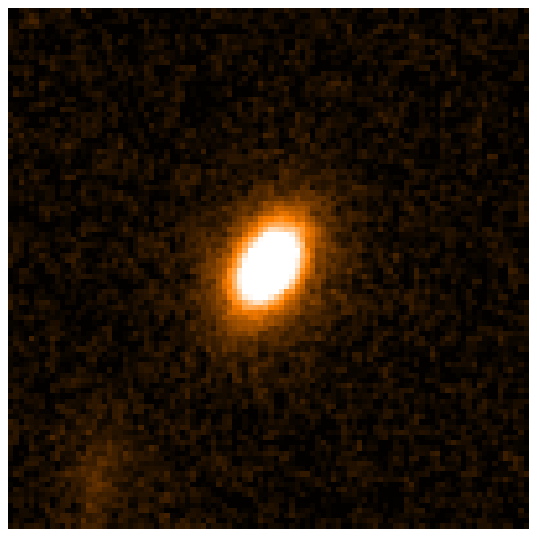}
\includegraphics[width=2.truecm]{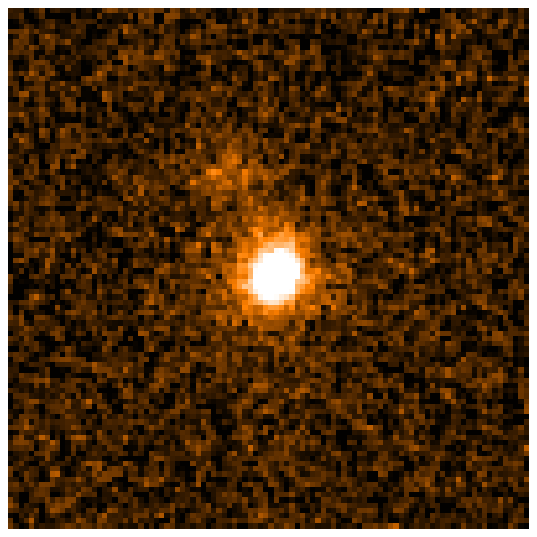}
\includegraphics[width=2.truecm]{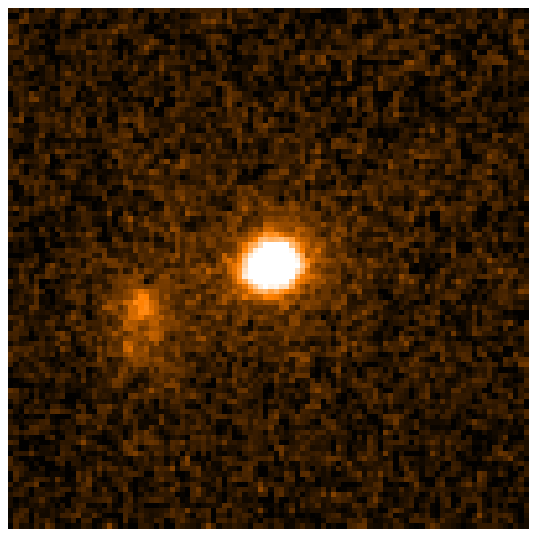}
\includegraphics[width=2.truecm]{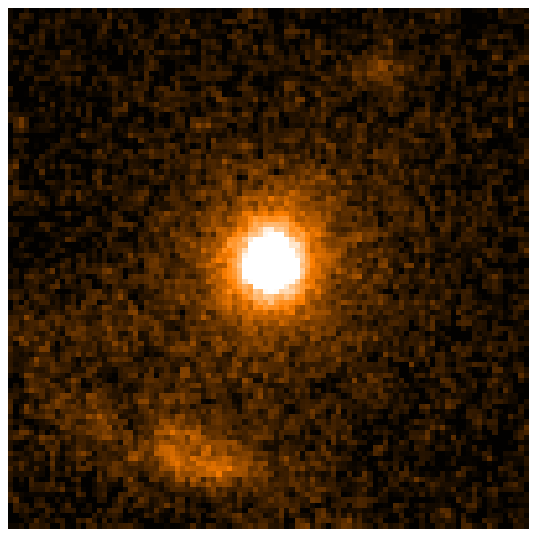}
\includegraphics[width=2.truecm]{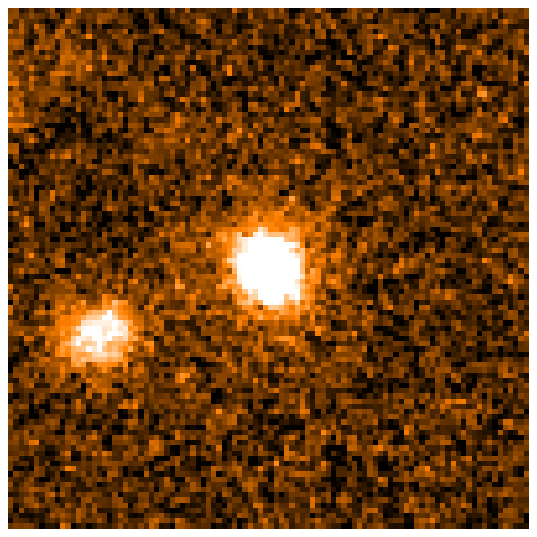}
\vskip 0.2truecm
\includegraphics[width=2.truecm]{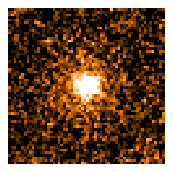}
\includegraphics[width=2.truecm]{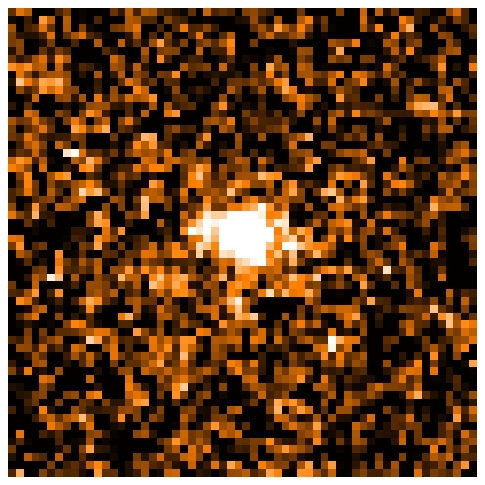}
\includegraphics[width=2.truecm]{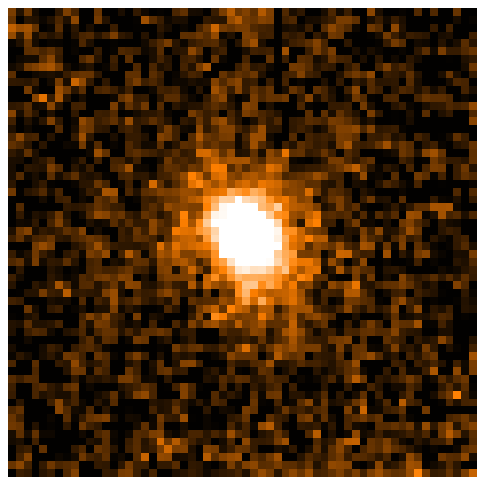}
\includegraphics[width=2.truecm]{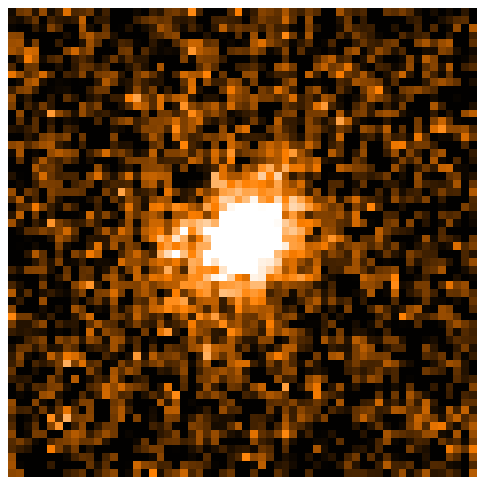}
\includegraphics[width=2.truecm]{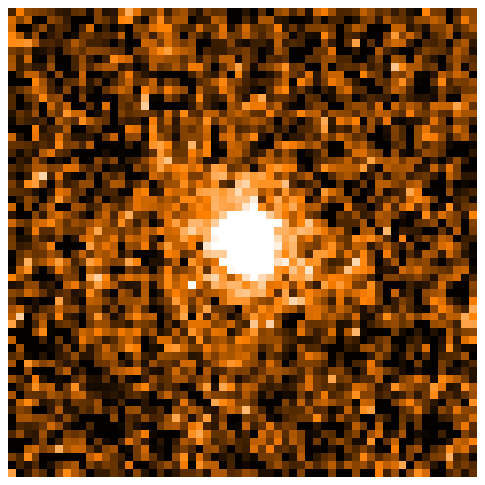}
\includegraphics[width=2.truecm]{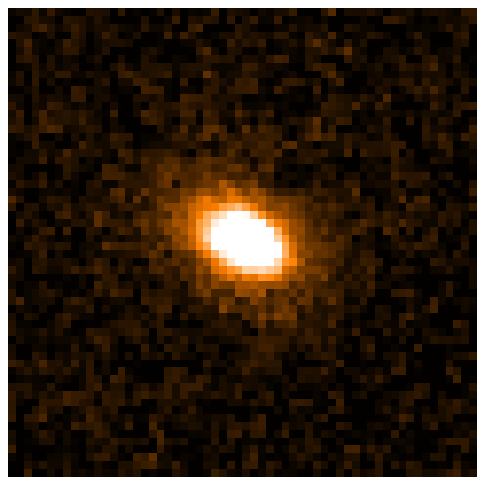}
\includegraphics[width=2.truecm]{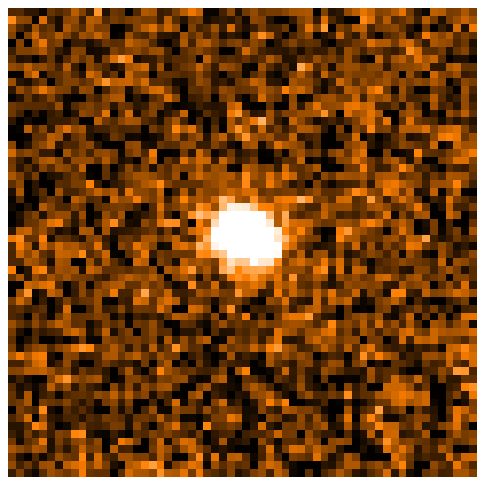}
\includegraphics[width=2.truecm]{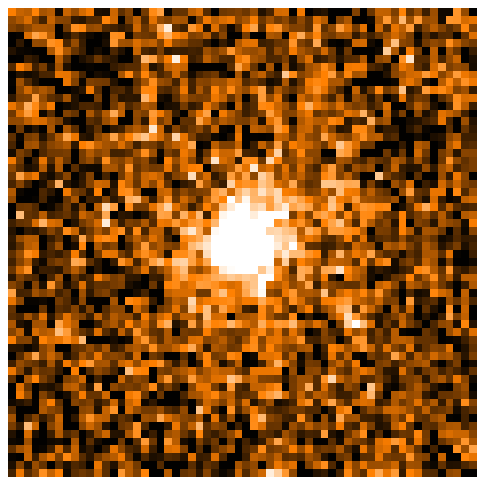}
\includegraphics[width=2.truecm]{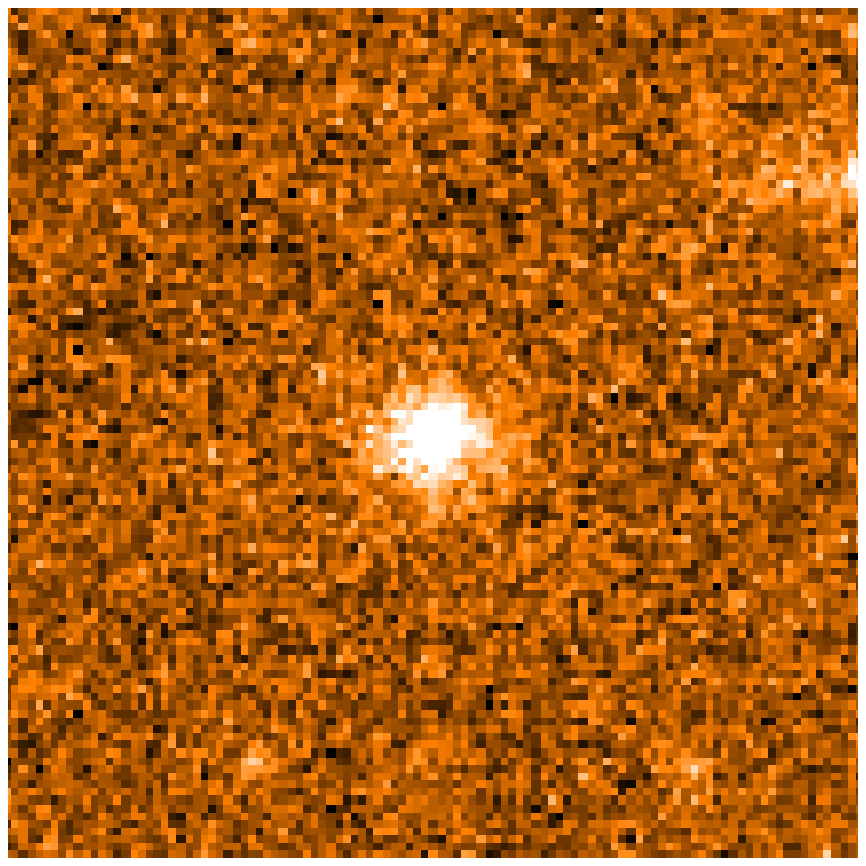}
\includegraphics[width=2.truecm]{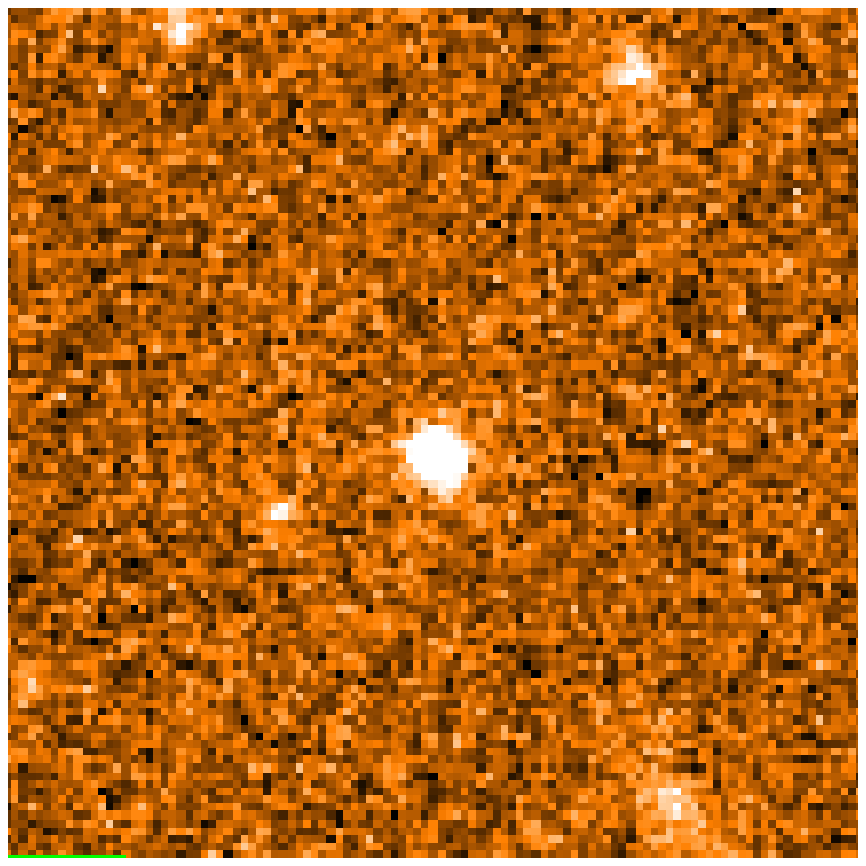}
\includegraphics[width=2.truecm]{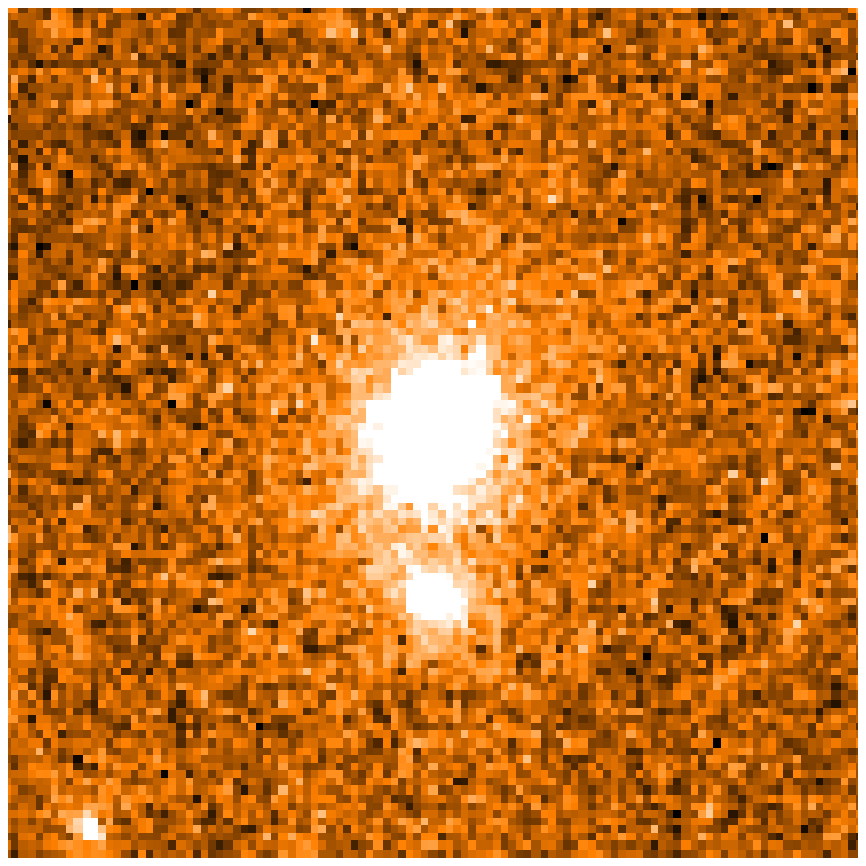}
\includegraphics[width=2.truecm]{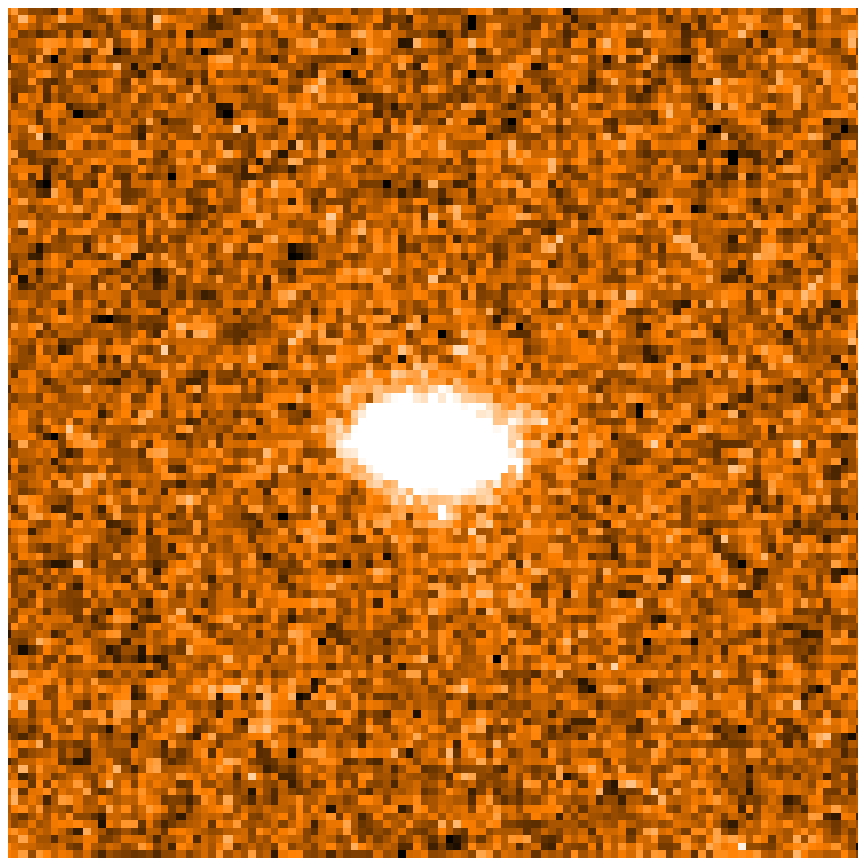}
\includegraphics[width=2.truecm]{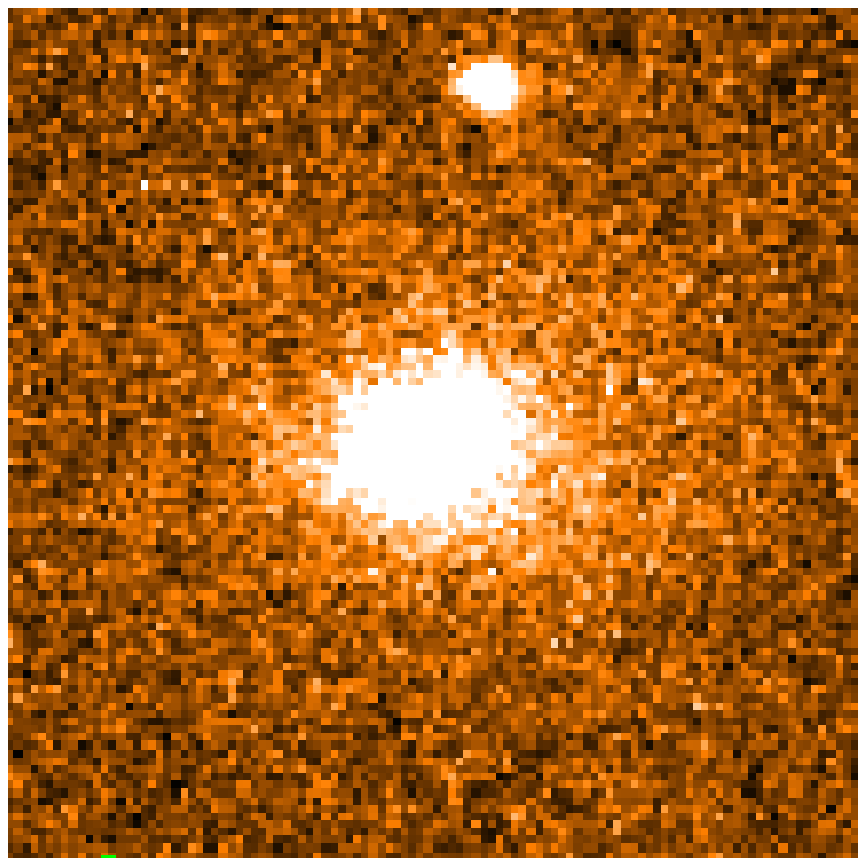}
\includegraphics[width=2.truecm]{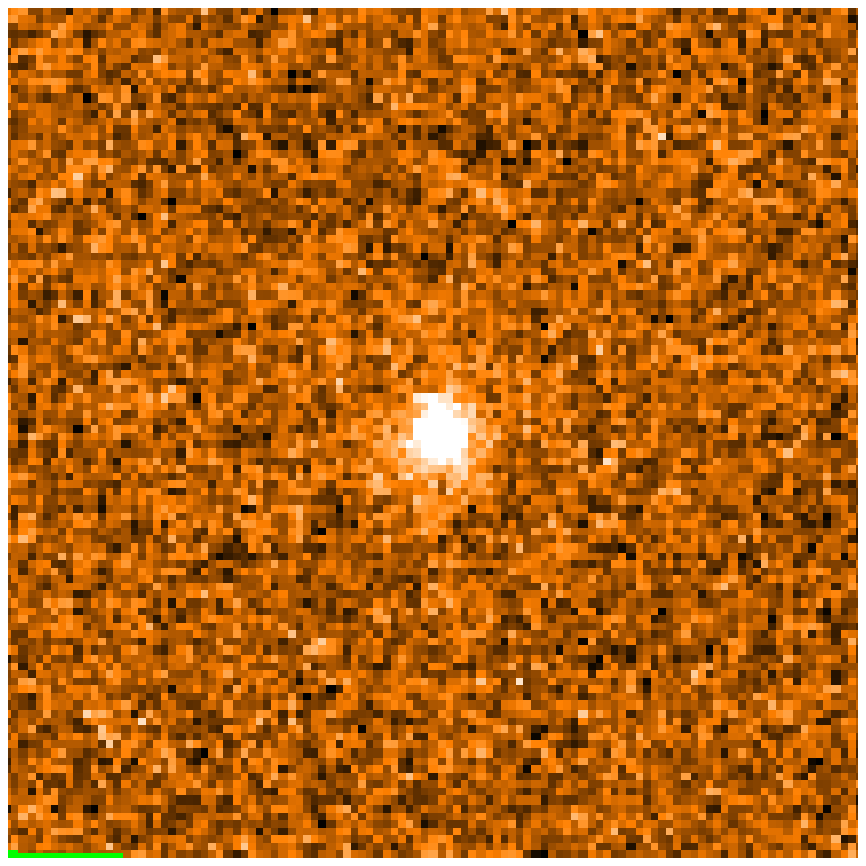}
\includegraphics[width=2.truecm]{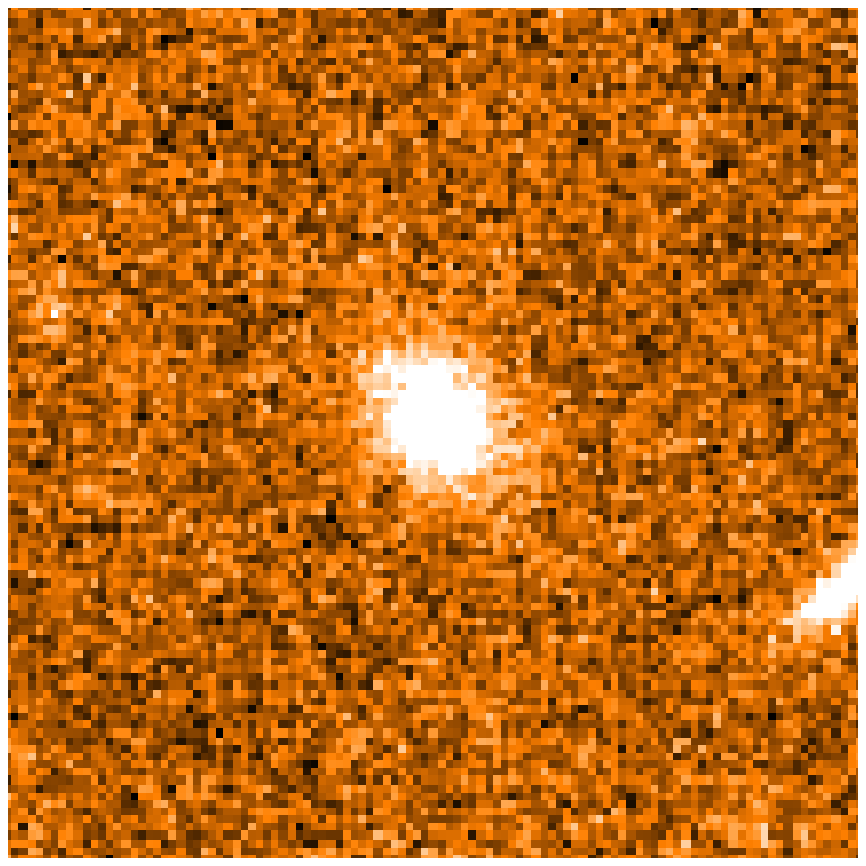}
\includegraphics[width=2.truecm]{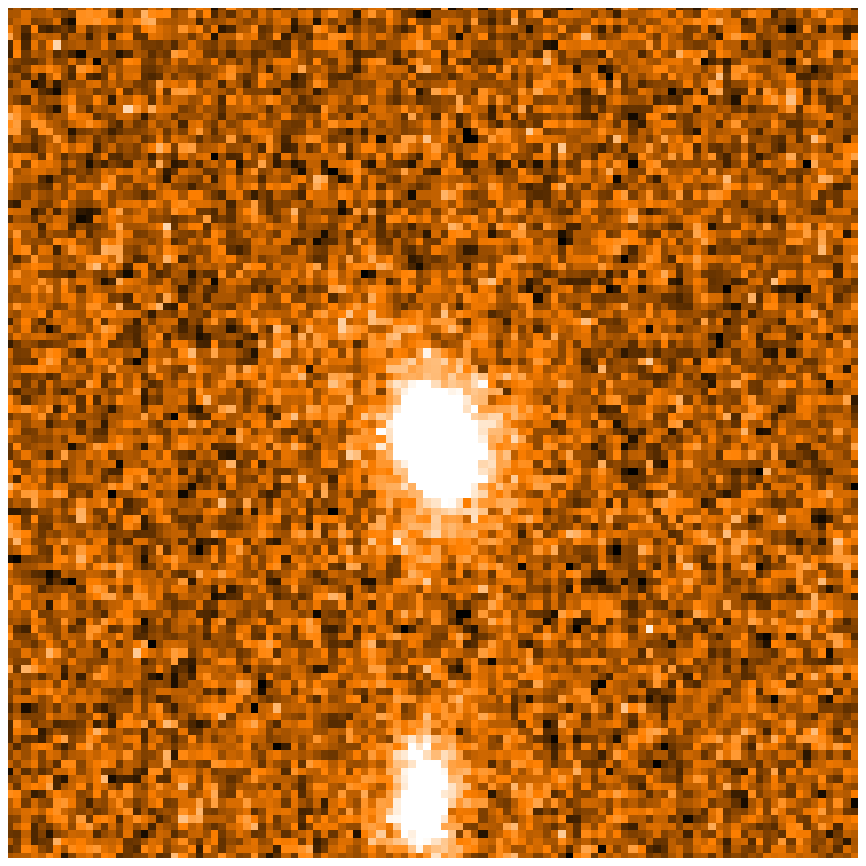}
\includegraphics[width=2.truecm]{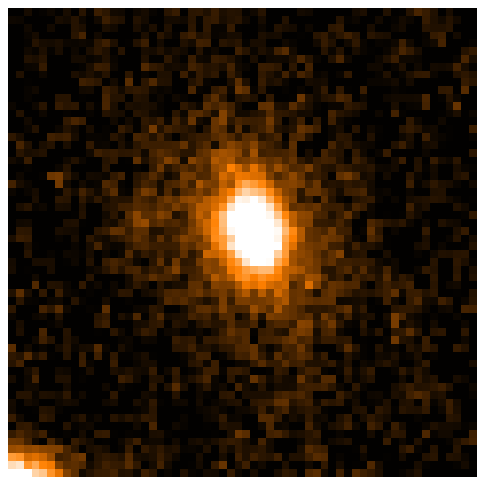}
\includegraphics[width=2.truecm]{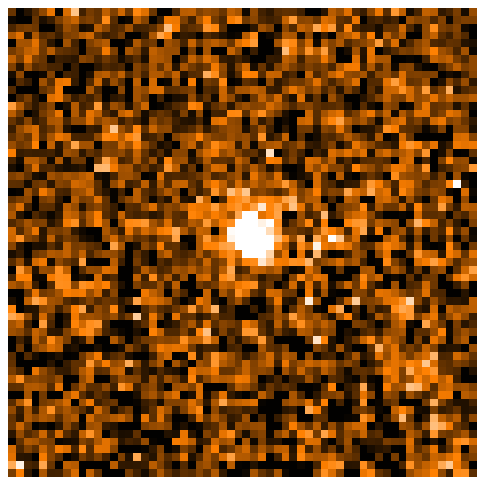}
\includegraphics[width=2.truecm]{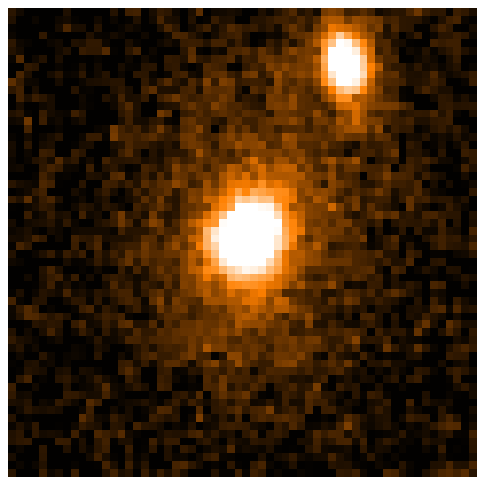}
\includegraphics[width=2.truecm]{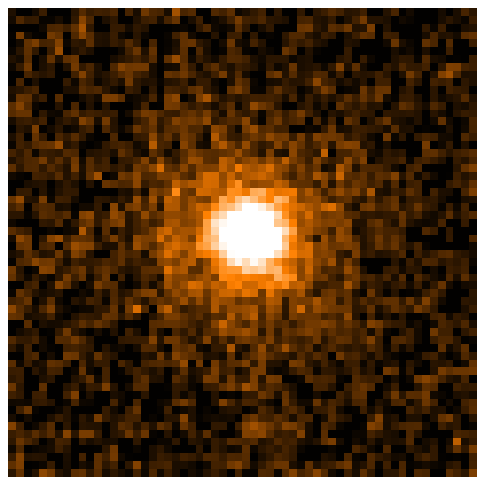}
\includegraphics[width=2.truecm]{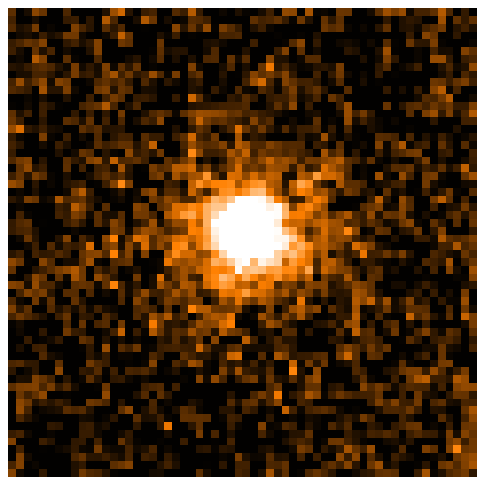}
\includegraphics[width=2.truecm]{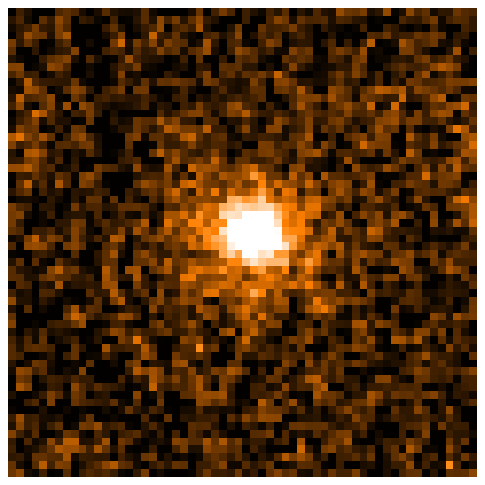}
\includegraphics[width=2.truecm]{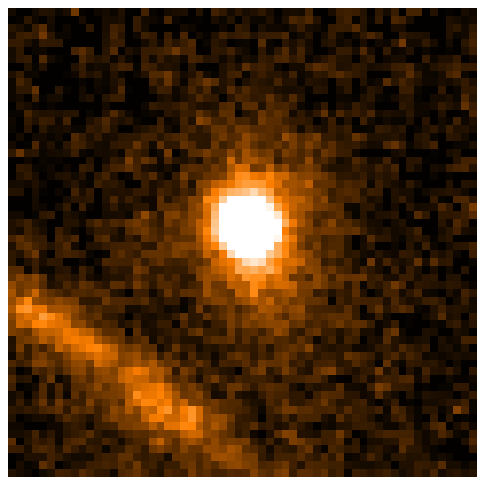}
\includegraphics[width=2.truecm]{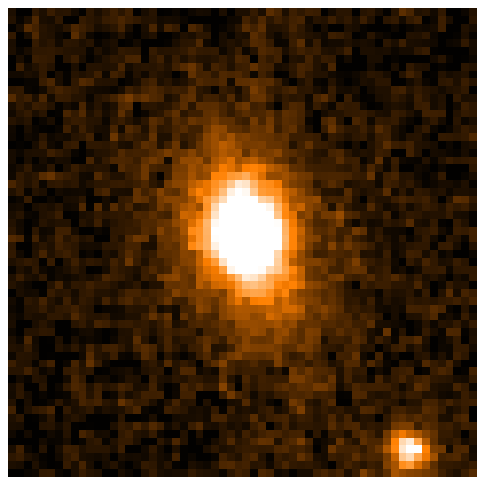}
\caption{F850LP band images for a representative sample of elliptical galaxies
in the field and cluster samples. Each row shows the 3$\times$3 arcsec width 
images for some elliptical galaxies in the
GOODS-S field (top row), in the XMMJ2235 cluster (second row), in the 
RDCSJ0848 cluster (third row), and in the XLSSJ0223 cluster (bottom row).
 }
\end{figure*}

In this appendix we report the structural and the physical parameters we
derived for the cluster sample (Table \ref{clsample})  and for the field 
sample (Table \ref{fisample}). 
Namely: Sersic index $n$, axial ratio $b/a$, apparent magnitude F850$_{fit}$ , and 
effective radius R$_e$ [kpc] as derived from the fitting to the surface brightness 
profile in the F850LP image; 
effective surface brightness $\langle \mu\rangle_e^B$,
age of the stellar population, and stellar mass $\mathcal{M}_*$ 
fitting (see \S 3); stellar mass within 1 kpc radius $\mathcal{M}_{1kpc}$ and
absolute magnitude M$_B$ in the B-band.
The error quoted for F850$_{fit}$ is the square root of the quadratic sum
of the photometric error with the Galfit error.{  The typical uncertainties (at one $\sigma$) affecting stellar masses and ages 
derived from the SED fitting  are 15\% and 20\% respectively 
(25\% for ages older than 3.5 Gyr).
These uncertainties take into account only photometric errors.}  
The term Ev$_{B}$ is the luminosity evolution  in the B band of the galaxy
from its redshift to $z=0$, according to its own age based on BC03 models.
In Fig. \ref{fig:thumb} we show the F850LP images for a representative sample of
elliptical galaxies in the field and cluster samples.

\begin{table*}
\caption{Morphological parameters of cluster elliptical galaxies.}
{\small
\centerline{
\begin{tabular}{rrrrrrcrrrrrr}
\hline
\hline
  ID &    RA      &        Dec   &$n_{850}$&$b/a$& F850$_{fit}$&R$_e^{F850}$ & $\langle\mu\rangle^B_e$&$Ev^B$ &age&
  log$\mathcal{M}_*$&log$\mathcal{M}_{1kpc}$&  M$_B$   \\
     &   [h:m:s]& [d:p:s]& &   &[mag]  & [kpc]              &[mag/arcsec$^2$]     & [mag]       & [Gyr]& [M$_\odot$]& [M$_\odot$]& [mag]\\
\hline
     &              &               &      &     &  XMM2235   & &      &     &    &    & &     \\ 
\hline
 358&    22:35:27.003&  -25:58:14.11& 6.0$\pm0.8$ &  0.6&  23.46$\pm$0.07&  1.5$\pm$0.2&  18.7$\pm$0.2&  3.11&  0.72&  10.08&    9.69&  -20.37\\ 
 595&    22:35:26.220&  -25:56:45.54& 5.6$\pm0.4$ &  0.9&  21.64$\pm$0.03&  3.5$\pm$0.5&  18.5$\pm$0.1&  1.47&  3.00&  11.56&   10.94&  -22.26\\ 
 684&    22:35:25.804&  -25:56:46.00& 6.0$\pm0.6$ &  0.6&  22.90$\pm$0.04&  1.4$\pm$0.2&  18.0$\pm$0.6&  2.06&  1.70&  10.56&   10.19&  -20.95\\ 
 692&    22:35:25.680&  -25:56:58.48& 4.1$\pm0.3$ &  0.7&  23.50$\pm$0.10&  1.5$\pm$0.2&  18.6$\pm$0.3&  2.40&  1.14&  10.56&   10.15&  -20.65\\ 
 837&    22:35:24.895&  -25:56:37.03& 6.0$\pm0.3$ &  0.7&  21.87$\pm$0.03&  7.8$\pm$0.1&  20.4$\pm$0.1&  1.63&  2.50&  11.47&   10.60&  -21.70\\ 
1284&    22:35:22.814&  -25:56:24.92& 4.3$\pm0.4$ &  0.9&  22.30$\pm$0.04&  2.4$\pm$0.3&  18.5$\pm$0.1&  2.65&  1.01&  11.05&   10.50&  -21.65\\ 
1539&    22:35:22.472&  -25:56:15.17& 3.5$\pm0.2$ &  0.6&  22.67$\pm$0.08&  1.5$\pm$0.2&  17.8$\pm$0.4&  2.02&  1.80&  11.01&   10.59&  -21.35\\ 
1740&    22:35:20.839&  -25:57:39.76& 3.3$\pm0.1$ &  0.6&  20.78$\pm$0.02& 12.8$\pm$2.0&  20.5$\pm$0.8&  1.36&  3.50&  11.93&   10.55&  -22.94\\ 
1747&    22:35:20.588&  -25:58:20.68& 4.8$\pm0.5$ &  0.6&  23.28$\pm$0.07&  1.7$\pm$0.2&  18.6$\pm$0.9&  2.65&  1.02&  10.55&   10.12&  -20.69\\ 
1758&    22:35:20.920&  -25:57:35.90& 2.9$\pm0.2$ &  0.8&  22.64$\pm$0.05&  2.5$\pm$0.3&  18.8$\pm$0.6&  2.02&  1.80&  11.14&   10.52&  -21.30\\ 
1782&    22:35:20.707&  -25:57:44.43& 3.6$\pm0.3$ &  0.6&  22.00$\pm$0.04&  3.4$\pm$0.5&  18.7$\pm$0.1&  1.95&  1.90&  11.57&   10.87&  -21.97\\ 
1790&    22:35:20.707&  -25:57:37.70& 4.4$\pm0.3$ &  0.6&  21.92$\pm$0.03&  2.4$\pm$0.3&  18.0$\pm$0.3&  1.21&  4.25&  11.60&   11.06&  -22.20\\ 
2054&    22:35:19.078&  -25:58:27.32& 4.4$\pm0.3$ &  0.7&  21.91$\pm$0.05&  4.2$\pm$0.6&  19.2$\pm$0.8&  2.06&  1.70&  11.22&   10.48&  -21.94\\ 
2147&    22:35:19.046&  -25:57:51.42& 5.3$\pm0.5$ &  0.7&  22.18$\pm$0.08&  5.3$\pm$0.8&  20.0$\pm$0.8&  2.40&  1.14&  11.01&   10.24&  -21.64\\ 
2166&    22:35:18.168&  -25:59:05.89& 3.0$\pm0.2$ &  0.6&  21.50$\pm$0.04&  1.4$\pm$0.1&  16.4$\pm$0.2&  2.71&  0.90&  10.99&   10.59&  -22.57\\ 
2429&    22:35:17.867&  -25:56:13.06& 2.1$\pm0.4$ &  0.6&  23.00$\pm$0.07&  0.9$\pm$0.1&  17.0$\pm$0.2&  3.11&  0.72&  10.58&   10.31&  -21.12\\ 
2809&    22:35:18.251&  -25:56:06.17& 3.2$\pm0.3$ &  0.6&  22.70$\pm$0.05&  1.4$\pm$0.2&  17.6$\pm$0.6&  3.25&  0.65&  10.73&   10.33&  -21.97\\ 
\hline
     &              &               &      &     &  RDCS0848   & &      &     &    &  &  &     \\ 
\hline
    1& 08:48:36.233&   44:53:55.42&  3.9$\pm0.3$& 0.7& 21.79$\pm$0.02&  7.8 $\pm$1.2 &  20.8$\pm$0.4 &  1.36&3.75 &11.31&  10.27 &-21.55 \\ 
    2& 08:48:36.160&   44:54:17.24&  6.3$\pm0.8$& 0.7& 21.08$\pm$0.02&  6.5 $\pm$0.7 &  19.8$\pm$0.3 &  2.35&1.43 &11.16&  10.36 &-22.24  \\ 
    3& 08:48:32.978&   44:53:46.61&  2.6$\pm0.4$& 0.5& 21.79$\pm$0.02&  1.3 $\pm$0.1 &  17.0$\pm$0.1 &  1.60&2.60 &11.01&  10.63 &-21.94 \\ 
    4& 08:48:35.978&   44:53:36.12&  4.4$\pm0.2$& 0.7& 20.38$\pm$0.01&  16.7$\pm$5.6 &  21.2$\pm$0.8 &  2.13&1.68 &11.25&   9.91 &-22.65 \\ 
    5& 08:48:32.434&   44:53:34.97&  3.6$\pm0.4$& 0.9& 22.68$\pm$0.04&  1.7 $\pm$0.2 &  18.6$\pm$0.3 &  2.35&1.43 &10.78&  10.33 &-21.49  \\
  606& 08:48:37.071&   44:53:33.99&  4.3$\pm0.4$& 0.6& 22.22$\pm$0.03&  2.4 $\pm$0.2 &  18.8$\pm$0.2 &  2.35&1.43 &10.70&  10.15 &-21.32 \\ 
  590& 08:48:34.069&   44:53:32.23&  2.8$\pm0.1$& 0.7& 22.44$\pm$0.03&  2.4 $\pm$0.1 &  19.0$\pm$0.2 &  1.92&2.30 &10.84&  10.23 &-21.18 \\ 
  568& 08:48:35.038&   44:53:30.83&  4.2$\pm0.3$& 0.4& 22.82$\pm$0.04&  1.1 $\pm$0.1 &  17.7$\pm$0.2 &  1.92&2.00 &10.57&  10.25 &-20.81 \\ 
  719& 08:48:33.031&   44:53:39.67&  6.0$\pm0.4$& 0.7& 22.73$\pm$0.04&  0.9 $\pm$0.1 &  17.2$\pm$0.2 &  1.42&3.50 &10.82&  10.54 &-20.67 \\ 
 1250& 08:48:37.341&   44:54:15.60&  2.2$\pm0.1$& 0.8& 23.20$\pm$0.06&  2.1 $\pm$0.3 &  19.6$\pm$0.2 &  2.32&1.28 &10.24&   9.65 &-20.38 \\ 
 1260& 08:48:36.160&   44:54:16.16&  3.9$\pm0.6$& 0.7& 23.75$\pm$0.07&  2.1 $\pm$0.4 &  20.3$\pm$0.6 &  2.98&0.71 & 9.70&   9.19 &-20.03 \\ 
  173& 08:48:34.058&   44:53:02.44&  3.2$\pm0.4$& 0.8& 23.63$\pm$0.08&  0.5 $\pm$0.1 &  16.9$\pm$0.4 &  1.21&4.25 &10.55&  10.40 &-20.02 \\ 
 1160& 08:48:32.768&   44:54:07.14&  4.6$\pm0.7$& 0.6& 22.54$\pm$0.05&  2.1 $\pm$0.3 &  18.9$\pm$0.4 &  1.58&3.00 &10.77&  10.27 &-20.89 \\ 
 657 & 08:48:32.442&   44:53:35.35&  2.4$\pm0.2$& 0.5& 22.12$\pm$0.04&  1.7 $\pm$0.2 &  18.0$\pm$0.3 &  2.35&1.43 &10.77&  10.28 &-21.52  \\
 626 & 08:48:32.390&   44:53:35.03&  4.2$\pm0.2$& 0.6& 21.45$\pm$0.02&  2.1 $\pm$0.3 &  17.8$\pm$0.3 &  2.35&1.43 &10.76&  10.25 &-21.48  \\
 471 & 08:48:29.685&   44:53:23.91&  4.6$\pm0.4$& 0.9& 23.33$\pm$0.07&  2.3 $\pm$0.3 &  19.8$\pm$0.4 &  2.13&1.61 &10.11&   9.59 &-20.04 \\
\hline
     &              &               &       &    &  XLSS0223   & &      &     &    &  &  &     \\ 
\hline
 406&    02:23:04.918&  -04:34:36.31&   2.4$\pm0.4$&  0.8 &   23.68$\pm$0.07&  1.8$\pm$0.3 & 19.5$\pm$0.4&  1.14&  4.50 &   10.93&   10.42&  -20.34 \\  
 537&    02:23:08.485&  -04:37:18.07&   3.3$\pm0.4$&  0.8 &   22.91$\pm$0.06&  1.8$\pm$0.3 & 18.8$\pm$0.3&  1.74&  3.25 &   10.96&   10.48&  -21.07 \\  
 651&    02:23:05.759&  -04:36:10.27&   2.9$\pm0.1$&  0.6 &   21.84$\pm$0.03&  3.5$\pm$0.2 & 19.4$\pm$0.1&  2.39&  1.70 &   10.94&   10.18&  -21.95 \\  
 962&    02:23:05.420&  -04:36:36.26&   3.8$\pm0.3$&  0.7 &   22.24$\pm$0.08&  4.2$\pm$0.5 & 20.3$\pm$0.3&  2.19&  2.50 &   10.87&   10.10&  -21.33 \\  
 972&    02:23:04.718&  -04:36:13.47&   8.9$\pm1.5$&  0.4 &   22.32$\pm$0.15&  4.1$\pm$1.5 & 20.6$\pm$0.8&  2.32&  1.28 &   10.54&    9.95&  -21.00 \\  
 983&    02:23:04.843&  -04:36:19.87&   3.3$\pm0.3$&  0.3 &   23.20$\pm$0.02&  0.6$\pm$0.2 & 16.5$\pm$0.9&  2.39&  1.70 &   10.48&   10.30&  -20.73 \\  
 994&    02:23:04.327&  -04:36:01.51&   3.9$\pm0.3$&  0.6 &   24.32$\pm$0.09&  0.5$\pm$0.6 & 17.6$\pm$2.6&  2.32&  1.28 &   10.01&    9.85&  -19.62 \\  
1090&    02:23:04.128&  -04:36:20.30&   3.4$\pm0.3$&  0.7 &   23.62$\pm$0.07&  1.3$\pm$0.3 & 18.9$\pm$0.5&  1.87&  3.00 &   10.53&   10.16&  -20.27 \\  
1101&    02:23:04.097&  -04:36:22.50&   3.5$\pm0.6$&  0.5 &   23.47$\pm$0.04&  0.8$\pm$0.2 & 17.5$\pm$0.7&  2.50&  1.14 &   10.21&    9.96&  -20.44 \\  
1142&    02:23:03.262&  -04:36:14.60&   5.4$\pm0.3$&  0.7 &   20.89$\pm$0.07&  9.1$\pm$1.1 & 21.0$\pm$0.3&  1.29&  4.50 &   11.50&   10.53&  -22.38 \\  
1144&    02:23:03.253&  -04:36:07.87&   2.3$\pm0.1$&  0.4 &   22.47$\pm$0.02&  1.3$\pm$0.1 & 17.8$\pm$0.1&  2.32&  1.28 &   10.71&   10.32&  -21.47 \\  
1151&    02:23:03.687&  -04:36:23.34&   3.8$\pm0.7$&  0.7 &   23.77$\pm$0.05&  0.8$\pm$0.3 & 17.8$\pm$0.8&  2.32&  1.28 &   10.24&    9.99&  -20.20 \\  
1171&    02:23:03.107&  -04:36:10.93&   5.0$\pm0.6$&  0.9 &   22.56$\pm$0.08&  2.5$\pm$0.4 & 19.4$\pm$0.4&  1.62&  3.50 &   10.89&   10.35&  -21.17 \\  
1175&    02:23:03.242&  -04:36:18.50&   3.0$\pm0.3$&  0.9 &   22.25$\pm$0.06&  3.0$\pm$0.3 & 19.2$\pm$0.2&  1.47&  3.25 &   11.02&   10.33&  -21.76 \\  
1184&    02:23:02.956&  -04:36:09.74&   3.2$\pm0.3$&  0.9 &   22.72$\pm$0.06&  2.2$\pm$0.3 & 19.1$\pm$0.3&  2.31&  1.80 &   10.70&   10.15&  -21.12 \\  
1188&    02:23:03.037&  -04:36:12.36&   1.6$\pm0.1$&  0.4 &   22.74$\pm$0.03&  0.7$\pm$0.2 & 17.1$\pm$0.8&  1.52&  3.75 &   10.77&   10.59&  -20.55 \\  
1199&    02:23:02.922&  -04:36:14.55&   4.9$\pm0.5$&  0.9 &   22.70$\pm$0.09&  3.4$\pm$0.6 & 20.4$\pm$0.4&  2.32&  1.28 &   10.53&    9.89&  -20.88 \\  
1263&    02:23:03.291&  -04:36:54.59&   5.0$\pm0.4$&  0.6 &   21.63$\pm$0.12&  9.1$\pm$1.8 & 21.7$\pm$0.4&  1.77&  2.50 &   11.05&   10.05&  -21.68 \\  
1302&    02:23:00.882&  -04:35:39.85&   2.9$\pm0.2$&  0.8 &   23.57$\pm$0.04&  1.0$\pm$0.2 & 18.0$\pm$0.5&  1.10&  4.75 &   10.87&   10.57&  -20.44 \\  
1370&    02:23:02.021&  -04:36:43.26&   3.8$\pm0.3$&  0.8 &   23.49$\pm$0.05&  0.8$\pm$0.2 & 17.7$\pm$0.6&  2.42&  1.61 &   10.08&    9.83&  -20.43 \\  
1448&    02:22:58.869&  -04:36:49.89&   6.5$\pm0.9$&  0.9 &   24.12$\pm$0.07&  1.0$\pm$0.6 & 19.2$\pm$1.2&  3.79&  0.51 &    9.47&    9.17&   -19.47 \\  
1630&    02:23:00.929&  -04:36:50.19&   5.3$\pm0.3$&  0.7 &   21.59$\pm$0.04&  3.3$\pm$0.3 & 19.2$\pm$0.2&  2.90&  1.43 &   10.75&   10.14&  -22.01 \\  
1711&    02:22:59.990&  -04:36:02.53&   6.0$\pm0.2$&  0.5 &   20.83$\pm$0.03&  4.7$\pm$0.3 & 19.3$\pm$0.2&  2.11&  1.68 &   11.20&   10.50&  -22.67 \\  
\hline
\end{tabular}
}}
%\tablefoot{Sersic index $n$, axial ratio $b/a$, apparent magnitude and 
%effective radius [kpc] as derived from the fitting to the surface brightness 
%profile in the F850LP image. 
%The error quoted for F850$_{fit}$ is the square root of the quadratic sum
%of the photometric error with the Galfit error.
%The term Ev$_{B}$ is the luminosity evolution  in the B band of the galaxy
%from its redshift to $z=0$, according to its own age based on BC03 models. 
%}
\label{clsample}
\end{table*}

\begin{table*}
\caption{Morphological parameters of elliptical galaxies in the field.}
{\small
\centerline{
\begin{tabular}{rrrrrrcrrrrrr}
\hline
\hline
  ID &    RA      &        Dec   &$n_{850}$&$b/a$& F850$_{fit}$&R$_e^{F850}$ & $\langle\mu\rangle^B_e$&$Ev^B$ &age&
  log$\mathcal{M}_*$&log$\mathcal{M}_{1kpc}$& M$_B$   \\
     &   [h:m:s]& [d:p:s]& &   &[mag]  & [kpc]              &[mag/arcsec$^2$]     & [mag]       & [Gyr]& [M$_\odot$]& [M$_\odot$]& [mag]\\
\hline
     &              &               &          &         &  GOODS-South   & &      &     &    &  &  &     \\ 
\hline
 4887 & 03:32:31.34 &  27:51:32.48  &    3.3$\pm0.3$ &  0.8 &  23.63$\pm$0.03 &  1.7$\pm$0.5  &   19.4$\pm$0.6  &  1.37 &   3.25 & 10.56 & 10.10 &  -20.02  \\   
14057 & 03:32:37.95 &  27:44:04.20  &    2.8$\pm0.3$ &  0.7 &  23.48$\pm$0.02 &  1.0$\pm$0.2  &   18.1$\pm$0.5  &  2.18 &   1.28 & 10.21 &  9.91 &  -20.32  \\   
17158 & 03:32:11.26 &  27:41:27.01  &    2.4$\pm0.2$ &  0.8 &  21.99$\pm$0.01 &  1.2$\pm$0.1  &   17.1$\pm$0.1  &  4.02 &   0.27 & 10.28 &  9.92 &  -21.86  \\   
12294 & 03:32:11.21 &  27:45:33.45  &    1.7$\pm0.2$ &  0.5 &  22.05$\pm$0.01 &  1.2$\pm$0.1  &   16.9$\pm$0.1  &  2.65 &   0.90 & 10.74 & 10.23 &  -21.83  \\   
14953 & 03:32:25.98 &  27:43:18.93  &    4.3$\pm0.4$ &  0.6 &  22.61$\pm$0.01 &  1.4$\pm$0.2  &   18.0$\pm$0.3  &   2.2 &   1.43 & 10.49 & 10.10 &  -21.06  \\   
12789 & 03:32:29.82 &  27:45:10.78  &    4.3$\pm0.5$ &  0.8 &  23.07$\pm$0.02 &  2.1$\pm$0.4  &   19.1$\pm$0.5  &  1.83 &   2.00 & 10.55 & 10.05 &  -20.60  \\   
13493 & 03:32:38.11 &  27:44:32.60  &    4.3$\pm0.3$ &  0.9 &  21.85$\pm$0.01 &  3.6$\pm$0.4  &   19.1$\pm$0.2  &  2.35 &   1.14 & 10.74 & 10.05 &  -22.05  \\   
12000 & 03:32:26.27 &  27:45:50.71  &    7.3$\pm1.2$ &  0.8 &  22.23$\pm$0.02 &  4.5$\pm$0.8  &   20.0$\pm$0.4  &   2.2 &   1.28 & 10.60 &  9.95 &  -21.24  \\   
 9702 & 03:32:35.79 &  27:47:34.77  &    4.2$\pm0.5$ &  0.9 &  23.42$\pm$0.03 &  1.2$\pm$0.3  &   18.4$\pm$0.6  &  1.83 &   2.00 & 10.41 & 10.06 &  -20.28  \\   
 3453 & 03:32:47.56 &  27:52:43.23  &    4.8$\pm0.2$ &  0.8 &  21.25$\pm$0.01 &  1.2$\pm$0.1  &   16.3$\pm$0.1  &  4.25 &   0.20 &  9.97 &  9.63 &  -22.12  \\   
 2907 & 03:32:50.22 &  27:53:12.26  &    5.4$\pm0.6$ &  0.6 &  23.44$\pm$0.02 &  1.1$\pm$0.2  &   17.9$\pm$0.5  &  1.76 &   2.20 & 10.51 & 10.19 &  -20.48  \\   
 4981 & 03:32:44.27 &  27:51:26.74  &    5.6$\pm0.4$ &  0.5 &  22.45$\pm$0.01 &  2.0$\pm$0.3  &   18.3$\pm$0.3  &  1.69 &   2.30 & 10.74 & 10.28 &  -21.37  \\   
 6791 & 03:32:50.19 &  27:50:01.04  &    3.5$\pm0.4$ &  0.6 &  23.16$\pm$0.02 &  0.4$\pm$0.1  &   15.5$\pm$0.4  &  2.81 &   0.81 & 10.24 & 10.12 &  -20.74  \\   
 9369 & 03:32:16.02 &  27:47:50.00  &    6.1$\pm0.8$ &  0.7 &  22.54$\pm$0.02 &  3.4$\pm$0.6  &   19.4$\pm$0.4  &  1.86 &   2.00 & 10.85 & 10.25 &  -21.50  \\   
11960 & 03:32:05.26 &  27:45:52.40  &    3.7$\pm0.2$ &  0.8 &  23.27$\pm$0.02 &  1.9$\pm$0.4  &   18.8$\pm$0.5  &  2.23 &   1.28 & 10.52 & 10.03 &  -20.83  \\   
12623 & 03:32:16.94 &  27:45:19.36  &    4.5$\pm0.4$ &  0.8 &  23.40$\pm$0.02 &  1.3$\pm$0.3  &   18.2$\pm$0.5  &  2.68 &   0.90 & 10.34 &  9.97 &  -20.64  \\   
11383 & 03:32:24.80 &  27:46:17.91  &    5.7$\pm0.6$ &  0.7 &  22.70$\pm$0.02 &  2.2$\pm$0.3  &   18.5$\pm$0.3  &  2.05 &   1.61 & 10.91 & 10.42 &  -21.41  \\   
12737 & 03:32:26.36 &  27:45:14.09  &    4.6$\pm0.4$ &  0.6 &  23.43$\pm$0.03 &  1.3$\pm$0.3  &   18.3$\pm$0.6  &  2.05 &   1.61 & 10.33 &  9.97 &  -20.47  \\   
10231 & 03:32:39.64 &  27:47:09.11  &    6.8$\pm0.4$ &  1.0 &  21.71$\pm$0.02 &  1.3$\pm$0.2  &   16.5$\pm$0.4  &  2.05 &   1.61 & 10.88 & 10.53 &  -21.64  \\   
17506 & 03:32:20.09 &  27:41:6.753  &    5.8$\pm0.4$ &  0.6 &  21.97$\pm$0.01 &  3.6$\pm$0.4  &   18.8$\pm$0.3  &  2.23 &   1.28 & 11.06 & 10.43 &  -22.25  \\   
17022 & 03:32:14.65 &  27:41:36.62  &    4.4$\pm0.3$ &  0.9 &  23.03$\pm$0.02 &  1.6$\pm$0.3  &   17.9$\pm$0.4  &  2.23 &   1.28 & 10.73 & 10.31 &  -21.37  \\   
 6989 & 03:32:46.11 &  27:49:53.47  &    4.2$\pm0.4$ &  0.6 &  23.65$\pm$0.02 &  0.8$\pm$0.2  &   17.1$\pm$0.5  &  1.72 &   2.30 & 10.41 & 10.16 &  -20.69  \\   
 4651 & 03:32:42.16 &  27:51:44.32  &    2.2$\pm0.1$ &  0.9 &  22.85$\pm$0.01 &  1.7$\pm$0.2  &   18.2$\pm$0.3  &  3.11 &   0.72 & 10.11 &  9.61 &  -21.33  \\   
14220 & 03:32:7.552 &  27:43:56.64  &    2.6$\pm0.2$ &  0.8 &  22.72$\pm$0.03 &  3.1$\pm$0.8  &   19.2$\pm$0.6  &  2.71 &   0.90 & 10.66 &  9.92 &  -21.62  \\   
 2659 & 03:32:23.90 &  27:53:26.22  &    4.1$\pm0.4$ &  0.8 &  23.50$\pm$0.03 &  2.1$\pm$0.7  &   18.9$\pm$0.7  &  1.63 &   2.50 & 10.60 & 10.09 &  -20.96  \\   
12505 & 03:32:6.812 &  27:45:24.35  &    3.4$\pm0.3$ &  0.5 &  23.10$\pm$0.02 &  1.1$\pm$0.2  &   17.3$\pm$0.3  &  1.78 &   2.20 & 10.72 & 10.39 &  -21.27  \\   
 8849 & 03:32:10.73 &  27:48:19.37  &    3.4$\pm0.6$ &  0.7 &  24.07$\pm$0.04 &  1.0$\pm$0.4  &   17.8$\pm$0.8  &  1.47 &   3.00 & 10.81 & 10.51 &  -20.63  \\   
12797 & 03:32:44.65 &  27:45:10.52  &    5.6$\pm0.4$ &  0.6 &  23.13$\pm$0.02 &  0.7$\pm$0.1  &   16.2$\pm$0.3  &  2.71 &   0.90 & 10.39 & 10.16 &  -21.38  \\   
16726 & 03:32:41.63 &  27:41:51.43  &    7.3$\pm0.9$ &  0.8 &  21.74$\pm$0.02 & 10.8$\pm$2.1  &   20.6$\pm$0.4  &  1.95 &   1.90 & 10.99 & 10.07 &  -22.19  \\   
16103 & 03:32:31.05 &  27:42:26.45  &    3.5$\pm0.2$ &  0.7 &  23.49$\pm$0.02 &  1.4$\pm$0.3  &   17.9$\pm$0.4  &  2.29 &   1.43 & 10.54 & 10.15 &  -21.21  \\   
 7616 & 03:32:44.67 &  27:49:24.21  &    2.4$\pm0.6$ &  0.6 &  24.00$\pm$0.03 &  1.0$\pm$0.3  &   17.5$\pm$0.6  &  1.81 &   2.20 & 10.66 & 10.36 &  -21.24  \\     
\hline
\end{tabular}
}}
%\tablefoot{Sersic index $n$, axial ratio $b/a$, apparent magnitude and 
%effective radius [kpc] as derived from the fitting to the surface brightness 
%profile in the F850LP image.   
%The term Ev$_{B}$ is the luminosity evolution in the B band of the galaxy
%from its redshift to $z=0$, according to its own age based on BC03 models. 
%}
\label{fisample}
\end{table*}

\section{Scaling relations: least square and orthogonal fitting}
\label{ap:scaling}
\begin{table*}
\begin{center}
\caption{Scaling relations of elliptical galaxies at $z\sim1.3$}
\begin{tabular}{llllll}
\hline
\hline
 Relation   &\hfill Least & Square fit &\hfill Orthogonal & fit & Notes\\ 
\hline
 $\langle\mu\rangle_e^B=\beta {\rm log}(R_e)+\alpha $ & $\beta=3.2(\pm0.2)$ & $\alpha=17.6(\pm0.1)$ & $\beta=4.4(\pm0.4)$ & $\alpha=17.3(\pm0.2)$ & Sec. 5 \\
 \\
 $R_e=b\times\mathcal{M}_*^a$     & $a=+0.50(\pm0.06)$ &log$(b)=-5.0(\pm0.7)$ & $a=0.67\pm0.08$ & log$(b)=-6.9\pm0.9$ & Sec. 6\\   
                                            & $a=+0.64(\pm0.09)$ &log$(b)=-6.6(\pm1.0)$ & $a=+1.0\pm0.1$ &log$(b)=-10.7\pm1.0$ & ($\mathcal{M}_*>m_t$)  \\
                                            & $a=-0.13(\pm0.2)$& log$(b)=+1.4(\pm2.0)$  & $a=-0.3\pm0.2$ &log$(b)=+3.4\pm2.0$  & ($\mathcal{M}_*<m_t$)  \\
\\
 $\Sigma_{Re}=b\times R_e^a$ & $a=-1.2(\pm0.1)$& log$(b)=+3.67(\pm0.05)$ &$a=-1.7(\pm0.2)$ &log$(b)=+3.83(\pm0.08)$ & Sec. 7.1 \\
\\
 $\Sigma_{1kpc}=b\times\mathcal{M}_*^a$ & $a=0.64(\pm0.06)$ & log$(b)=-3.2(\pm0.6)$ & $a=0.73(\pm0.08)$& log$(b)=-4.1(\pm0.9)$ & Sec. 7.2 \\
                                        & $a=0.50(\pm0.07)$ & log$(b)=-1.5(\pm0.8)$ & $a=0.65(\pm0.09)$& log$(b)=-3.2(\pm1.0)$ &($\mathcal{M}_*>m_t$) \\
                                        & $a=1.07(\pm0.10)$ & log$(b)=-7.7(\pm1.3)$ & $a=1.20(\pm0.30)$& log$(b)=-9.1(\pm2.9)$ &($\mathcal{M}_*<m_t$) \\
\hline
\end{tabular}
\tablefoot{$m_t\simeq3\times10^{10}$ M$_\odot$ is the transition mass (see Sec. 6).}
\end{center}
\label{relations}
\end{table*}

For the sake of completeness, in this appendix we summarize the best fitting 
relations we obtained with the classical least squares method, which minimizes 
the sum of the square of the residuals in the dependent variable Y from the line,
and those obtained using the Person's orthogonal regression line, which 
treats the variables symmetrically, minimizing the squares of the perpendicular 
distances from the point to the line.
The best fitting parameters are listed in Table \ref{relations} where,
for each relation indicated in the first column, we report the values obtained 
with the least square fitting (second and third columns) and those obtained with
the orthogonal fitting (fourth and fifth columns). 
In the last column we report the section in which the relations are 
derived and discussed and, where appropriate, the range of fitting.

The two methods provide  slopes
that differ at $\sim2\sigma$ for the Kormendy relation and the 
$\Sigma_{R_e}$-R$_e$ relation, while for the size-mass and the 
$\Sigma_{1kpc}$-$\mathcal{M}_*$ the slopes  are consistent within $1\sigma$.
It is worth noting that, in the presence of boundary conditions along the Y axis,
as in the case of the fitting at masses larger or lower than $m_t$, the orthogonal 
fit is biased toward steeper slopes.
This can be easily seen in Fig. 7 where the data close
to the boundary affect asymmetrically the orthogonal regression lines in the 
two mass ranges.
Boundary conditions should be applied orthogonally to the fit or treated 
through iterative routines, procedures that go beyond the scope of this work. 
Our aim is to show the presence of two different regimes in the size-mass
relation and in the $\Sigma_{1kpc}$-$\mathcal{M}_*$ relation at masses
lower and higher than a mass $\mathcal{M}_*\simeq2-3\times 10^{10}$ M$_\odot$.   
It is important to note that, for the reasons above, the true slope of the relations 
will lie in between the one provided by the least square method, that can be considered as
a lower limit, and the one provided by the orthogonal fit, that can be
considered as an upper limit.

\end{appendix}

\end{document}